\documentclass[fleqn,usenatbib]{mnras}

\usepackage[T1]{fontenc}

\DeclareRobustCommand{\VAN}[3]{#2}
\let\VANthebibliography\thebibliography
\def\thebibliography{\DeclareRobustCommand{\VAN}[3]{##3}\VANthebibliography}

\usepackage{graphicx}	% Including figure files
\usepackage{amsmath}	% Advanced maths commands
\usepackage{verbatim}
\usepackage{multirow}
\usepackage{multicol}
\usepackage{newtxtext}
\usepackage[varvw]{newtxmath}

\usepackage{comment}
\usepackage{xspace}
\usepackage{enumitem}
\usepackage{graphicx}
\usepackage{rotating}
\usepackage{color}
\usepackage{tabularx}
\usepackage{float}
\usepackage{svg}

\newcommand{\HbtoHa}        {\relax\ifmmode{{ \mathcal{N}}_{\rm H\alpha \to H\beta}\xspace} \else {$\mathcal{C}_{\rm H\alpha\to H\beta}$}\expandafter\xspace\fi}

\newcommand{\hii}{\mbox{H\,{\sc ii}}\xspace}
\newcommand{\hi}{\mbox{H\,{\sc i}}\xspace}
\newcommand{\lya}        {Ly$\alpha$\xspace}
\newcommand{\Ha}        {H$\alpha$\xspace}
\newcommand{\Hb}        {H$\beta$\xspace}
\newcommand{\NHI}        {\relax\ifmmode{{ N}_{\rm HI}\xspace} \else {${ N}_{\rm HI}$}\expandafter\xspace\fi}
\newcommand{\NHItwo}        {\relax\ifmmode{{ N}_{\rm HI, 2s}\xspace} \else {${ N}_{\rm HI, 2s}$}\expandafter\xspace\fi}
\newcommand{\NHIone}        {\relax\ifmmode{{ N}_{\rm HI, 1s}\xspace} \else {${ N}_{\rm HI, 1s}$}\expandafter\xspace\fi}
\newcommand{\kms}        {\ifmmode{\rm \,km\,s^{-1}}\else\,km\,s$^{-1}$\xspace\fi}
\newcommand{\unitNHI}    {\ifmmode{\rm \,cm^{-2}}\else\,cm$^{-2}$\xspace\fi}  
\newcommand{\vexp}       {\relax\ifmmode {v_{\rm exp}} \else {$v_{\rm exp}$}\expandafter\xspace\fi}
\newcommand{\vran}       {\relax\ifmmode {v_{\rm ran}} \else {$v_{\rm ran}$}\expandafter\xspace\fi}
\newcommand{\vth}       {\relax\ifmmode {v_{\rm th}} \else {$v_{\rm th}$}\expandafter\xspace\fi}
\newcommand{\sigran}     {\relax\ifmmode {\sigma_{\rm Ran}} \else {$\sigma_{\rm Ran}$}\expandafter\xspace\fi}
\newcommand{\sigr}     {\relax\ifmmode {\sigma_{\rm R}} \else {$\sigma_{\rm R}$}\expandafter\xspace\fi}
\newcommand{\sigsrc}     {\relax\ifmmode {\sigma_{\rm Src}} \else {$\sigma_{\rm Src}$}\expandafter\xspace\fi}
\newcommand{\massbh}     {\relax\ifmmode {M_{\rm BH}} \else {$M_{\rm BH}$}\expandafter\xspace\fi}

\newcommand{\tauth}     {\relax\ifmmode { \tau_{\rm Th} } \else {$ \tau_{\rm Th}$}\expandafter\xspace\fi}

\newcommand{\loglevel}     {\relax\ifmmode { \log N_{\rm 2s}/N_{\rm 1s} } \else {$\log N_{\rm 2s}/N_{\rm 1s} $}\expandafter\xspace\fi}
\newcommand{\level}     {\relax\ifmmode { N_{\rm 2s}/N_{\rm 1s} } \else {$ N_{\rm 2s}/N_{\rm 1s} $}\expandafter\xspace\fi}

\title[Radiative Transfer in LRDs]{
Impact of Resonance, Raman, and Thomson Scattering on Hydrogen Line Formation in Little Red Dots
}

\author[Chang et al.]{
Seok-Jun Chang,$^{1}$\thanks{E-mail: sjchang@mpa-garching.mpg.de}
Max Gronke,$^{2,1}$
Jorryt Matthee$^{3}$
and Charlotte Mason$^{4,5}$
\\
$^{1}$ Max-Planck-Institut f\"{u}r Astrophysik, Karl-Schwarzschild-Stra$\beta$e 1, 85748 Garching b. M\"{u}nchen, Germany \\
$^{2}$ Astronomisches Rechen-Institut, Zentrum f\"{u}r Astronomie, Universit\"{a}t Heidelberg, Mönchhofstra$\beta$e 12-14, 69120 Heidelberg, Germany\\
$^{3}$ Institute of Science and Technology Austria (ISTA), Am Campus 1, 3400 Klosterneuburg, Austria \\
$^{4}$ Cosmic Dawn Center (DAWN) \\
$^{5}$ Niels Bohr Institute, University of Copenhagen, Jagtvej 128, DK-2200, Copenhagen N, Denmark \\
}

\date{Accepted XXX. Received YYY; in original form ZZZ}

\pubyear{\the\year{}}

\begin{document}
\label{firstpage}
\pagerange{\pageref{firstpage}--\pageref{lastpage}}
\maketitle

% Abstract of the paper
\begin{abstract}
Little Red Dots (LRDs) are compact sources at $z>5$ discovered through JWST spectroscopy. Their spectra exhibit broad Balmer emission lines ($\gtrsim1000\rm~km~s^{-1}$), alongside absorption features and a pronounced Balmer break -- evidence for a dense, neutral hydrogen medium, in which the $n = 2$ state is significantly populated. When interpreted as arising from AGN broad-line regions, inferred black hole masses from local scaling relations exceed expectations given their stellar masses, challenging models of early black hole--galaxy co-evolution. However, radiative transfer effects in dense media may also impact the formation of hydrogen emission lines.
We model three scattering processes shaping hydrogen line profiles: resonance scattering by hydrogen in the $n=2$ state, Raman scattering of UV radiation by ground-state hydrogen, and Thomson scattering by free electrons. Using 3D Monte Carlo radiative transfer simulations, we examine their imprint on line shapes and ratios.
Resonance scattering produces strong deviations from Case B flux ratios, clear differences between H$\alpha$ and H$\beta$, and encodes gas kinematics in line profiles but cannot broaden H$\beta$ due to conversion to Pa$\alpha$. While Raman scattering can yield broad wings, scattering of the UV continuum is disfavored given the absence of strong FWHM variations across transitions. Raman scattering of higher Lyman-series emission can produce H$\alpha$/H$\beta$ wing width ratios of $\gtrsim1.28$, agreeing with observations.
Thomson scattering can reproduce the observed $\gtrsim1000~\rm km\, s^{-1}$ wings under plausible conditions -- e.g., $T_{\rm e} \sim 10^4$~K and $N_{\rm e}\sim10^{24}\rm~cm^{-2}$ -- and lead to black hole mass overestimates by factors $\gtrsim10$. Our results provide a framework for interpreting hydrogen lines in LRDs and similar systems.
\end{abstract}

% Select between one and six entries from the list of approved keywords.
% Don't make up new ones.

%%%%%%%%%%%%%%%%%%%%%%%%%%%%%%%%%%%%%%%%%%%%%%%%%%

%%%%%%%%%%%%%%%%% BODY OF PAPER %%%%%%%%%%%%%%%%%%
\section{Introduction}

Little Red Dots (LRDs) are compact sources in the early universe at $z > 5$ identified by JWST spectroscopy \citep{Harikane23,Kocevski23,Greene24,Matthee2024,Akins24}. They exhibit broad emission features with widths exceeding 1000~\kms in hydrogen and helium lines \citep{Juodzbalis2025,Brazzini2025,Rusakov2025,Torralba2025, DEugenio2025}, typically interpreted as signatures of active galactic nuclei (AGN). Many LRDs show a ``v-shaped'' spectrum with a blue UV continuum and a red UV to optical slope \citep[e.g.][]{Akins24,Ji2025}, often a Balmer break \citep{BWang24,Setton24}.  While LRDs often lack traditional AGN indicators, such as strong X-ray \citep{Maiolino24,Sacchi25}, radio emission \citep{Gloudemans25}, or an infrared excess \citep{Williams23,Xiao25}, other faint indicators such as Fe~II emission \citep{Lambrides2024} or high ionization emission lines (\citealp{Tang2025}, cf. \citealp{Lambrides2024}) have been identified. This rather unusual combination of spectral properties challenges standard AGN interpretations and the applicability of standard calibrations of bolometric luminosity.

LRDs appear to host unexpectedly massive supermassive black holes (SMBHs) \citep{Pacucci2023,Inayoshi2025b}. The broad emission features in \Ha and \Hb allow estimates of black hole mass \massbh using single epoch virial methods \citep{Greene2005,Reines2015,Woo2015}, which are commonly applied to nearby AGNs. Using this approach, recent studies \citep{Juodzbalis2024a, Juodzbalis2025, Inayoshi2025, Taylor2025,Taylor2025b} have estimated $\massbh \sim 10^{8-9}~M_{\odot}$ for LRDs. However, it remains unclear whether Balmer line widths are reliable tracers of \massbh in these systems (see also lower \massbh by X-ray \citealp{Ananna2024}). \citet{Rusakov2025} suggests that the broad wings are shaped by electron (Thomson) scattering, based on their observed exponential profiles, and derive SMBH masses that are $\sim 10-100$ times smaller when using Balmer line luminosities instead of widths. Additionally, \citet{Naidu2025} proposes that resonance scattering of \Hb photons in a dense neutral medium contributes to the line broadening, offering an alternative explanation to a virialized broad-line region.
Thus, while potentially crucial radiative transfer effects in LRDs are discussed in the literature, a systematic theoretical study investigating those effects is still missing.

Spectra of LRDs frequently exhibit not only broad Balmer emission lines but also complex profiles, including absorption features, P-Cygni-like shapes in the \Ha and \Hb lines, and significant Balmer breaks. These features point to the presence of gas that is optically thick to Balmer transitions, likely due to an abundant population of hydrogen atoms in the $n = 2$ state \citep{Matthee2024, Juodzbalis2024a,Ji2025, Maiolino2025,Naidu2025,DEugenio2025a, DEugenio2025b,Inayoshi2025,deGraaff2025}. Interestingly, \Ha and \Hb often show distinct line profiles: \Ha frequently shows a P-Cygni-like shape with blueshifted absorption and a redshifted emission peak, while \Hb tends to exhibit deeper absorption and a less prominent red peak. In contrast, Paschen lines typically lack absorption features \citep{Juodzbalis2025,Brazzini2025}. These observational trends suggest that radiative transfer effects (i.e., resonance scattering of Balmer lines) may play a key role in shaping the emergent line profiles.

While resonance scattering is typically invoked to explain the frequency and spatial diffusion of Lyman-$\alpha$ (Ly$\alpha$) in neutral hydrogen environments \citep{Neufeld1990,Dijkstra2019}, similar processes can occur for Balmer lines in environments like LRDs, where a significant $n = 2$ population exists. 
Thus, modeling the resonance scattering of Balmer photons is crucial for understanding hydrogen line formation in these systems.
However, importantly, the atomic physics of Balmer resonance scattering differs from that of Ly$\alpha$. For example, electrons in the atomic hydrogen excited by \Hb photons from $n = 2$ to $n = 4$ can decay via multiple de-excitation channels to $n = 3$, $n = 2$, or $n = 1$, with each branch leading to different line formations, while \Ha involves transitions from $n = 3$ to $n = 2$. As a result, \Hb photons are more likely to be converted into Pa$\alpha$ via cascades to the $n = 4 \to 3$ transition, due to multiple scatterings. These distinctions highlight the necessity for radiative transfer models that incorporate both these multiple branches of Balmer line scattering and the physical and kinematic properties of the scattering medium.

Other scattering processes can also produce broad hydrogen line profiles. Thomson scattering by free electrons can generate symmetric, exponential wings around emission lines, with widths scaling with the electron thermal speed ($\sim 500~\kms$ at $T = 10^4~\rm K$). Such wings have been observed and studied in Type IIn supernovae \citep{Dessart2009}, symbiotic stars \citep{Sekeras2012,Chang2018b}, AGNs \citep{Weymann1970,Sunyaev1980,Lee1999,Laor2006}, and recently in LRDs (\citealp{Rusakov2025} cf. \citealp{Brazzini2025}).

Raman scattering, where UV photons near Lyman series wavelengths are inelastically scattered by ground-state hydrogen atoms, can also generate extremely broad, often asymmetric features around Balmer lines like \Ha and \Hb \citep{Chang2015}. Raman wings have been observed and investigated in symbiotic stars \citep{Nussbaumer1989,Yoo2002,Chang2018b,Lee2018}, dense star-forming regions \citep{Dopita2016,Henney2020}, and AGNs \citep{Chang2015, Kokubo2024_Raman}, and are proposed for LRDs as well \citep{Kokubo2024}.
These scattering mechanisms offer alternative explanations for the broad Balmer components in LRDs, independent of virialized motion in a classical AGN broad line region. 
Even if scattering is not the dominant mechanism, its contribution complicates the interpretation of line widths and, therefore, affects standard black hole mass estimation methods, as the observed broadening may additionally originate from scattering processes rather than purely kinematic broadening.\\

Given the complex line shapes and multiple broadening mechanisms, we investigate how radiative transfer effects influence hydrogen emission line profiles in LRDs, focusing on \Ha and \Hb. We perform Monte Carlo radiative transfer simulations including three key processes: resonance scattering of \Ha and \Hb, Thomson scattering by free electrons, and Raman scattering by neutral hydrogen. Section~\ref{sec:simulation} details the physics of each mechanism, the model geometry, and the simulation setup. Section~\ref{sec:results} presents the emergent spectra and explores how each process shapes the line profiles. In Section~\ref{sec:discussion}, we examine observable signatures such as line profiles and explore the impact of scattering on the estimation of black hole mass.

\begin{figure*}
    \centering
    \includegraphics[width=0.98\textwidth]{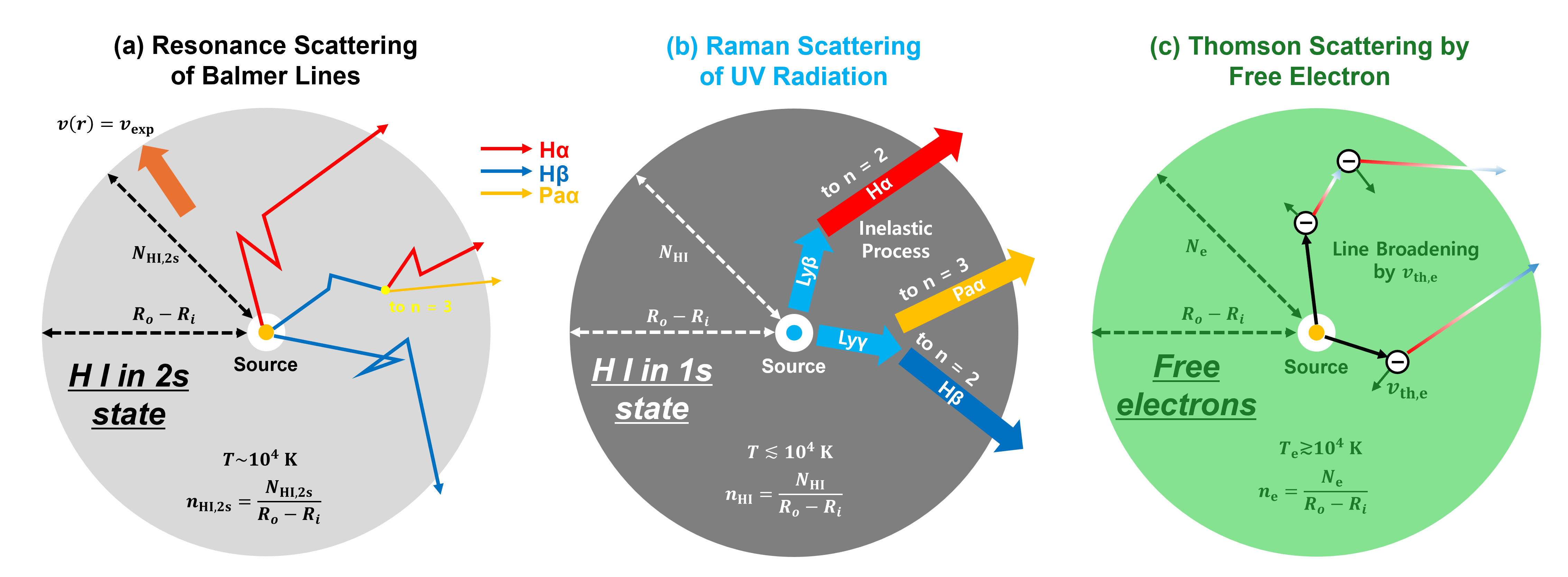}
    \caption{Schematic illustration of the model geometry. A central point source emits either \Ha and \Hb photons (orange points for resonance and Thomson scattering) or a UV continuum (a blue point for Raman scattering), surrounded by a spherical scattering region. The sphere is divided into three wedges representing the different scattering processes: (a) Resonance scattering of Balmer lines, \Ha and \Hb, in a spherical \hi region populated with 2s-state hydrogen atoms, characterized by a column density \NHItwo, a random motion \sigr, and an expansion velocity \vexp. (b) Raman scattering of UV photons near the Lyman series by ground-state \hi atoms with a column density \NHI, producing broad optical and IR features (e.g., H$\alpha$, H$\beta$, Pa$\alpha$). (c) Thomson scattering of Balmer line photons by free electrons in a spherical \hii region, characterized by electron temperature $T_e$ and electron density $n_e$, producing symmetric broadening proportional to the electron’s thermal speed $v_{\rm th,e}$.}
    \label{fig:scattering_geometry}
\end{figure*}

\section{Method}\label{sec:simulation}

To understand the formation of Balmer line profiles in the Little Red Dots (LRDs), we conduct Monte Carlo radiative transfer (RT) simulations focusing on three scattering processes: (i) resonance scattering of Balmer lines, \Ha and \Hb, by hydrogen atoms in the metastable $2s$ state, (ii) Raman scattering of UV continuum photons by ground-state hydrogen atoms, and (iii) Thomson scattering of Balmer line photons by free electrons. Each mechanism can contribute to the broadening and shaping of emission lines, producing features such as P-Cygni profiles, double peaks, or broad wings in the spectra.

In order to explore the impact of scattering processes on observables, our simulations adopt a simplified spherical geometry, where a central point source is embedded within a uniform scattering medium. For each process, we use appropriate Monte Carlo radiative transfer codes to track photon propagation and scattering interactions, accounting for the specific cross-sections, branching ratios, and physical conditions relevant to each scattering mechanism. The following sections describe the setup, physics, and simulation methods for each scattering process in detail.

\subsection{Radiative Transfer Framework and Geometry}\label{sec:geometry}

\begin{table*}
    \centering
    \caption{Key parameters for each scattering process. The ranges represent typical values used in our simulations.}
    \label{tab:parameters}
    \begin{tabular}{llc}
        \hline
        Type of Scattering & Parameter & Typical Range \\
        \hline
        \multirow{4}{*}{Resonance} 
        & \hi column density in the 2s state \NHItwo & $10^{12} - 10^{16}$ cm$^{-2}$ \\
        & Expansion velocity \vexp & $0 - 500$ km\,s$^{-1}$ \\
        & Random motion of \hi gas \sigr & $13.6$ (hydrogen thermal motion at $T=10^4\rm K$)$- 100$ km\,s$^{-1}$ \\
        & Width of intrinsic emission \sigsrc & $0 - 400$ km\,s$^{-1}$ \\
        \hline
        \multirow{1}{*}{Raman} 
        & \hi column density \NHI & $10^{20} - 10^{23}$ cm$^{-2}$ \\
        \hline
        \multirow{3}{*}{Thomson} 
        & Electron temperature $T_e$ & $10^4 - 10^6$ K \\
        & Thomson optical depth $\tau_{\rm Th}$ & $0.01 - 10$ \\
        & Width of intrinsic emission \sigsrc & $50 - 500$ km\,s$^{-1}$ \\
        \hline
    \end{tabular}
\end{table*}

To investigate the formation of hydrogen emission lines in LRDs, we consider radiative transfer through different scattering processes within a simplified geometry. Our model consists of a central isotropic point source embedded in a uniform scattering medium, as illustrated in Fig.~\ref{fig:scattering_geometry}. The central source emits either hydrogen line photons (for resonance and Thomson scattering) or a flat UV continuum (for Raman scattering).

The geometry of the scattering medium is spherical, characterized by an outer radius $R$ and an inner radius set at $0.01R$. The number density of hydrogen atoms or electrons is uniform within this region. For each scattering mechanism, we apply relevant physical parameters that control the resulting line profiles:
\begin{itemize}
    \item For \textbf{resonance scattering} of Balmer lines, H$\alpha$ and H$\beta$, we consider the neutral hydrogen column density in the 2s state, $N_{\mathrm{HI},2s}$, the random velocity dispersion $\sigma_R$, and an optional bulk radial outflow velocity $v_{\mathrm{exp}}$. The velocity field parameters $\sigma_R$ and $v_{\mathrm{exp}}$ influence the frequency shifts and the formation of double-peaked profiles or P-Cygni features.
    
    \item For \textbf{Raman scattering} of the UV continuum, we adopt a flat incident spectrum covering the Lyman series region from Ly$\beta$ to Ly$\delta$. The process depends primarily on the total \hi column density $N_{\mathrm{HI}}$, as the broadening of Raman-scattered features is dominated by the interaction cross-section, branching probabilities near the Lyman transitions, and the energy conservation during Raman scattering (i.e., Raman broadening). We neglect the effects of bulk or turbulent motions, as they have a negligible impact on the broadening of Raman features compared to the intrinsic energy shift from the inelastic scattering.
    
    \item For \textbf{Thomson scattering} of Balmer lines, the scattering region is an ionized \hii sphere characterized by the electron density $n_e$, temperature $T_e$, and the resulting Thomson optical depth $\tau_{\mathrm{Th}} = \sigma_{\mathrm{Th}} n_e R$. The broadening of the scattered line profiles arises primarily from the thermal motions of electrons $\vth = \sqrt{2kT/m_e} \approx 548 \kms$ at $T_e = 10^4$\, K. Similar to Raman scattering, we neglect the effect of bulk outflows or turbulence, as these do not significantly impact the Thomson scattering process due to the high thermal speed of electrons and wavelength-independent scattering cross section.
\end{itemize}

The radiative transfer simulations for each process are performed using dedicated Monte Carlo codes, as detailed in Sections~\ref{sec:resonance}, \ref{sec:raman}, and \ref{sec:thomson}. The input parameters for each scattering process are summarized in Table~\ref{tab:parameters}.

\subsection{Resonance Scattering of Balmer Lines}
\label{sec:resonance}

\begin{table} 
\caption{Atomic parameters of H$\alpha$ and H$\beta$ to calculate their scattering cross section in Equation~\ref{eq:cross_section}. The first row shows the oscillator strength from $2s$ to n$p$ state.
The second and third rows represent the damping term and line center frequency.}
\centering
 \begin{tabular}{lcccccc}
\hline
 & \Ha ($2s\to3p)$ & \Hb ($2s\to4p$) \cr
\hline
$f$ & 0.4360 &  0.1028 \cr
$\Gamma$ & $4.4 \times 10^7$ $s^{-1}$& $8.4 \times 10^6$ $s^{-1}$\cr
$\nu_0$ & $4.5600 \times 10^{14}$ $s^{-1}$& $6.1694 \times 10^{14}$ $s^{-1}$\cr
\hline
\label{tab:atomic_data}
 \end{tabular}
%     \\
%            \footnotesize
%        $*$: the oscillator strength from $2s$ to n$p$ state 
\end{table}

\begin{table}
\caption{Branching ratio of permitted transitions at $3p$ and $4p$ states.}
\centering
 \begin{tabular}{lcccccc}
\hline
Initial State & $3p$ & $3p$ &  $4p$ &  $4p$&  $4p$\cr
\hline
Final State & $ 1s$    & $ 2s$ & $1s$    & $2s$   &  $3s,\, 3d$    \cr    
\hline
Branching ratio &  0.882 &  0.118  &  0.839 &  0.119 & 0.042 \cr
Branching ratio w/o $1s$ &  - &  1 &  - &  0.739 & 0.261 \cr
\hline 
\label{tab:branching}
 \end{tabular}
\end{table}

Resonance scattering in Balmer lines is expected in environments where the $n = 2$ population of hydrogen is significantly enhanced. In LRDs, the presence of Balmer absorption features and a strong Balmer break is indicative of this condition. Since \Ha and \Hb photons can be trapped and experience multiple scatterings within optically thick gas, their emergent profiles are sensitive to the velocity structure and geometry of the medium. Modeling this process is essential to explain their P-Cygni-like profiles and the asymmetric absorption observed in Balmer lines, particularly the different behaviors of \Ha and \Hb.

\subsubsection{Physical Mechanism}

Resonance scattering of Balmer lines occurs when hydrogen atoms in the $2s$ state absorb photons near the H$\alpha$ and H$\beta$ transition energies, exciting electrons to the $n=3$ or $n=4$ states, respectively. The subsequent de-excitation of the electron can result in the re-emission of a photon at the same wavelength or, through branching, produce photons in other lines such as Pa$\alpha$. This process is analogous to Ly$\alpha$ scattering (see, e.g., review by \citealp{Dijkstra2019}), but for Balmer transitions.

\begin{figure}
    \centering
    \includegraphics[width=0.49\textwidth]{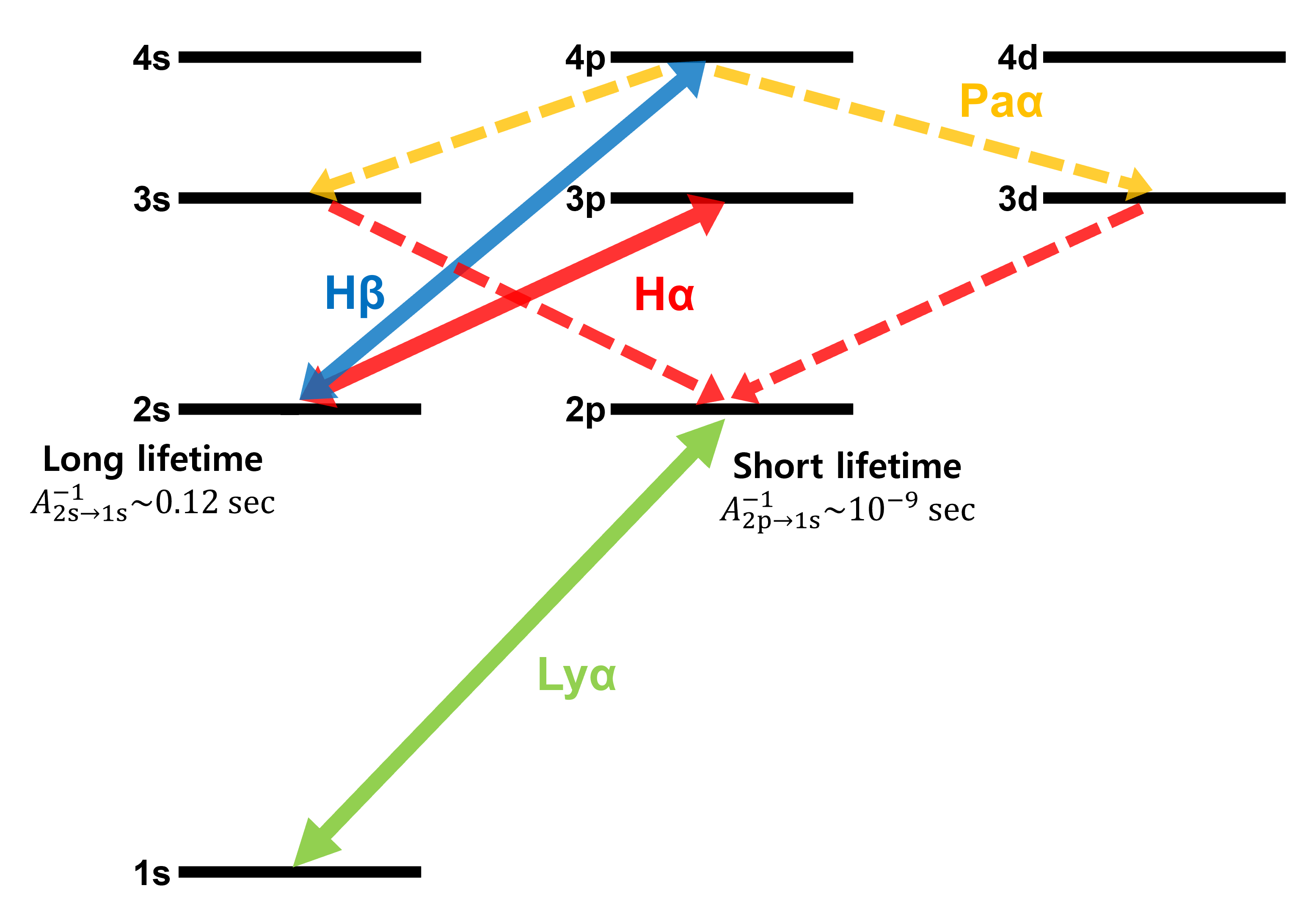}
    \caption{Energy levels of atomic hydrogen in $n = 1-4$. Solid arrows indicate the 'resonant transitions' involved in resonance scattering (see Section~\ref{sec:resonance}), while dashed arrows show other possible transitions for H$\alpha$, H$\beta$, and Pa$\alpha$ (omitting the higher Lyman transitions; `case B' assumption).}
    \label{fig:energy_level}
\end{figure}

The key role of the $2s$ state arises from its long lifetime (approximately 0.1 s), which is orders of magnitude longer than the lifetimes of the $np$ excited states ($\sim10^{-9}$ s). This long lifetime enables a significant population of hydrogen atoms to accumulate in the $2s$ state, providing a reservoir for resonance scattering. The atomic structure of hydrogen and relevant transitions are illustrated in Fig.~\ref{fig:energy_level}.

The scattering cross-section for a Balmer line is described by a Voigt profile:
\begin{equation}\label{eq:cross_section}
    \sigma_\nu = \frac{\sqrt{\pi} e^2}{m_e c} \frac{f}{\Delta \nu_D} H(a, x),
\end{equation}
where $f$ is the oscillator strength, $\Delta \nu_D$ is the Doppler width, and $H(a,x)$ is the Voigt function:
\begin{equation}
    H(a, x) = \frac{a}{\pi} \int_{-\infty}^{\infty} \frac{e^{-y^2}}{(x - y)^2 + a^2} dy.
\end{equation}
Here, $a = \Gamma / (4\pi \Delta \nu_D)$ is the natural width parameter, and $x = (\nu - \nu_0)/\Delta \nu_D$ is the dimensionless frequency offset. The thermal Doppler width is $\Delta \nu_D = \nu_0 v_{\mathrm{th}}/c$, where $v_{\mathrm{th}} = \sqrt{2kT/m_H}$ is the thermal speed of hydrogen at a temperature $T$.
As the total random motion $\sigr = \sqrt{\vth^2 + v_{\rm turb}^2}$, assuming the microturbulent approximation with the turbulent speed $v_{\rm turb}$, 
we set the thermal speed \vth to be the random motion of hydrogen \sigr in our simulation.

Adopting the atomic data in Table~\ref{tab:atomic_data}, The optical depths at line center for H$\alpha$ and H$\beta$ are given by:
\begin{align}\label{eq:optical_depth}
    \tau_{0, \mathrm{H}\alpha} &\approx 33.4 \left( \frac{N_{\mathrm{HI},2s}}{10^{14} \, \mathrm{cm}^{-2}} \right) \left( \frac{\sigma_R}{12.8 \, \mathrm{km\,s^{-1}}} \right)^{-1}, \\
    \tau_{0, \mathrm{H}\beta} &\approx 5.83 \left( \frac{N_{\mathrm{HI},2s}}{10^{14} \, \mathrm{cm}^{-2}} \right) \left( \frac{\sigma_R}{12.8 \, \mathrm{km\,s^{-1}}} \right)^{-1}.
\end{align}

\subsubsection{Branching Ratios and Line Conversion}

When an H$\alpha$ photon excites a $2s$ electron to the $3p$ state, the electron can decay to either the $n = 1$ ($1s$) or $n = 2$ ($2s$) state, emitting Ly$\beta$ or H$\alpha$, respectively (see also Fig.~\ref{fig:energy_level}). The branching ratio to the $1s$ state is significantly larger (0.882) than that to the $2s$ state (0.118), as shown in Table~\ref{tab:branching}. However, under the Case B assumption, the $3p \rightarrow 1s$ transition is neglected, as Ly$\beta$ photons are typically reabsorbed by the surrounding \hi medium with the $1s$ state. As the population of the $n=1$ state is much higher than that of the $n=2$ state, if the gas is optically thick for Balmer lines, it is also optically thick for Lyman series. Thus, the Case B assumption is natural in the scattering medium for resonance scattering.
Consequently, H$\alpha$ photons predominantly remain H$\alpha$ throughout the scattering process.

For H$\beta$ photons, excitation to the $4p$ state allows decay to the n = 1 ($1s$), $n = 2$ ($2s$), or $n = 3$ ($3s/3d$) states. The raw branching ratios for these transitions are 0.839, 0.119, and 0.042, respectively. When the $1s$ transition is suppressed under the Case B assumption, the relative branching ratio to the n = 3 state increases to 0.261. Therefore, the conversion of H$\beta$ photons into Pa$\alpha$ via the $4p \rightarrow 3s/3d$ transition is not negligible and must be considered in the simulation for the optically thick medium. This process also generates H$\alpha$ photons at the scattering location, as the Pa$\alpha$ cascade involves a subsequent $3s/3d \rightarrow 2p$ transition (see Fig.~\ref{fig:energy_level}).

\subsubsection{Monte Carlo Simulation}\label{sec:monte_carlo_resonance}

The resonance scattering simulations are performed using the 3D Monte Carlo code \texttt{RT-scat} \citep{Chang2023,Chang2024} in which discrete photon packages are followed through real and frequency space until they escape the simulation domain, at which point the emergent properties are recorded. Specifically, we follow these steps:
\begin{enumerate}
    \item Define the spherical geometry characterized by a \hi column density in the 2s state \NHItwo, a random motion \sigr, and an expansion velocity \vexp; the radial bulk motion $v(r)$ is constant (i.e., $v(r) = \vexp$). %is proportional to the distance from the central source $r$ (i.e., $v(r) = \vexp r/R$).
    \item Generate isotropic \Ha and \Hb photons from the central source, with an intrinsic Gaussian width \sigsrc.
    \item For each photon, determine the scattering location. Using a dimensionless free path of the photon, $\tau_f = -\ln{r}$, where $r$ is a random number drawn uniformly in $[0,1]$, a free path of a photon $l_f$ is given by $\tau_f/\sigma_{\nu}$; $\sigma_{\nu}$ is a scattering cross section (Eq.~\ref{eq:cross_section}). The frequency $\nu$ is adopted as the frequency in the rest frame of the cell, where the photon penetrates.
    \item Determine the scattering outcome:
    \begin{itemize}
        \item For \Ha, photons always remain \Ha after scattering.
        \item For \Hb, the final state is selected by the branching ratios. If a random number $r$ drawn uniformly in $[0,1]$ is smaller than the branching ratio to the $n=2$ state, $0.739$, the photon remains \Hb. Otherwise, it converts to Pa$\alpha$ with an accompanying \Ha photon at the same location (with the wavelength at the line center in the atom rest frame). Because of the low number of \hi atoms in the $n=3$ state, the Pa$\alpha$ photon immediately escapes, while the newly generated \Ha photon continues to propagate in the spherical \hi region.
    \end{itemize}
    \item Compute the new photon direction with the isotropic distribution and frequency shift by a random motion \sigr. While the incident and scattered frequencies are identical in the rest frame of atomic hydrogen, which scatters a photon, the scattered frequency in the rest frame of the cell is changed due to the Doppler shift caused by the random motion of the hydrogen.
    \item Repeat steps (iii)--(v) until the photon escapes the grid. When collecting photons, the photon's frequency is converted to the frequency in the observed frame, taking into account the kinematics at the last scattering location.
    \item Back to a step (ii) until the total number of intrinsic photons becomes $10^8$.
\end{enumerate}
A visual representation of this process is shown in the left panel of Fig.~\ref{fig:scattering_geometry}.
The detailed Monte-Carlo method of \texttt{RT-scat} is described in Section 2.5 of \cite{Chang2024}.
We will show simulated results of resonance scattered \Ha, \Hb, Pa$\alpha$ profiles in Section~\ref{sec:result_resonance}.

\subsection{Raman Scattering}
\label{sec:raman}
\begin{figure}
    \centering
    \includegraphics[width=0.40\textwidth]{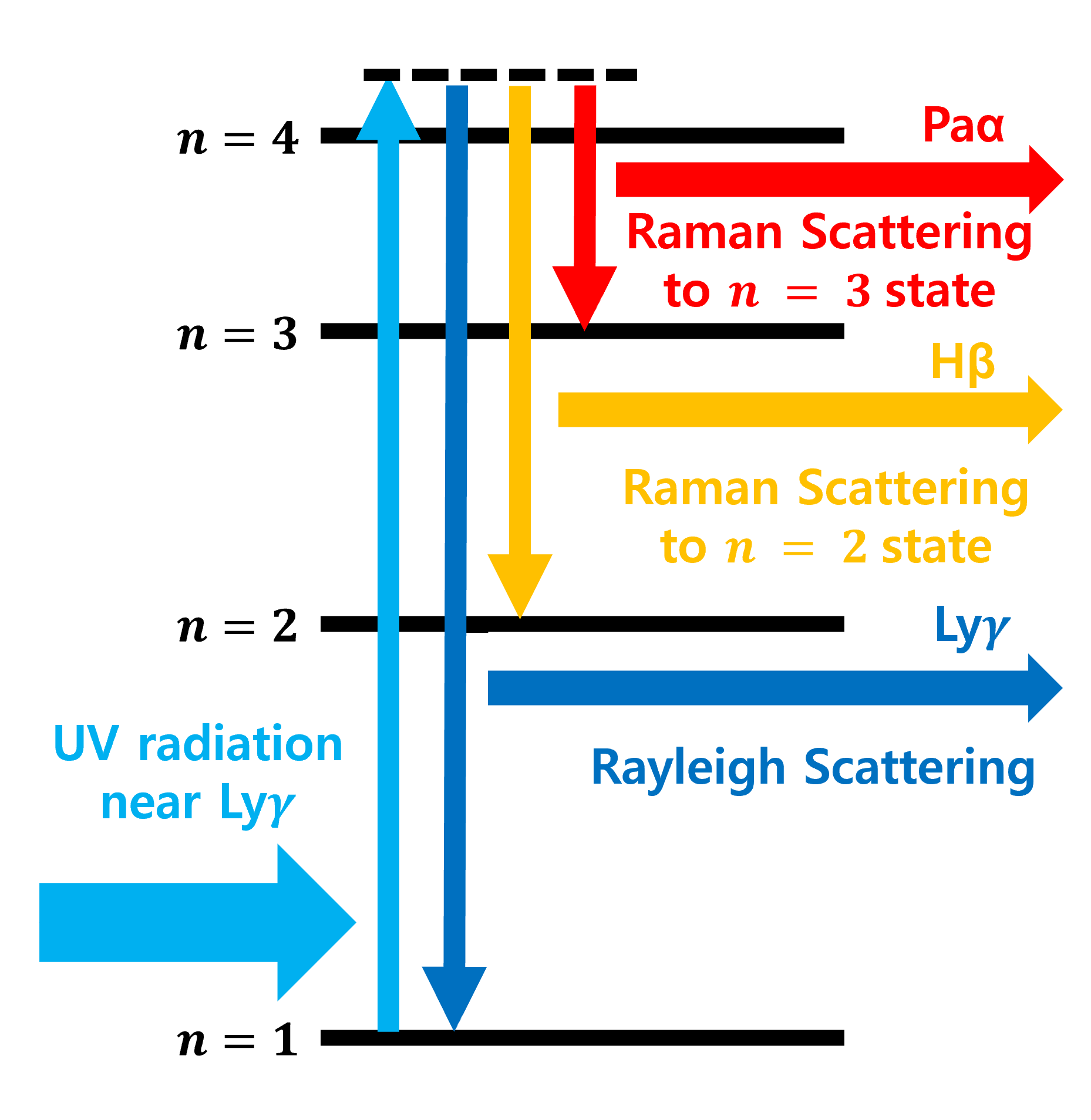}
    \caption{Schematic illustration of Raman scattering of UV radiation near Ly$\gamma$. The black solid lines represent the energy levels of atomic hydrogen. The light blue line indicates the incident UV radiation near Ly$\gamma$. The blue arrow represents Rayleigh scattering, in which the electron de-excites to the $n=1$ state, emitting Ly$\gamma$. The red and orange arrows represent Raman scattering to the $n=2$ and $n=3$ states, producing H$\beta$ and Pa$\alpha$, respectively.}
    \label{fig:Raman_scattering}
\end{figure}

Raman scattering generates extremely broad wings around Balmer lines, due to the inelastic scattering of UV photons near Lyman-series wavelengths by neutral hydrogen. While this process requires strong UV radiation fields and large \NHI, both conditions are plausible in LRDs, which show V-shaped SEDs and deep Balmer breaks. If Lyman emission lines or a strong UV continuum exist, Raman scattering can significantly reshape the Balmer line profiles and even mimic AGN-like broad components.

\subsubsection{Physical Mechanism}\label{sec:Raman_scattering_mechanism}

Raman scattering with \hi atom is an inelastic process in which a UV photon near the hydrogen Lyman series is scattered by a ground-state hydrogen atom, producing an optical or infrared photon through de-excitation to $n > 1$ states. If the electron returns to the ground state ($n=1$) after scattering, the process is elastic, and the photon retains its original energy; this is known as Rayleigh scattering. In contrast, if the electron de-excites to higher energy levels ($n > 1$), the scattered photon undergoes an energy shift corresponding to the difference between the Lyman transition energy and the new state. Specifically, UV continuum photons near Ly$\beta$ form scattered features around \Ha through Raman scattering to the $n=2$ state. Fig.~\ref{fig:Raman_scattering} illustrates how Raman scattering of a Ly$\gamma$ photon to $n=2$ and $n=3$ states results in the conversion of the UV photon to \Ha and Pa$\alpha$, respectively. 

Computing the probability of Raman scattering occurring as a function of a wavelength in the far-UV regime is non-trivial.
While the scattering probability for Rayleigh and Raman scatterings is given by the total cross section, the individual transition probabilities are given by the Branching ratios, which are the ratio between the cross section for Raman scattering divided by the total scattering cross section.
Thus, the sum of the branching ratios to the $n = 1$ state (Rayleigh scattering) and $n > 1$ states (Raman scattering) is unity.
Fig.~\ref{fig:Raman_cross_section} shows the total scattering cross section and branching ratios to each $n$ state.

This process involves an energy shift corresponding to the Lyman transition energy, given by
\begin{equation}\label{eq:raman_energy}
    E_{\mathrm{inc}} = E_{\mathrm{sc}} + E_{\mathrm{Lyman}},
\end{equation}
where $E_{\mathrm{inc}}$ is the energy of the incident UV photon, $E_{\mathrm{sc}}$ is the energy of the scattered photon, and $E_{\mathrm{Lyman}}$ is the energy of the corresponding Lyman transition. For instance, following the example to Pa$\alpha$ depicted in Fig.~\ref{fig:Raman_scattering} (light blue to red arrow), $E_{\rm inc}\approx E_{\rm Ly\gamma}$, $E_{\rm Lyman}=E_{\rm Ly\beta}$, and $E_{\rm sc}\approx E_{\rm Pa\alpha}$.

The scattered wavelength is computed correspondingly as
\begin{equation}\label{eq:Raman_wavelength}
    \frac{1}{\lambda_{\mathrm{sc}}} = \frac{1}{\lambda_{\mathrm{inc}}} - \frac{1}{\lambda_{\mathrm{Lyman}}},
\end{equation}
where $\lambda_{\mathrm{Lyman}}$ is the wavelength of the relevant Lyman transition (e.g., Ly$\beta$, Ly$\gamma$).

This scattering process leads to an emergent broad line. To first order, the emergent line shape follows the corresponding cross section (cf. Fig.~\ref{fig:Raman_cross_section}). However, the broadening of Raman-scattered features arises not only from the natural cross-section dependence but also from the energy shift. This can be illustrated by the differential relation between the incident and scattered wavelengths, which is given by
\begin{equation}
    d\lambda_{\mathrm{sc}} = \left( \frac{\lambda_{\mathrm{sc}}}{\lambda_{\mathrm{inc}}} \right)^2 d\lambda_{\mathrm{inc}},
    \label{eq:Raman_broad}
\end{equation}
where the factor $\left( \lambda_{\mathrm{inc}} / \lambda_{\mathrm{sc}} \right)^2$ represents the additional broadening from the Raman scattering process. As the derivative of the velocity is given by $dv = d\lambda/\lambda$, this factor becomes $\lambda_{\mathrm{sc}}/\lambda_{\mathrm{inc}}$ in velocity space, which is $\sim 6.4$ and $5$ for \Ha and \Hb, respectively.

Due to the cross-section dependence, the overall broadening of Raman features depends primarily on the total \hi column density \NHI, with higher \NHI leading to broader features with widths exceeding several thousand~\kms \citep{Lee1998,Chang2015, Chang2018b,Kokubo2024_Raman}. Unlike resonance scattering, the effects of the bulk velocity \vexp or the random motion \sigr are negligible compared to the intrinsic broadening from the energy shift, assuming the continuum radiation as an input source. 
However, if UV emission lines, such as He II lines at 1025, 972, and 949 \AA\, and O VI resonance lines at 1032 and 1038 \AA (rather than the continuum) undergo Raman scattering, the kinematics of the \hi region and the profile of input radiation \citep{Schmid1989,Harries1996,Choi2020a, Chang2020, Heo2021, Chang2023_Raman} can affect the formation of Raman-scattered features.

We will show results of Raman scattered continuum photons in Section~\ref{sec:result_Raman} and discuss Raman-scattered features of the emission lines of the higher Lyman series (Ly$\beta$ and Ly$\gamma$) in Appendix~\ref{sec:Raman_emission}.

\subsubsection{Monte Carlo Simulation}
The Raman scattering simulations are performed using the Monte Carlo code \texttt{STaRS} \citep{Chang2020}, following these steps:
\begin{enumerate}
    \item Define the spherical geometry with uniform \hi column density \NHI.
    \item Generate a flat UV continuum between 940--1060 \AA\ from the central source or Gaussian emission of Lyman series.
    \item For each photon, determine the scattering location using the total scattering cross section (the black solid line in Fig.~\ref{fig:Raman_cross_section}).
    The details to calculate a free path are described in (iii) of Section~\ref{sec:monte_carlo_resonance}.
    \item Select the de-excitation channel (i.e., final state of the electron) based on branching ratios computed from the Kramers--Heisenberg formula. If the random number $r$ from 0 to 1 is smaller than the branching ratio to the $n > 1$ state, the photon Raman scatters and converts to an optical or IR photon (e.g., \Ha from Ly$\beta$, and \Hb and Pa$\alpha$ from Ly$\gamma$). Otherwise, the photon undergoes Rayleigh scattering and remains a UV photon.
    \item Compute the scattered wavelength $\lambda_{\mathrm{sc}}$ using Eq.~\ref{eq:Raman_wavelength}, and generate the new photon direction following the Thomson scattering phase function  (Eq.~\ref{eq:phase_function_thomson}) as Raman scattering is off-resonance scattering \citep{Karzas1961}. If the photon Raman scatters, it is considered to escape the grid immediately.
    \item Repeat steps (iii)--(v) until the photon escapes the grid or undergoes Raman scattering.
    \item Back to a step (ii) until the total number of intrinsic photons becomes $10^8$.
\end{enumerate}

The detailed Monte-Carlo method of \texttt{STaRS} is described in Section 2 of \cite{Chang2020}.
We will investigate the formation of the broad wings around hydrogen lines by Raman scattering and estimate their widths in Section~\ref{sec:result_Raman}.

\subsection{Thomson Scattering}
\label{sec:thomson}

Thomson scattering is particularly relevant in the presence of ionized gas in LRDs. The exponential wings observed in many Balmer profiles suggest the influence of electron scattering, as proposed in previous work \citep{Laor2006, Rusakov2025}. Importantly, Thomson scattering produces symmetric broadening across all lines when the emission region is embedded within the scattering medium. Unlike resonance or Raman processes, Thomson scattering can introduce significant line broadening without requiring large gas velocities. This poses a challenge for interpreting Balmer line widths as purely virial in origin, motivating detailed modeling of this mechanism in the LRD context.

\subsubsection{Physical Mechanism}

Thomson scattering is an elastic process in which photons scatter with free electrons, without any change in photon energy in the electron’s rest frame. The scattering cross section is independent of wavelength \citep{Rybicki1986} and given by
\begin{equation}\label{eq:Thomson_cross_section}
    \sigma_{\mathrm{Th}} \approx 6.65 \times 10^{-25} \, \mathrm{cm}^2.
\end{equation}
The angular distribution of scattered photons is described by the Thomson phase function:
\begin{equation}\label{eq:phase_function_thomson}
    P(\theta) = \frac{3}{16\pi} (1 + \cos^2 \theta),
\end{equation}
where $\theta$ is the scattering angle between the incident and scattered directions. With this scattering phase function, the probability of forward and backward scattering is greater than that of scattering in the perpendicular direction.

The broadening of the scattered photon spectrum arises from the thermal motion of electrons. The thermal speed of electrons is
\begin{equation}\label{eq:v_elec}
    v_{\mathrm{th,e}} = \sqrt{\frac{2 k T_e}{m_e}} \approx 548 \, \mathrm{km \, s^{-1}} \, \left( \frac{T_e}{10^4 \, \mathrm{K}} \right)^{1/2},
\end{equation}
where $T_e$ is the electron temperature. Since the scattering cross section is wavelength-independent, the resulting broad features from Thomson scattering appear around any emission lines from the same emission region with a similar width in velocity space, if these lines are emitted from the same region.

\subsubsection{Monte Carlo Simulation}

Since Thomson scattering is an elastic process in the electron's rest frame, like resonance scattering, the Thomson scattering simulations are performed using the Monte Carlo code \texttt{RT-scat} \citep{Chang2024} with the wavelength-independent cross section, following these steps:
\begin{enumerate}
    \item Define the spherical geometry with uniform electron density $n_e$, electron temperature $T_e$, and compute the Thomson optical depth $\tau_{\mathrm{Th}} = \sigma_{\mathrm{Th}} n_e R$.
    \item Generate isotropic photons from the central source, with an intrinsic Gaussian width \sigsrc.
    Due to the wavelength independence of the Thomson scattering cross section, the generated photon represents any emission from the central source (e.g., \Ha and \Hb).
    \item For each photon, determine the scattering location using the Thomson cross section $\sigma_{\rm Th}$ in Eq.~\eqref{eq:Thomson_cross_section} (details are described in (iii) of Section~\ref{sec:monte_carlo_resonance}).
    \item Compute the new photon direction using the Thomson phase function. The scattered wavelength is identical to the incident one in the rest frame of the electron. However, the scattered wavelengths vary in the observed frame. With a 3D velocity from a Maxwell-Boltzmann distribution ${\bf v_{e}} = v_{\rm th,e} \hat{\bf v} $ at an electron temperature $T_e$, a scattered wavelength in the observed frame $\lambda_s$ is given by
    \begin{equation}
        \lambda_s \approx \lambda_i \left[ 1 + \frac{v_{\rm th,e}}{c} \hat{\bf v} \cdot (\hat{\bf k_i}-\hat{\bf k_s}) \right],
    \end{equation}
    where $\hat{\bf k_i}$ and $\hat{\bf k_s}$ are unit vectors of incident and scattered photons, and 
    $\lambda_i$ is an incident wavelength in the observed frame; the range of $\hat{\bf v} \cdot(\hat{\bf k_i}-\hat{\bf k_s})$ term is $[-1,1]$.
    \item Repeat steps (iii)--(iv) until the photon escapes the grid.
    \item Back to a step (ii) until the total number of intrinsic photons becomes $10^8$.
\end{enumerate}

The detailed Monte-Carlo method of \texttt{RT-scat} is described in Section 2.5 of \cite{Chang2024} (also see \citealp{Laor2006,Kim2007,Chang2018b,Choe2023}).
We will explore Thomson-scattered features for various electron temperatures $T_e$ and optical depths $\tau_{\mathrm{Th}}$, and investigate which environment forms an exponential profile.

\section{Results}\label{sec:results}

In this section, we present the simulated spectral features of hydrogen emission lines arising from resonance, Raman, and Thomson scattering processes. Our radiative transfer models predict how each mechanism imprints distinct signatures on Balmer lines, such as H$\alpha$ and H$\beta$. We systematically explore the effects of key physical parameters for each scattering process.
In Section~\ref{sec:result_resonance}, we investigate the multi-transition effects in the $n=2$ and $n=3$ states involved in \Hb scattering and the formation of \Ha and \Hb spectra. 
In Section~\ref{sec:result_Thomson}, we explore the formation of Thomson scattered features for various electron temperatures and optical depths. 
In Section~\ref{sec:result_Raman}, we investigate the broad wings around hydrogen emission lines through Raman scattering of UV radiation near the Lyman series.

\subsection{Resonant scattering of Balmer lines} \label{sec:result_resonance}

\begin{figure}
    \centering
    \includegraphics[width=0.48\textwidth]{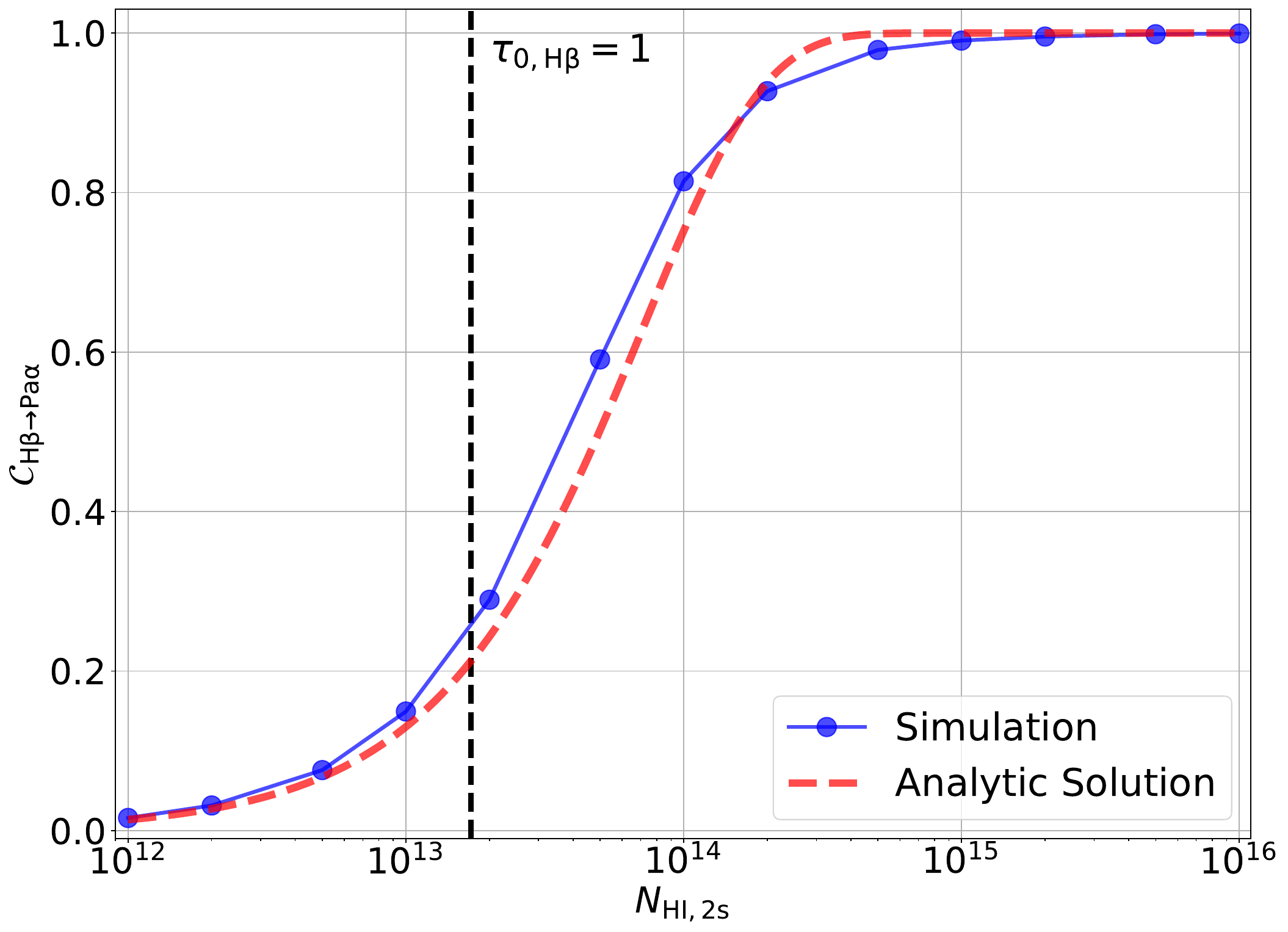}
\caption{
    Conversion rate from H$\beta$ to Pa$\alpha$ through inelastic \Hb scattering from $n = 2 \to 4 \to 3$, $\mathcal{C}_{\rm H\beta \to Pa\alpha}$, as a function of \NHItwo. The blue dots show the simulation results for the monochromatic case, while the red dashed line represents the analytic solution derived in Eq.~\ref{eq:Hb_to_Ha}. The vertical dashed line represents the \NHItwo corresponding to an optical depth of unity at the H$\beta$ line center ($\tau_{0,H\beta} = 1$). The conversion rate $\mathcal{C}_{\rm H\beta \to Pa\alpha}$ increases with \NHItwo, as higher optical depths lead to more scatterings and chances of branching via the $4p \to n = 3\ (3s/3d)$ transition.
}
    \label{fig:Hb_to_Ha}
\end{figure}

\begin{figure}
    \centering
    \includegraphics[width=0.48\textwidth]{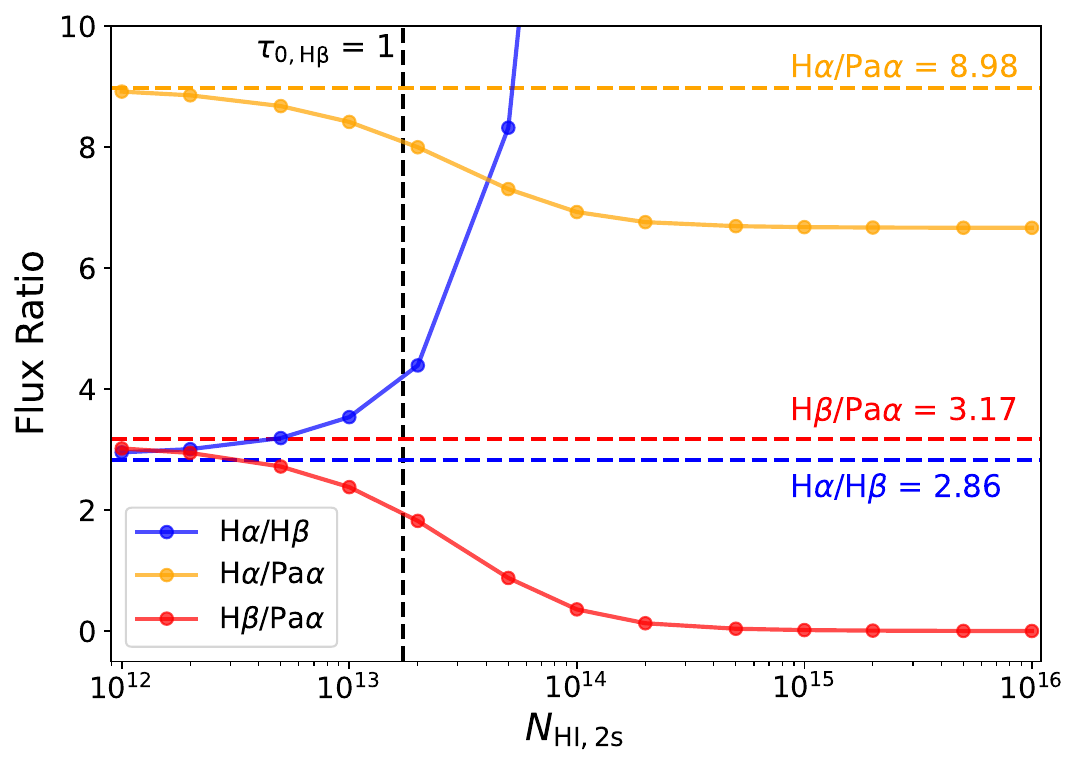}
\caption{
    Emergent Flux ratios experiencing \Hb scattering of H$\alpha$/H$\beta$ (blue), H$\beta$/Pa$\alpha$ (red), and H$\alpha$/Pa$\alpha$ (orange) as functions of \NHItwo. The ratio variations originate from H$\beta$ scattering, as shown in Fig.~\ref{fig:Hb_to_Ha}. The vertical dashed line marks the \NHItwo corresponding to an optical depth of unity at the H$\beta$ line center ($\tau_{0,H\beta} = 1$). The horizontal dashed lines indicate the line ratios expected under Case B recombination.
}
    \label{fig:ratio}
\end{figure}

\subsubsection{Pa$\alpha$ and H$\alpha$ from H$\beta$ Scattering} \label{sec:monochromatic_balmer}

While H$\beta$ photons undergo scattering, they can convert to Pa$\alpha$ and H$\alpha$ photons through transitions from the $4p$ state to the $3s$ and $3d$ states, with a branching ratio BR3 of 0.261 (see Fig.~\ref{fig:energy_level} and Table~\ref{tab:branching}). To quantify this process, we adopt a monochromatic source emitting H$\beta$ photons at the line center and estimate the conversion rate of H$\beta$ to Pa$\alpha$ via scattering to $n=3$ states, defined as
\begin{equation}
   \mathcal{C}_{\rm H\beta\to Pa\alpha} = \frac{ \Phi_{\rm H\beta\to Pa\alpha} }{ \Phi_{\rm H\beta, in} },
\end{equation}
where $\Phi_{\rm H\beta, in}$ is the number of intrinsic H$\beta$ photons, and $\Phi_{\rm H\beta\to Pa\alpha}$ is the number of escaping Pa$\alpha$ photons generated by H$\beta$ scattering. Fig.~\ref{fig:Hb_to_Ha} shows $\mathcal{C}_{\rm H\beta\to Pa\alpha}$ as a function of \NHItwo.

$\mathcal{C}_{\rm H\beta\to Pa\alpha}$ increases with increasing \NHItwo in Fig.~\ref{fig:Hb_to_Ha} because higher optical depths lead to more scatterings, providing more opportunities for H$\beta$ photons to convert to Pa$\alpha$ and H$\alpha$ via the $4p \to 3s/3d$ transition. To understand this trend, we derive an analytic solution for $\mathcal{C}_{\rm H\beta\to Pa\alpha}$.

The effective optical depth for the $n=3$ transition in a spherical geometry is ${\rm BR3} \, \tau_{\rm 0, H\beta}$, where BR3 is the branching ratio to the n = 3 state, $\sim 0.26$. The number of escaping H$\beta$ photons consists of those that either escape directly or undergo scattering into the $2s$ state. In other words, they are photons that escape without converting via the $n=3$ transition. The number of escaping H$\beta$ photons is thus
\begin{equation}
    \Phi_{\rm H\beta} = \Phi_{\rm H\beta, in} \, e^{-{\rm BR3}\, \tau_{\rm 0, H\beta}}.
\end{equation}
%where $\Phi_{\rm in}$ is the total number of input photons. 
The number of Pa$\alpha$ photons produced via scattering is $\Phi_{\rm in}-\Phi_{\rm H\beta}$, i.e.,
\begin{equation}
    \Phi_{\rm H\beta\to Pa\alpha} = \Phi_{\rm H\beta,in} \left(1 - e^{-{\rm BR3}\, \tau_{\rm 0, H\beta}}\right).
\end{equation}
Thus, the conversion rate is
\begin{equation} \label{eq:Hb_to_Ha}
   \mathcal{C}_{\rm H\beta\to Pa\alpha} = \frac{ \Phi_{\rm H\beta\to Pa\alpha} }{ \Phi_{\rm H\beta,in} } = 1 - e^{-{\rm BR3}\, \tau_{\rm 0, H\beta}}
   %\frac{1}{e^{{\rm BR3}\, \tau_{\rm 0, H\beta}} - 1}.
\end{equation}
Fig.~\ref{fig:Hb_to_Ha} shows that the analytic solution and the simulation results match well. For $\tau_{\rm 0, H\beta} > 0.5$, the simulation slightly exceeds the analytic estimate due to multiple scatterings, which increase the effective path length via random walks.

Fig.~\ref{fig:ratio} presents the variation of line flux ratios—H$\alpha$/H$\beta$, H$\beta$/Pa$\alpha$, and H$\alpha$/Pa$\alpha$—as functions of \NHItwo, assuming intrinsic line ratios consistent with Case B recombination. As H$\beta$ photons begin to convert into Pa$\alpha$ and H$\alpha$ at \NHItwo $\gtrsim 10^{13} \unitNHI$, the H$\alpha$/H$\beta$ ratio increases with \NHItwo. This trend is similar to the common Balmer decrement due to the dust extinction, but it originates from H$\beta$ scattering. 
However, note that the intrinsic spectra used in Fig.~\ref{fig:ratio} are $\delta$-functions at the respective line centers. If, on the other hand, the incident radiation has a broad intrinsic width and photons can escape more easily directly from the central source without significant interaction, the variation in line ratios can be significantly smaller.

\subsubsection{Formation of the H$\alpha$ Spectrum}

\begin{figure*}
    \centering
    \includegraphics[width=0.33\textwidth]{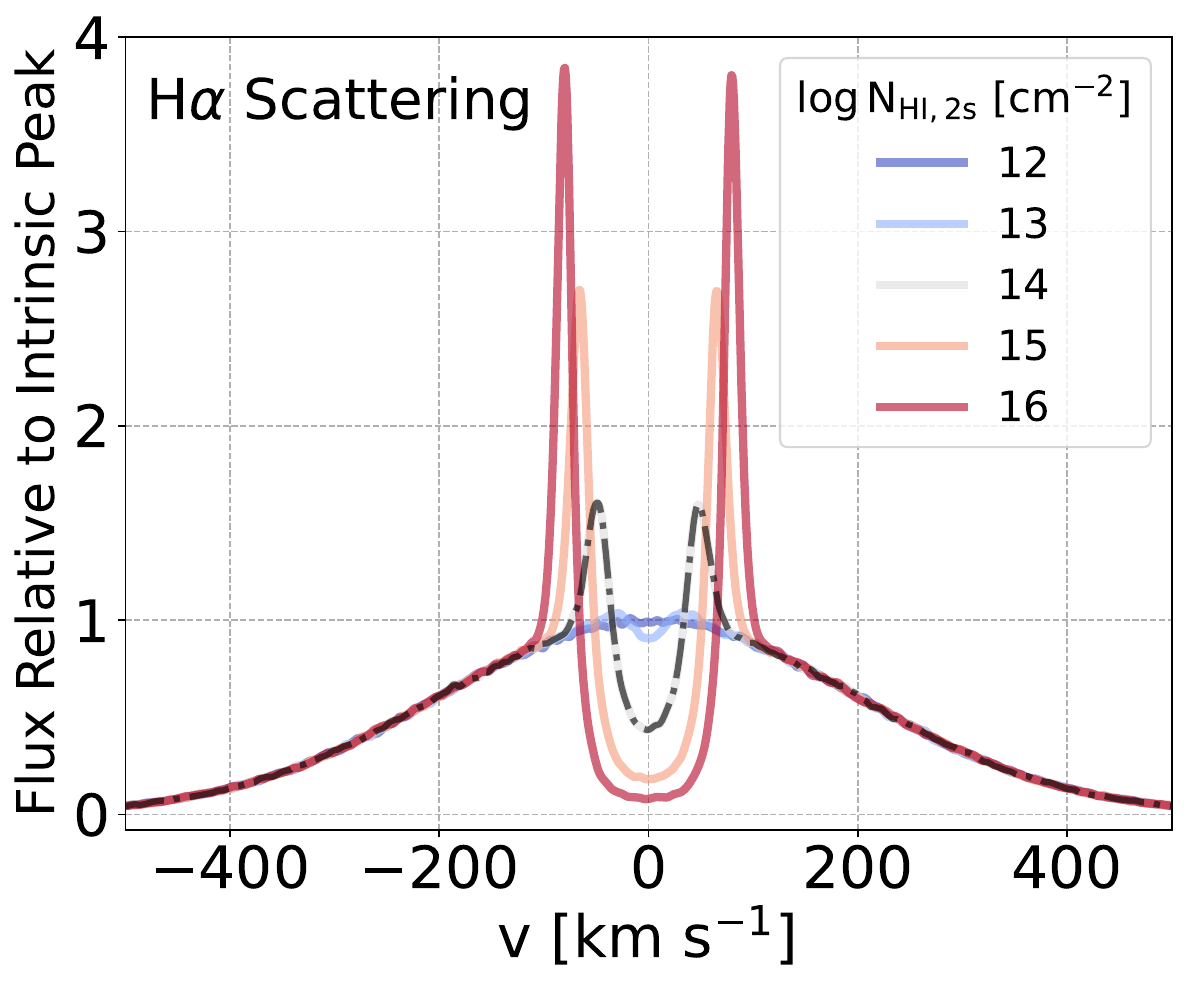}
    \includegraphics[width=0.33\textwidth]{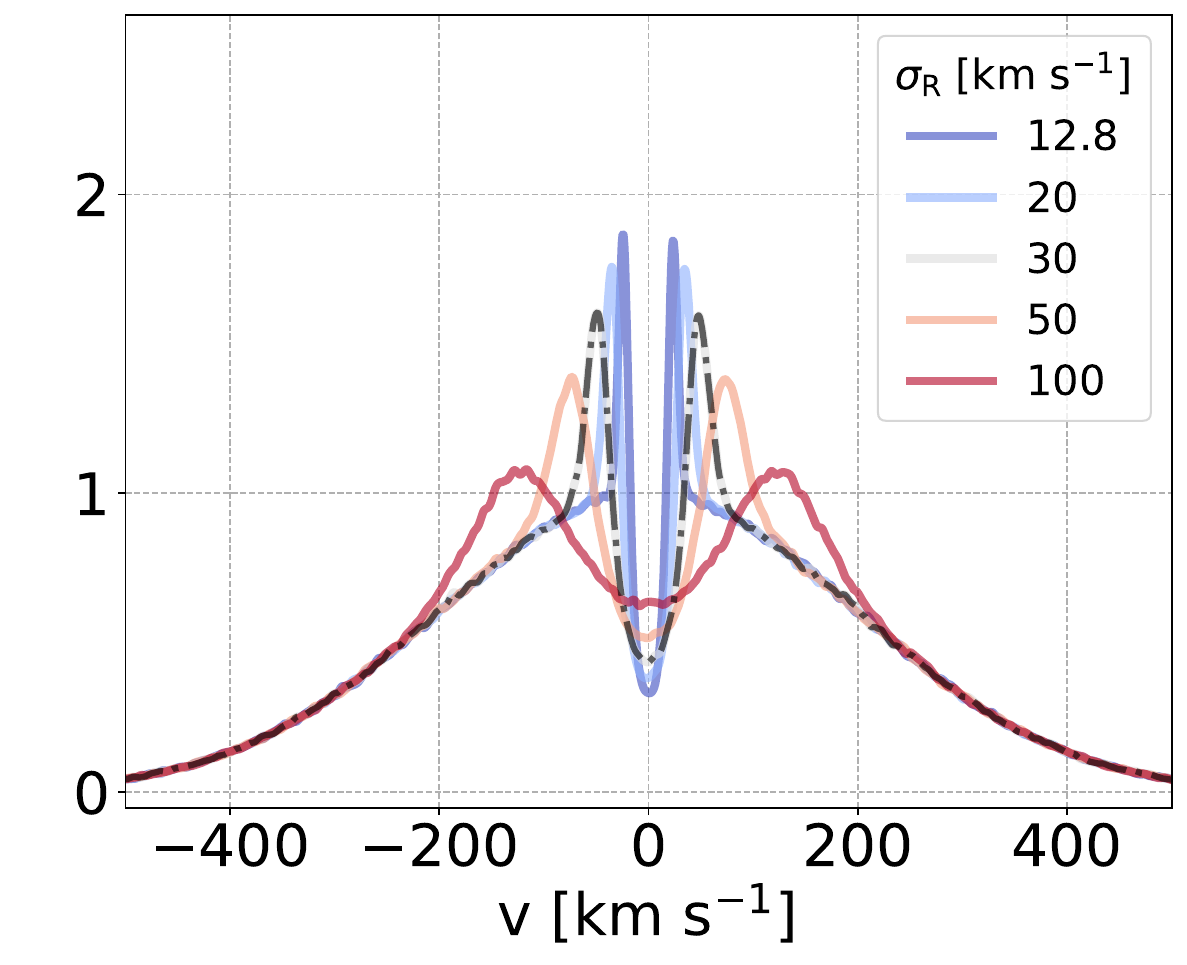}
    \includegraphics[width=0.33\textwidth]{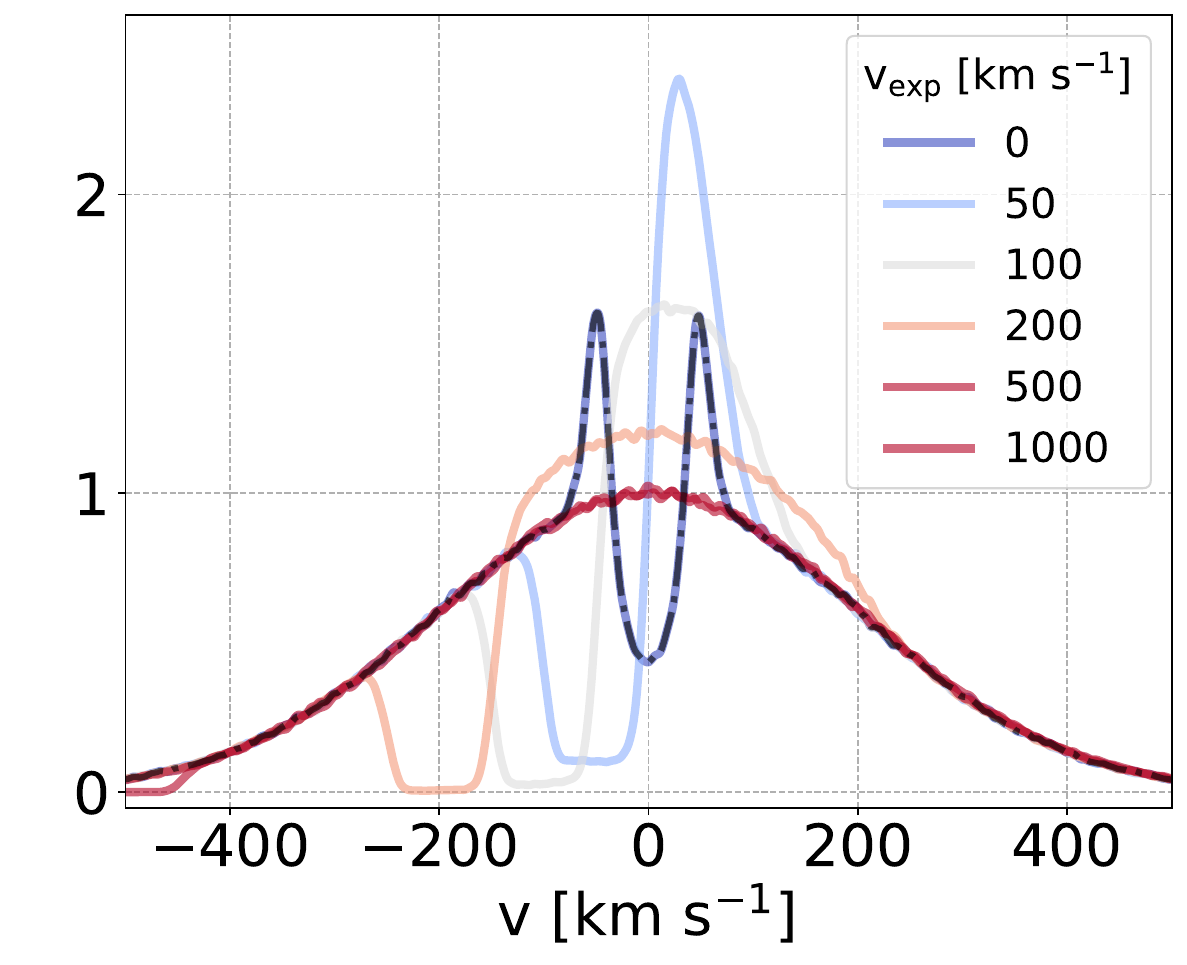}
    \caption{
    Spectra of resonance-scattered \Ha for different values of \NHItwo (left panel), random velocities \sigr (center panel), and bulk outflow speeds \vexp (right panel). The spectra are normalized by dividing the peak flux of intrinsic radiation; thus, the $y$-axis is the relative flux.
    Fiducial parameters are set to \NHItwo $=10^{14} \unitNHI$, \vexp $= 0 \kms$, and \sigr $= 30 \kms$. The intrinsic Gaussian emission profile width \sigsrc is fixed at 200 \kms. Spectral variations closely resemble those of Ly$\alpha$ under similar optical depth conditions, although typical Ly$\alpha$ optical depths are orders of magnitude larger.
    }
    \label{fig:spec_Ha}
\end{figure*}

\begin{figure*}
    \centering
    \includegraphics[width=0.33\textwidth]{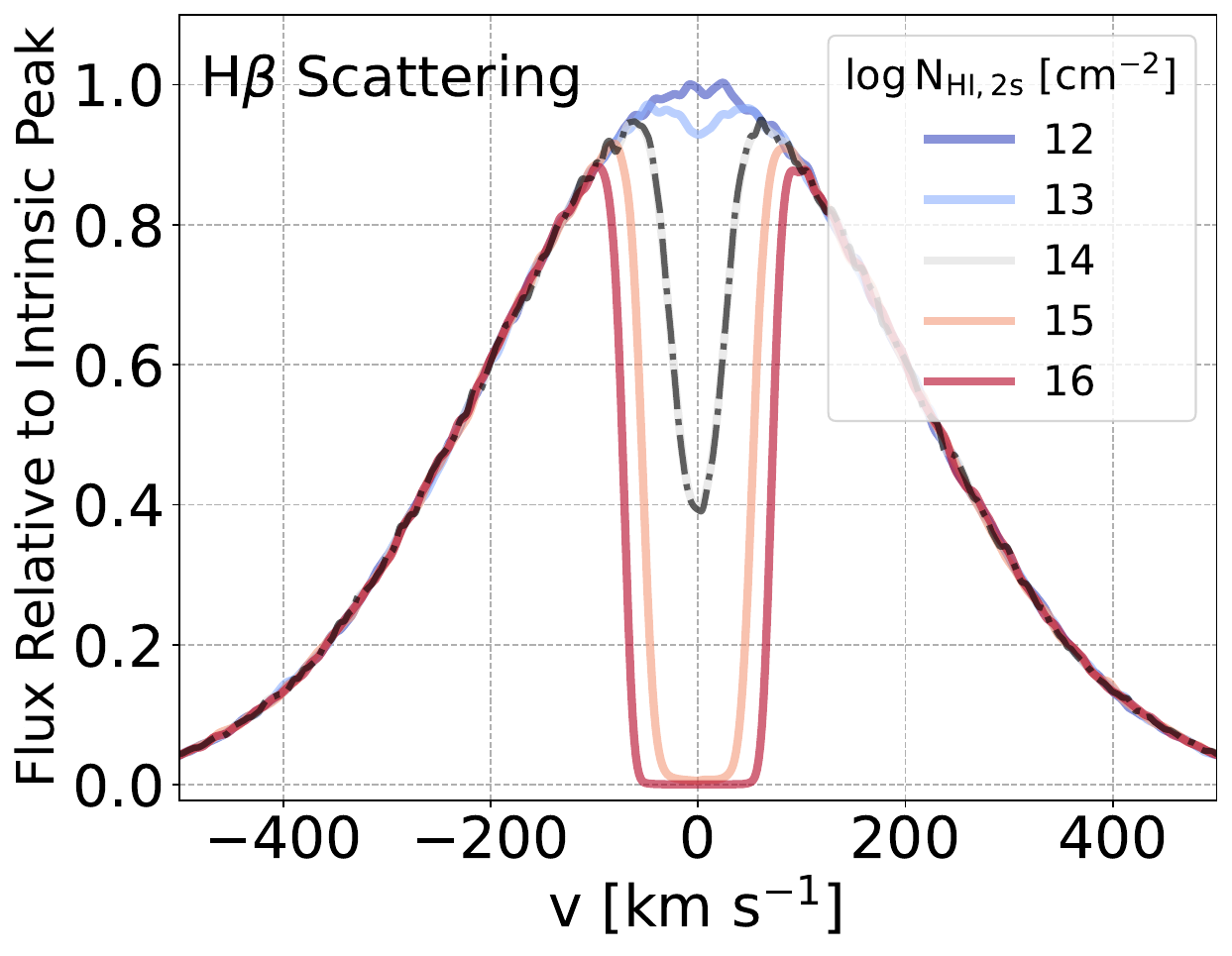}
    \includegraphics[width=0.33\textwidth]{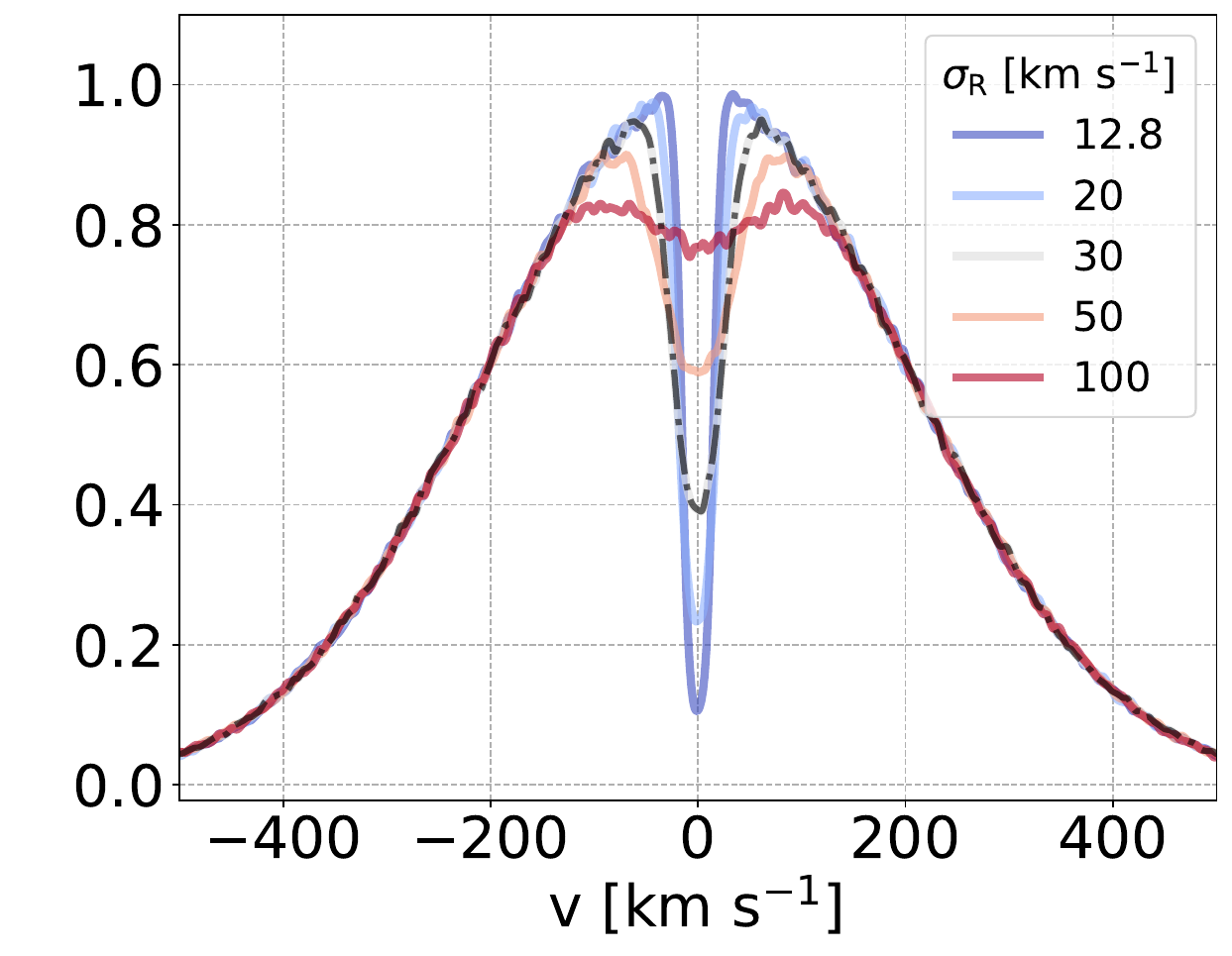}
    \includegraphics[width=0.33\textwidth]{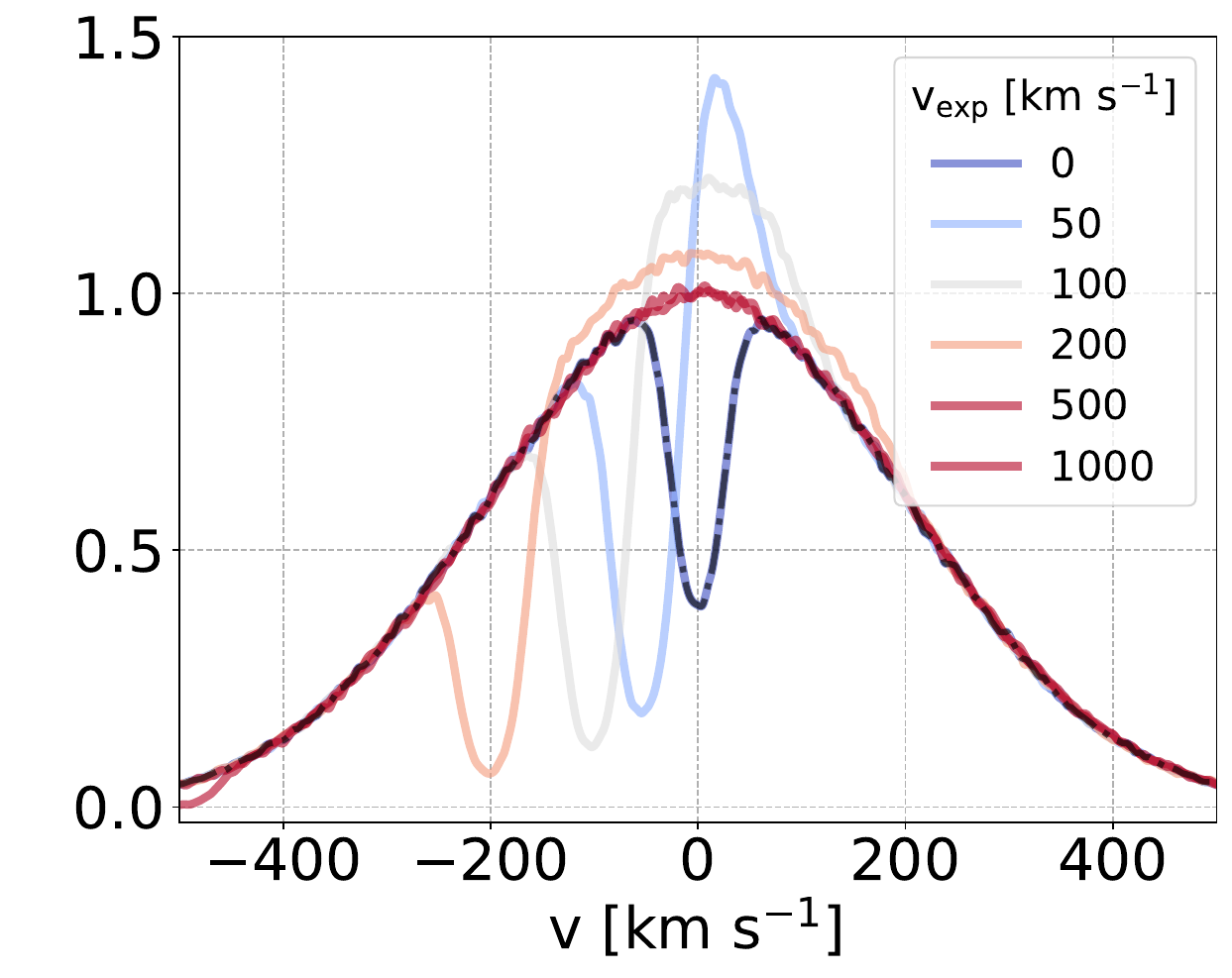}
    \caption{
    Spectra of resonance-scattered \Hb for various \NHItwo (left), \sigr (center), and \vexp (right). 
    The parameters of the fiducial case (black dot-dashed line) and the range of parameters for each panel are the same as those in Fig.~\ref{fig:spec_Ha}.
    The spectra are normalized by dividing the peak of intrinsic radiation at $v = 0 \kms$; $y$-axis represents the relative flux. 
    Unlike the double-peaked structures prominently visible in the H$\alpha$ spectra in the left and center panels of Fig.~\ref{fig:spec_Ha}, the H$\beta$ line generally shows a central absorption-like feature due to significant photon conversion to Pa$\alpha$ via scattering into $n=3$ states.
}
\label{fig:spec_Hb}
\end{figure*}

In this section, we examine the emergent H$\alpha$ spectra for a Gaussian input profile with an intrinsic width of $\sigsrc = 200 \kms$.
Considering that H$\alpha$ radiative transfer involves only the $2s-3p$ transition -- analogous to Ly$\alpha$ radiative transfer with the $1s-2p$ transition -- we anticipate spectral behavior similar to that of Ly$\alpha$ -- however, typically at much lower optical depths. Fig.~\ref{fig:spec_Ha} illustrates H$\alpha$ spectra for varying \NHItwo, \sigr, and \vexp using an intrinsic spectral width of \sigsrc = 200 \kms.

In the left panel of Fig.~\ref{fig:spec_Ha}, as \NHItwo increases, the double-peaked profiles near the line center become more pronounced, and the separation between the peaks widens due to an increased number of scatterings. The central panel shows that increasing the random motion of the scattering medium \sigr leads to broader double peaks, reflecting enhanced frequency diffusion driven by greater random gas motions.

In the right panel of Fig.~\ref{fig:spec_Ha}, an outflow velocity smaller than \sigsrc (i.e., \vexp = 50 and 100 \kms) shows a P-Cyni profile composed of an absorption-like feature blueward of line center and a red peak. For larger outflow velocities ($\vexp \geq 500\kms$), scattering becomes negligible because the medium effectively becomes optically thin for the intrinsic photons. 

In summary, the spectral behavior of \Ha is similar to that of Ly$\alpha$, albeit \lya is typically studied at much larger column densities $N_{\rm HI}\gtrsim 10^{17}\unitNHI$. Leaving aside the doublet structure, the emergent \Ha spectrum resembles the one typically found for metal resonance lines observed in outflowing environments (e.g., MgII \citealp{Prochaska2011,Chang2024}).

\subsubsection{H$\beta$ Spectrum}\label{sec:result_resonance_Hb}

\begin{figure}
    \centering
    \includegraphics[width=0.85\linewidth]{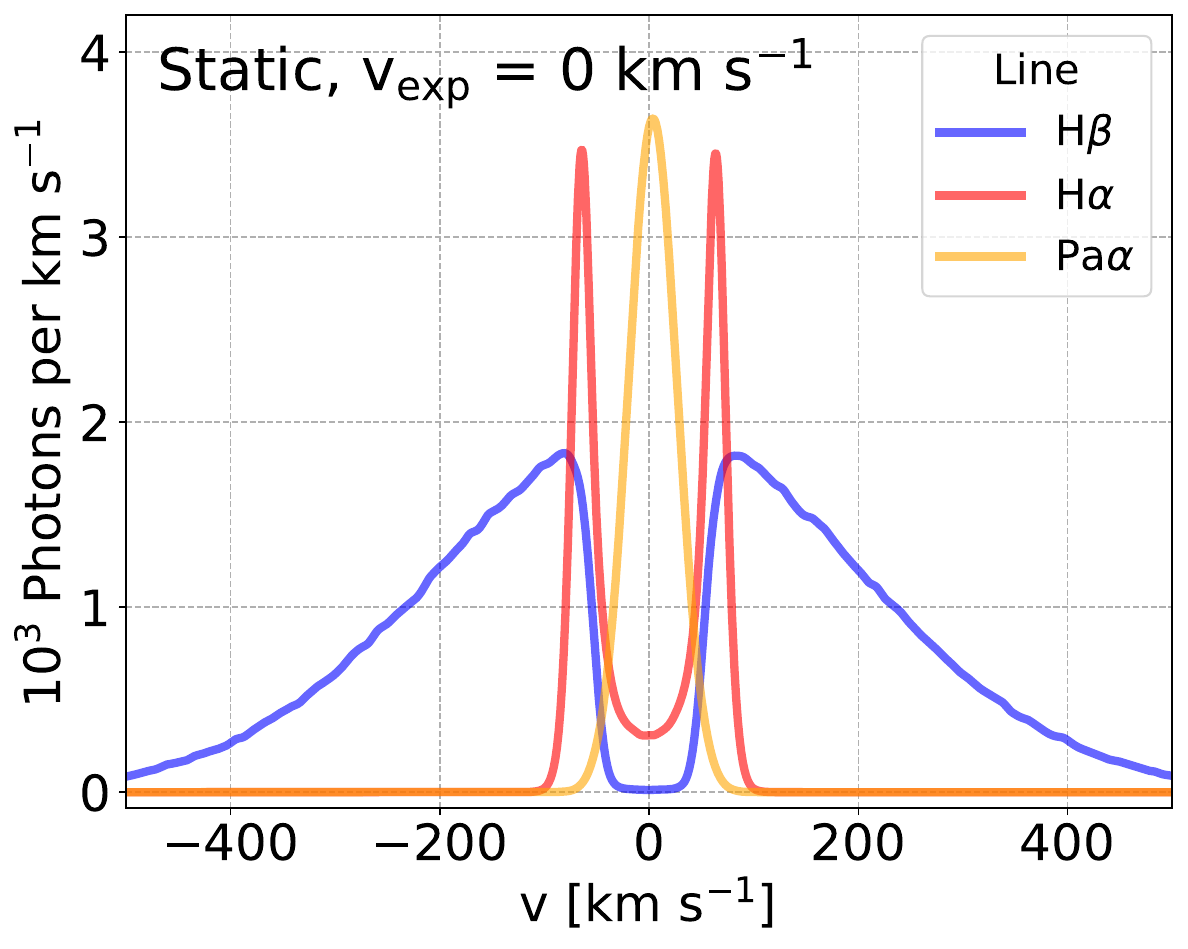}\\
    \includegraphics[width=0.85\linewidth]{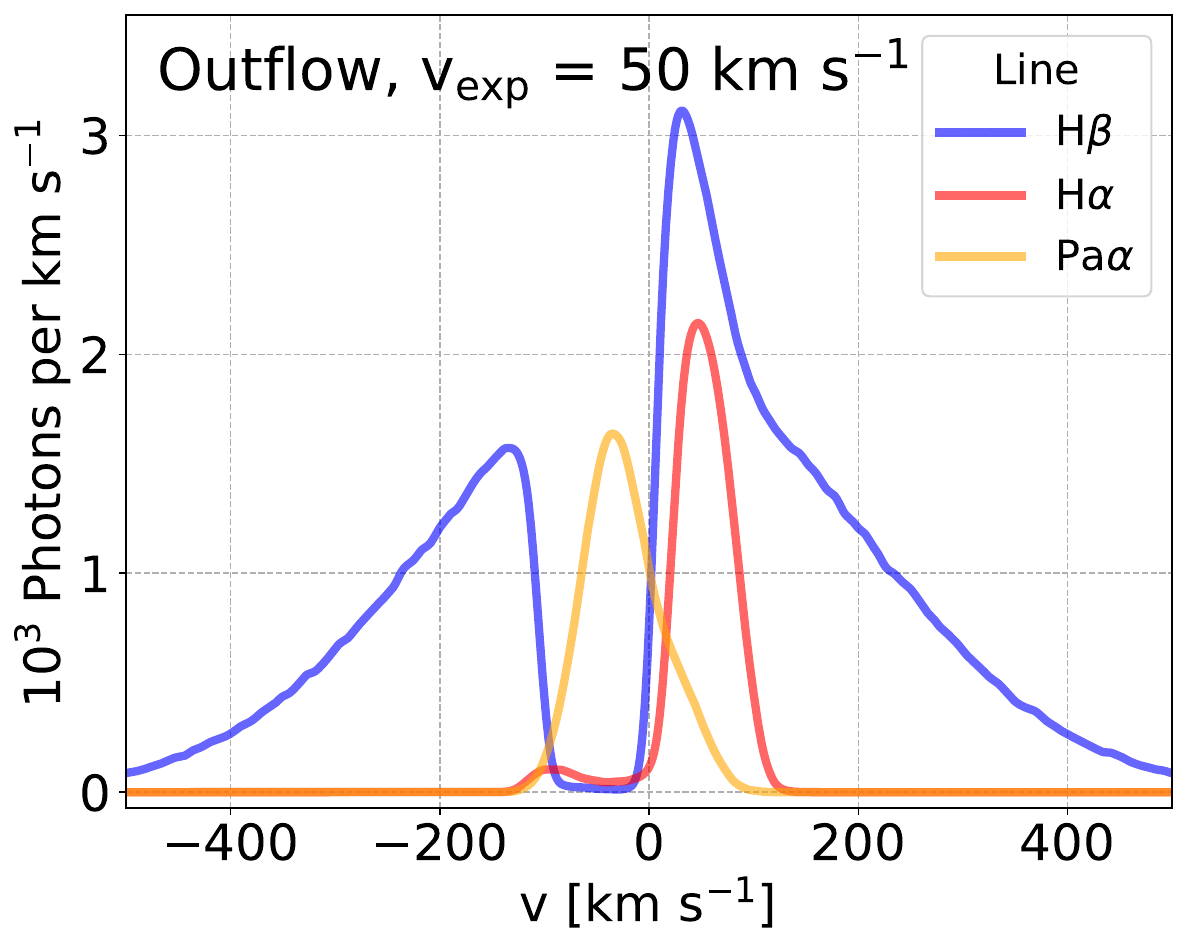}
    \caption{
    Simulated spectra as a consequence of resonantly scattered H$\beta$ radiative transfer, including $4p \to n = 2$ and $4p \to n = 3$ transitions and assuming only \Hb Gaussian emission with $\sigsrc = 200 \kms$. 
    The colors of solid lines represent simulated spectra of \Hb (blue), \Ha (red), and Pa$\alpha$.  
    \Ha and Pa$\alpha$ spectra are composed of photons produced by \Hb scattering to the $n=3$ state, while \Hb spectra are composed of scattered and directly escaping \Hb.
    $y$-axis is the number of photons per \kms to show how many photons convert from \Hb to Pa$\alpha$ and \Ha.
    \NHItwo and \sigr are fixed at $10^{15} \unitNHI$ and 30 \kms, respectively.
    The top panel shows spectra for a static medium (\vexp = 0 \kms), while the bottom panel presents results for an outflowing medium (\vexp = 50 \kms).
}
    \label{fig:spec_Hb_Ha_Paa}
\end{figure}

\begin{figure}
    \centering
    \includegraphics[width=0.90\linewidth]{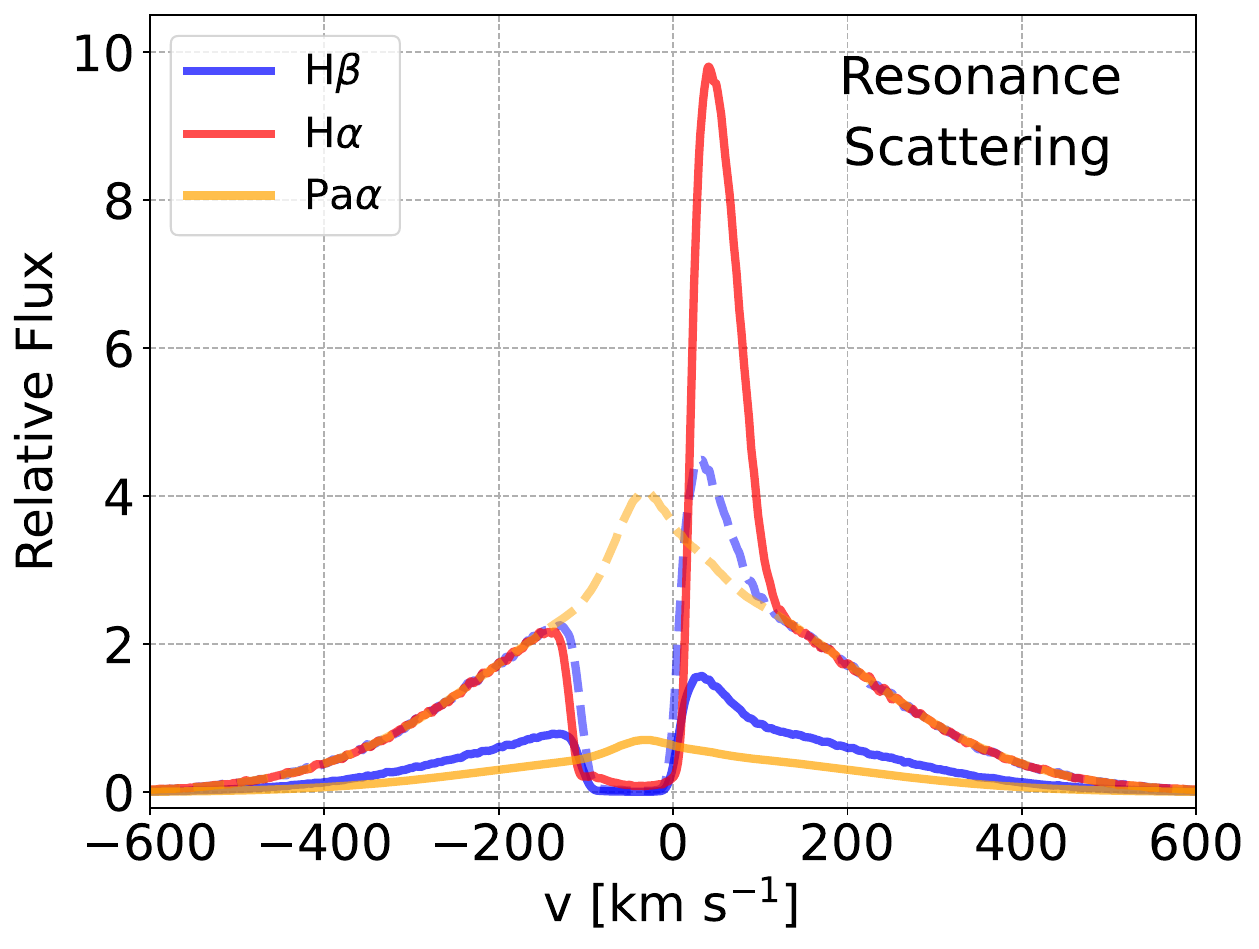}
\caption{
    Emergent spectra of \Ha, \Hb, and P$\alpha$ after radiative transfer of \Ha and \Hb resonance scattering through an outflow medium with $\vexp = 50\,\kms$ and $\NHItwo=10^{15}\,\unitNHI$.
    The widths of intrinsic profiles of \Ha, \Hb, and Pa$\alpha$, \sigsrc, are fixed at 200 \kms, assuming the flux ratios between lines under case B recombination. 
    The blue solid line is identical to the H$\beta$ spectrum in the bottom panel of Fig.~\ref{fig:spec_Hb_Ha_Paa}. 
    The red (orange) solid line shows the integrated H$\alpha$ (Pa$\alpha$) spectrum combining the emergent spectrum of intrinsic emission and those produced via H$\beta$ scattering to the n =3 state in bottom panel of Fig.~\ref{fig:spec_Hb_Ha_Paa}.
    While the \textit{solid lines} show the correct (observable) flux ratios, the \textit{dashed curves} are normalized to the wings, illustrating their relative spectral shapes.
    }
    \label{fig:observed_spectra}
\end{figure}

In this section, we consider the intrinsic H$\beta$ emission from a central source, assuming a width of $\sigsrc = 200 \kms$. The radiative transfer of H$\beta$ involves transitions from the $4p$ state to both the $2s$ state and the $3s/3d$ states. The scattering characteristics of H$\beta$ differ from those of H$\alpha$ and Ly$\alpha$ because of the transition to the $3s/3d$ states, as discussed in Section~\ref{sec:monochromatic_balmer}.

Fig.~\ref{fig:spec_Hb} shows H$\beta$ spectra for various \NHItwo, \sigr, and \vexp. In the left and central panels for a static medium ($\vexp=0\,\kms$), the spectra do not exhibit clear double-peaked features like those observed in H$\alpha$ spectra in Fig.~\ref{fig:spec_Ha}. Instead, the profiles look like simple Voigt absorption profiles centered at the intrinsic wavelength. This occurs because most H$\beta$ photons rarely experience sufficient scattering events to form a double-peaked profile before converting into Pa$\alpha$. 
As discussed in Section~\ref{sec:monochromatic_balmer}, when the optical depth at H$\beta$ exceeds $\sim 10$, most H$\beta$ photons undergo scattering through the $4p \rightarrow 3s/3d$ transition, converting into Pa$\alpha$ and H$\alpha$ photons (cf. Fig.~\ref{fig:energy_level}).
The branching ratio to $n=3$ states $\sim 0.26$ is non-negligible, implying that it is statistically unlikely for H$\beta$ photons to remain in the H$\beta$ state after more than about four scatterings ($\sim 1/{\rm BR3}$).

In the right panel of Fig.~\ref{fig:spec_Hb}, for the slow outflow case smaller than \sigsrc = 200 \kms (\vexp$=50$ and $100\,\kms$), a red peak emerges due to the scattering in the outflowing medium. This behavior contrasts with the static case because the frequency shifts in an outflow allow some scattered H$\beta$ photons to escape before conversion to Pa$\alpha$. Nevertheless, since conversion to Pa$\alpha$ is still efficient, the red peak in H$\beta$ remains weaker compared to that seen in the corresponding H$\alpha$ spectra in the right panel of Fig.~\ref{fig:spec_Ha}.
We will discuss the differences between \Ha and \Hb spectra in the following section.

\subsubsection{Comparing \Ha, \Hb, and Pa$\alpha$ Spectra}
Fig.~\ref{fig:spec_Hb_Ha_Paa} illustrates the impact of multiple-transition scattering of H$\beta$ photons on the emergent line profiles of H$\beta$, H$\alpha$, and Pa$\alpha$, assuming only \Hb emission as input radiation. In the top panel for the static medium ($\vexp =0\,\kms$), the \Hb spectrum exhibits a central absorption-like feature without clear double peaks, indicating significant photon conversion into Pa$\alpha$ via the $4p \rightarrow n=3$ transition. The  Pa$\alpha$ spectrum emerges as a symmetric Gaussian profile, shaped by the random motion \sigr of the medium. Simultaneously, this conversion generates additional H$\alpha$ photons, which are emitted at the line-center wavelength and subsequently scatter, forming a double-peaked profile.

In the bottom panel of Fig.~\ref{fig:spec_Hb_Ha_Paa}, the presence of the outflow ($\vexp =50\kms$) significantly modifies the spectral shape. The H$\beta$ line exhibits a prominent P-Cygni profile. The partial photons on the blue side of the line center are scattered into the $n=3$ states due to the outflowing medium, converting into Pa$\alpha$ photons that escape as a blueshifted emission. The simultaneously generated H$\alpha$ photons continue to scatter, forming a redshifted peak. 
The effect of \Hb scattering in the outflowing medium creates distinct observational signatures, blueshifted Pa$\alpha$ and redshifted H$\alpha$ emission features.

To investigate the observational impact of H$\beta$ scattering, we assume the central source emits intrinsic Gaussian profiles for H$\alpha$, H$\beta$, and Pa$\alpha$ with intrinsic width $\sigsrc =200 \kms$ and relative flux ratios following Case B recombination. Since the $n=3$ population of neutral hydrogen is negligible due to its short lifetime, intrinsic Pa$\alpha$ photons escape directly without scattering, resulting in a symmetric Gaussian profile.

Fig.~\ref{fig:observed_spectra} presents the observable emergent spectra (i.e., considering both the intrinsic as well as converted radiation) of H$\alpha$, H$\beta$, and Pa$\alpha$ in the outflowing case. Both H$\alpha$ and H$\beta$ exhibit P-Cygni profiles, with the red peak of H$\alpha$ stronger than that of H$\beta$, due to additional photons produced via the $n = 3$ transition. The Pa$\alpha$ profile has an enhanced blueward peak due to the direct conversion from H$\beta$ scattering, as shown in Fig.~\ref{fig:spec_Hb_Ha_Paa}.\\

In summary, even though the optical depths of H$\alpha$ and H$\beta$ are of similar magnitudes due to the same \NHItwo column density and comparable cross-sections (Eq.~\eqref{eq:optical_depth}, their radiative transfer processes differ substantially. 
While H$\alpha$ radiative transfer closely mirrors that of \lya in a low optical depth regime (i.e.,  dominated by resonance scattering without line conversion), H$\beta$ photons cannot undergo a large number of scatterings due to the large branching ratio into the $n=3$ state.
As a result, the H$\alpha$ line can be significantly reshaped due to resonant radiative transfer, and H$\beta$ only shows absorption-like features (with additional Pa$\alpha$ emergent at the corresponding velocity offset.
We will discuss the potential impact of resonant scattering on observed Balmer line profiles in Sections~\ref{sec:discussion_resonance} and \ref{sec:Thomson_Resonance}.

\begin{figure*}
    \centering
    \includegraphics[width=0.49\textwidth]{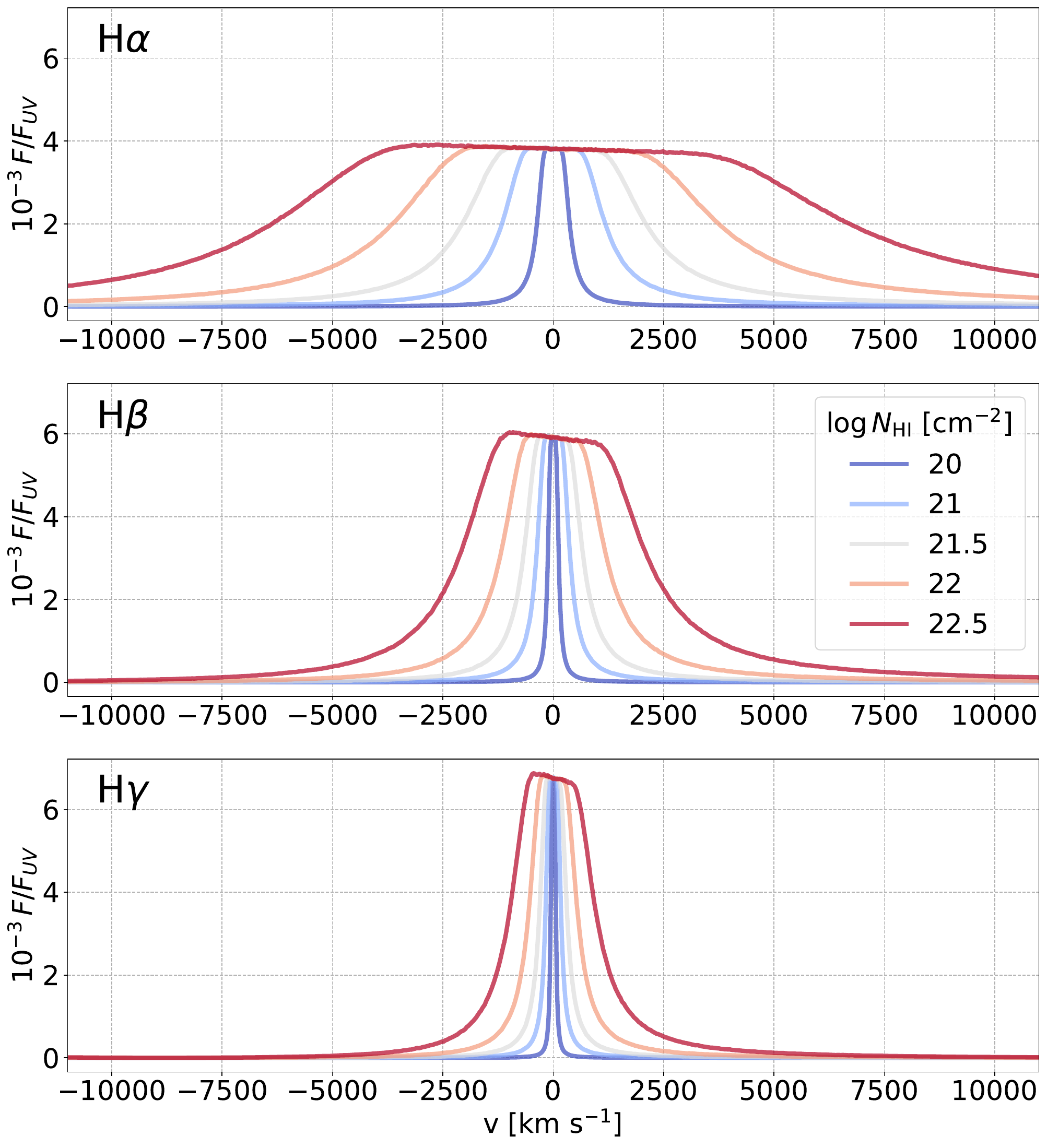}
    \includegraphics[width=0.49\textwidth]{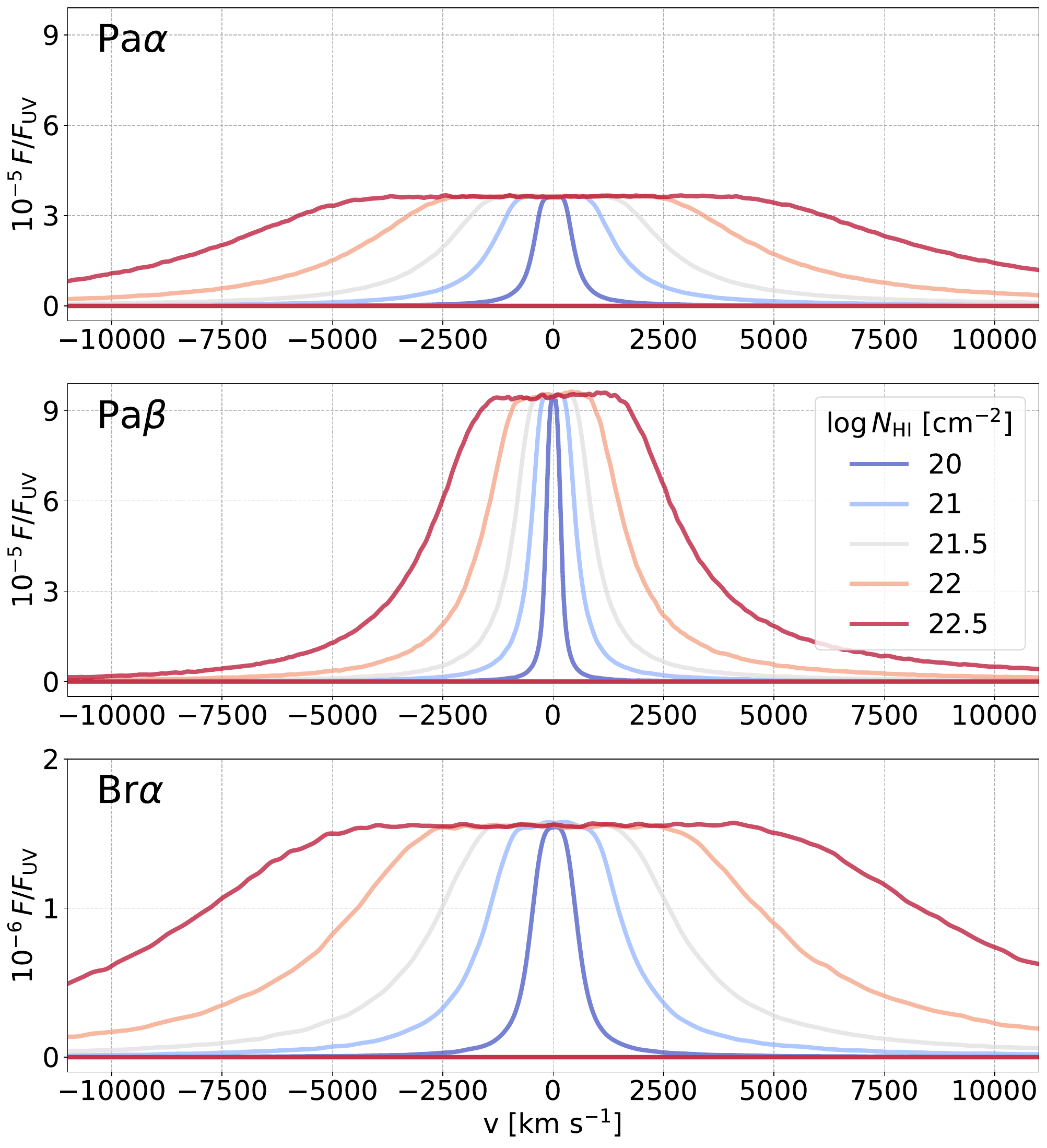}
    \caption{
        Raman-scattered features of hydrogen lines for various \hi column densities, $\NHI =  10^{20-22.5}\, \unitNHI$.
        The vertical axis $F/F_{\rm UV}$ denotes the flux of the Raman feature normalized by the incident UV continuum near the corresponding Lyman series line.
        Specifically, $F_{\rm UV}$ refers to the continuum around Ly$\beta$, Ly$\gamma$, and Ly$\delta$ for \Ha, \Hb, and H$\gamma$, respectively.
        \textit{Left:} Raman features in Balmer lines: H$\alpha$ (top), H$\beta$ (middle), and H$\gamma$ (bottom).
        \textit{Right:} Raman features in infrared transitions: Pa$\alpha$ (top), Pa$\beta$ (middle), and Br$\alpha$ (bottom).
        As \NHI increases, the wings broaden due to Raman scattering of UV continuum photons over a wider frequency range.
    }
    \label{fig:spec_Raman}
\end{figure*}

\subsection{Raman Scattering} \label{sec:result_Raman}

\begin{figure}
    \centering
    \includegraphics[width=0.48\textwidth]{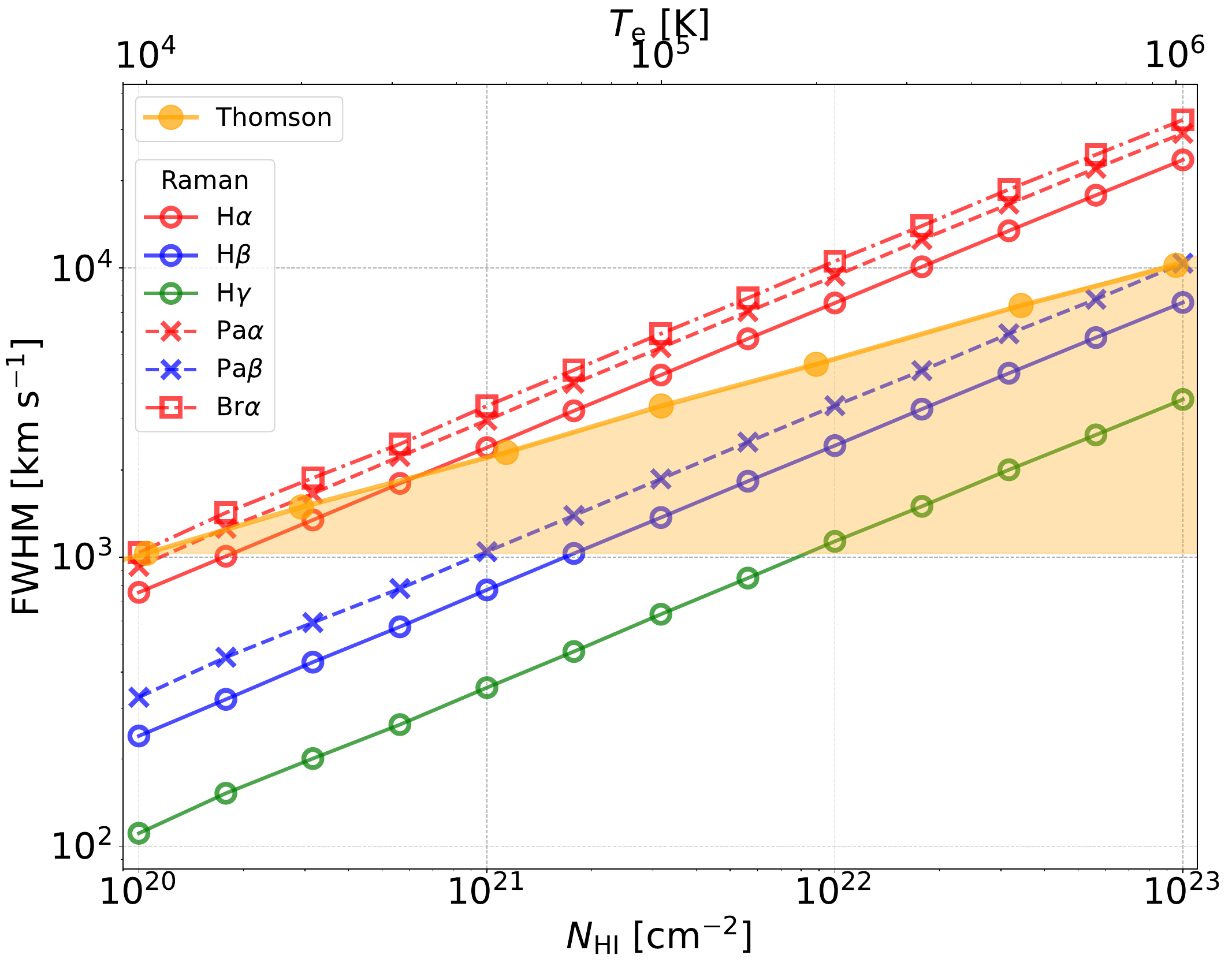}
    \caption{
        Full width at half maximum (FWHM) of Raman-scattered features.
        FWHM is plotted as a function of $N_{\rm HI} = 10^{20-23}\, \unitNHI$.
        The red, blue, and green solid lines represent Raman features of Balmer lines H$\alpha$, H$\beta$, and H$\gamma$, respectively, while dashed and dot-dashed lines correspond to Paschen and Brackett lines.
        The orange solid line represents FWHM of Thomson-scattered features as a function of $T_{\rm e}$ at fixed optical depth $\tau_{\rm Th} = 1$.
        Unlike Thomson scattering -- which yields the same width for all hydrogen lines due to its wavelength-independent cross section -- Raman scattering produces broader H$\alpha$ wings than H$\beta$ or H$\gamma$, driven by differences in scattering cross section and Raman wavelength shift.
    }
    \label{fig:width_broad_wing}
\end{figure}

Fig.~\ref{fig:spec_Raman} shows the Raman scattered features around various hydrogen lines, H$\alpha$, H$\beta$, H$\gamma$, Pa$\alpha$, Pa$\beta$, and Br$\alpha$ from the UV continuum near Ly$\beta$, Ly$\gamma$ and Ly$\delta$. As \NHI increases, Raman features get broader, reflecting the fact that a wider range of UV continuum photons near the Lyman series undergo Raman scattering to form broad features around hydrogen lines \citep{Chang2015, Chang2018b}.

For example, UV continuum near Ly$\beta$ produces broad wings around H$\alpha$ via Raman scattering into the $n=2$ state. Similarly, Ly$\gamma$ photons scatter into both $n=2$ and $n=3$ states, generating Raman features near H$\alpha$ and Pa$\alpha$, respectively. Ly$\delta$ photons also contribute via Raman scattering into the $n=4$ state, producing Br$\alpha$ features. At high \NHI ($>10^{21}\,\unitNHI$), the incident flat continuum is fully imprinted onto the Raman-scattered features, resulting in a flattened spectral shape near each line center. The slight slopes in these flat regions originate from the energy conservation inherent in the Raman scattering process ( Eq.~\ref{eq:Raman_broad}).

The left panel of Fig.~\ref{fig:spec_Raman} shows that the Raman H$\alpha$ feature is broader than those of H$\beta$ and H$\gamma$ at the same $N_{\rm HI}$.
This trend is driven by two factors: (1) the larger Raman scattering cross section near Ly$\beta$ compared to Ly$\gamma$ and Ly$\delta$ (see Fig.~\ref{fig:Raman_cross_section}), and (2) the different wavelength-broadening factors in Eq.~\ref{eq:Raman_broad}.
As a result, the $\alpha$ line of each hydrogen series (Balmer, Paschen, Brackett) exhibits the broadest Raman wings relative to the higher-order transitions, assuming a flat input continuum.
These spectral trends offer a potential observational diagnostic: broader wings of Balmer lines (e.g., H$\alpha$ compared to H$\beta$) may indicate a Raman-scattered wing.
We discuss these observational signatures of Raman scattering further in Section~\ref{sec:discussion_raman}.

We further compare the widths and profile shapes of Raman-scattered features around hydrogen lines.
In Fig.~\ref{fig:width_broad_wing}, when \NHI $=10^{20-23} \unitNHI$, the width of Raman H$\alpha$ features spans $\sim$1,000 to 30,000\,\kms, scaling approximately as $\NHI^{1/2}$. Moreover, the width varies across hydrogen lines depending on their origin of Lyman series; H$\alpha$, H$\beta$, and H$\gamma$ arise from Raman scattering of Ly$\beta$, Ly$\gamma$, and Ly$\delta$, respectively. 
Because Ly$\beta$ has both a higher scattering cross section and a larger Raman wavelength shift (see Fig.~\ref{fig:Raman_cross_section} and Eq.~\ref{eq:Raman_broad}), H$\alpha$ wings are significantly broader -- typically by a factor of $\sim$3 -- compared to H$\beta$.

In summary, Raman scattering produces broader features at higher \NHI due to increased optical depth and interaction range with the UV continuum. Continuum Raman-scattered features show the different widths of each hydrogen line.
If H$\alpha$ wings are significantly broader than H$\beta$, it suggests a contribution from Raman scattering of the UV continuum.
However, this width contrast is not significant when Ly$\beta$ and Ly$\gamma$ appear as intrinsic emission features in the input radiation, rather than a flat continuum.
We discuss this case in more detail in Section~\ref{sec:discussion_raman} and Appendix~\ref{sec:Raman_emission}.

\begin{figure}
    \centering
    \includegraphics[width=0.48\textwidth]{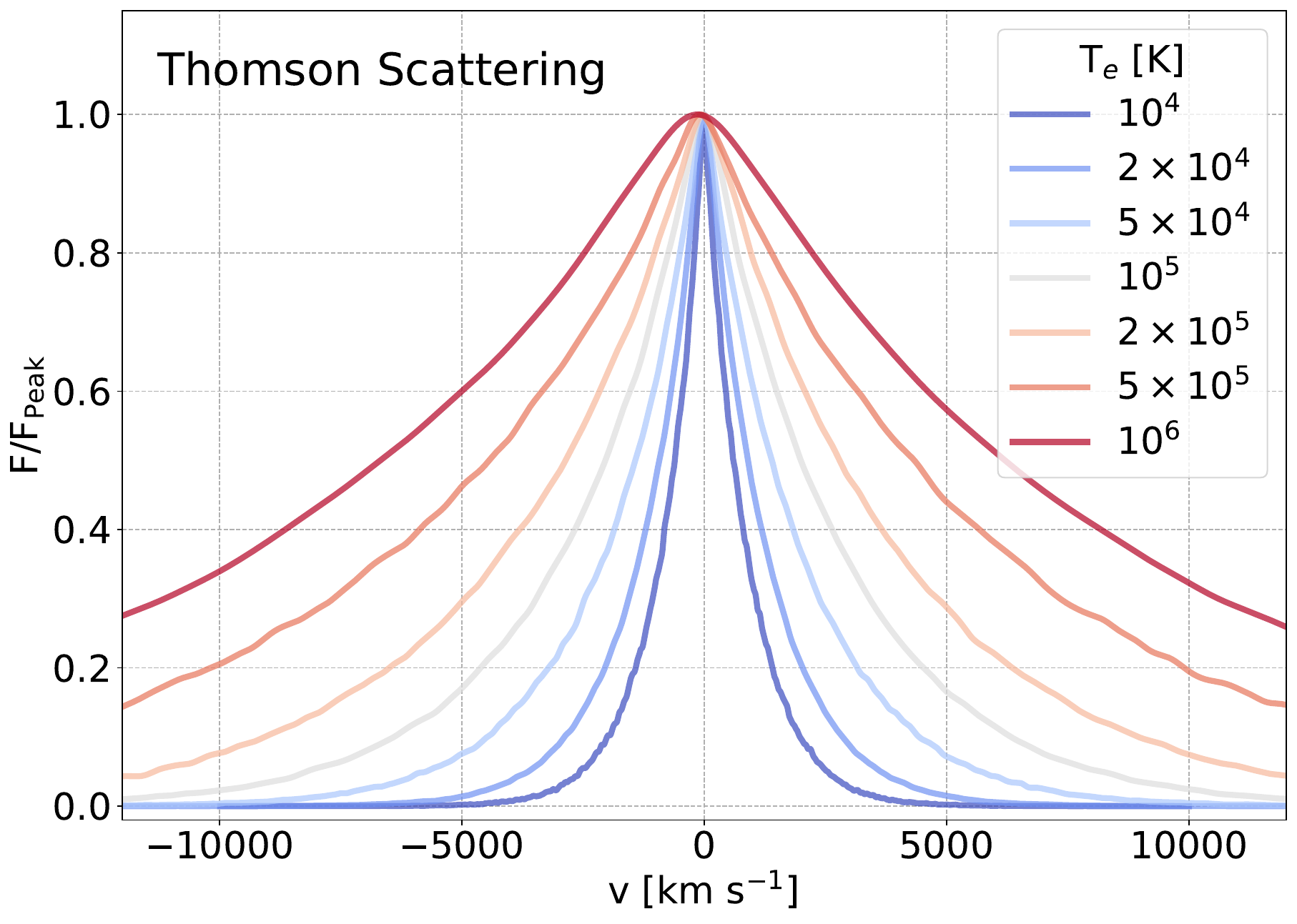} 
    \includegraphics[width=0.48\textwidth]{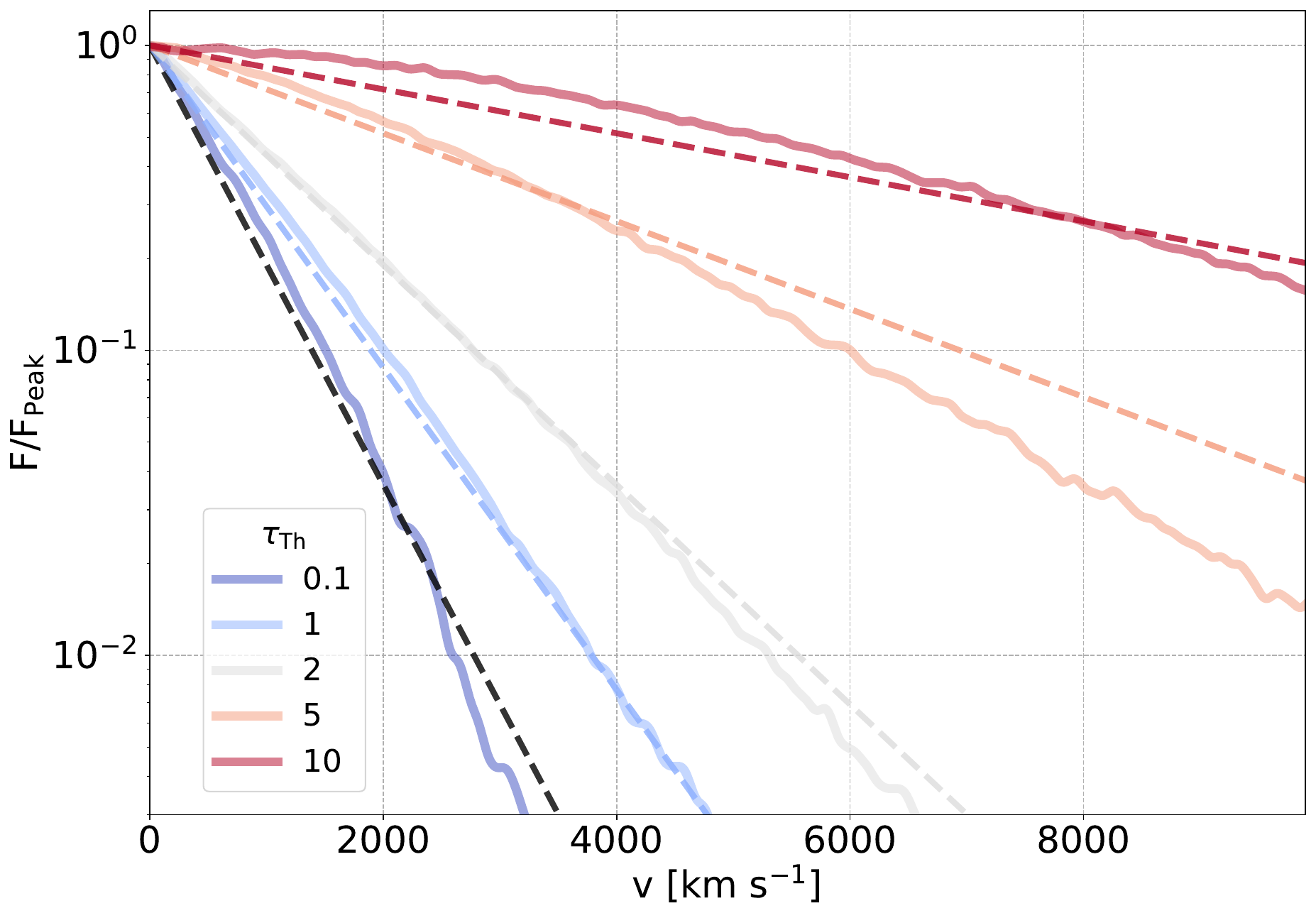} 
    \caption{
        Spectra composed of Thomson-scattered photons, normalized by the peak flux at the line center.
        \textit{Top:} Thomson-scattered profiles for various electron temperatures $T_{\rm e} = 10^{4-6}\, \rm K$ at fixed optical depth $\tau_{\rm Th} = 1$, plotted in linear scale.
        The width of the line wings increases with $T_{\rm e}$, corresponding to higher electron thermal speeds.
        \textit{Bottom:} Thomson-scattered profiles in the red wing of the line center at fixed $T_{\rm e} = 10^4\, \mathrm{K}$ for various optical depths $\tau_{\rm Th} = 0.1{-}10$, plotted in logarithmic scale.
        Solid lines show Monte Carlo simulation results, and dashed lines represent exponential fits. At low optical depths ($\tau_{\rm Th} < 1$), profiles follow an exponential shape, while at higher depths ($\tau_{\rm Th} \gtrsim 2$), multiple scatterings distort the profile from the exponential form.
    }
    \label{fig:spec_Thomson}
\end{figure}

\begin{figure}
    \centering
    \includegraphics[width=0.48\textwidth]{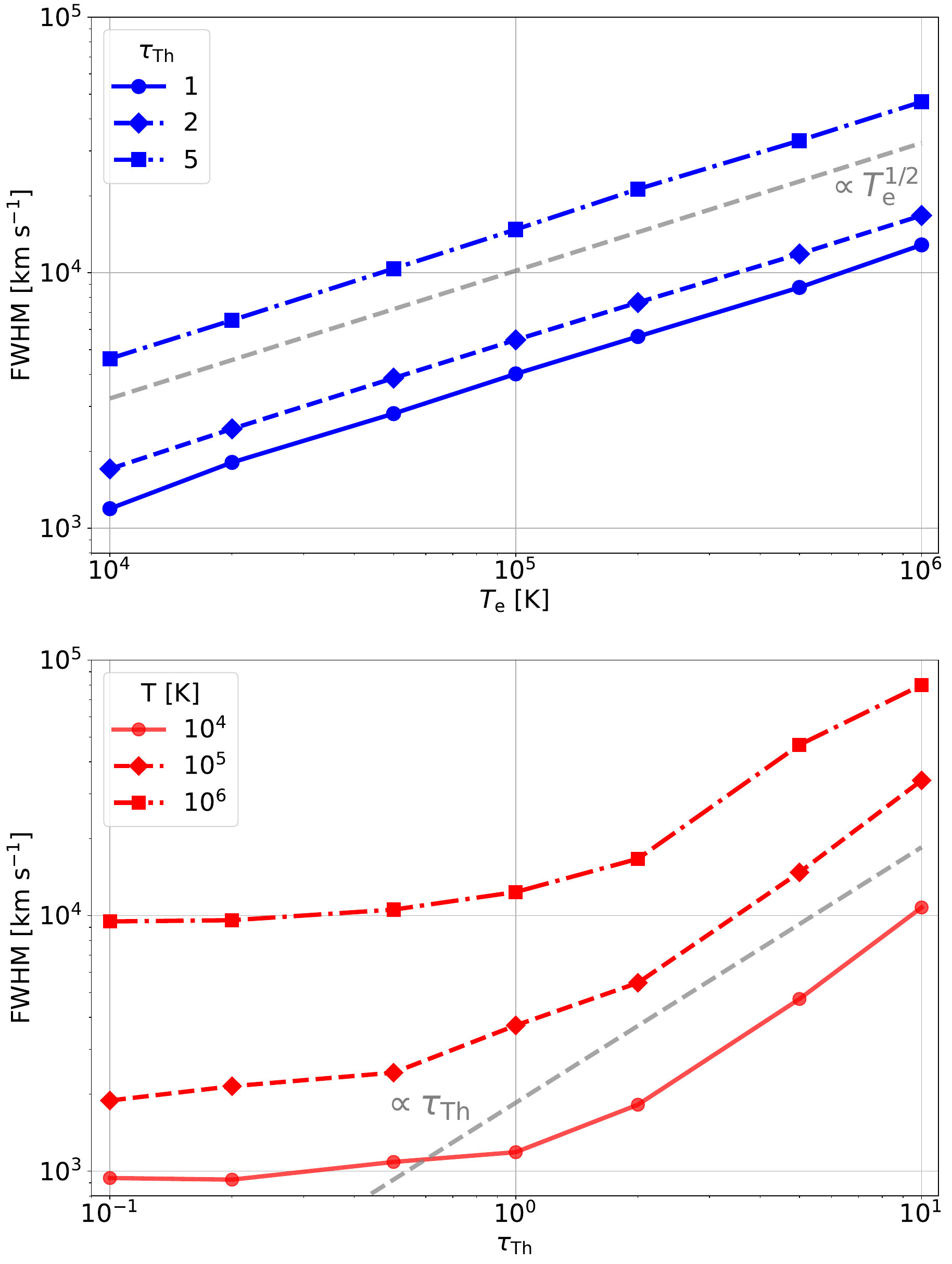} 
    \caption{
        Full width at half maximum (FWHM) of Thomson-scattered features.
        \textit{Top:} FWHM as a function of electron temperature $T_{\rm e}$ for different optical depths: $\tau_{\rm Th} = 1$ (solid), 2 (dashed), and 5 (dot-dashed).
        The gray dashed line indicates the scaling ${\rm FWHM} \propto T_{\rm e}^{1/2}$, consistent with thermal Doppler broadening.
        \textit{Bottom:} FWHM versus $\tau_{\rm Th}$ for fixed $T_{\rm e} = 10^4$ (solid), $10^5$ (dashed), and $10^6\, \rm K$ (dot-dashed).
        In the optically thick regime ($\tau_{\rm Th} > 1$), FWHM increases with $\tau_{\rm Th}$ due to multiple scatterings.
        In the optically thin case ($\tau_{\rm Th} < 1$), FWHM remains nearly constant and is dominated by single scattering.
    }
    \label{fig:Thomson_width}
\end{figure}

\subsection{Thomson Scattering} \label{sec:result_Thomson}

Fig.~\ref{fig:spec_Thomson} shows simulated results of Thomson-scattered line profiles for various electron temperatures $T_{\rm e} = 10^{4-6}\,$K and Thomson optical depth $\tau_{\rm Th} = 0.1-10$.
Since the dependence on the input radiation is negligible if $\sigsrc < v_{\rm e, th}$ (i.e., in most cases), we assume a width of the intrinsic emission $\sigsrc=50 \kms$ to focus on the broad wing features due to Thomson scattering.
In the left panel of Fig.~\ref{fig:spec_Thomson}, the width of the scattered wings increases with increasing $T_{\rm e}$, following the trend in electron thermal velocity $v_{\rm th,e}$ in Equation~\ref{eq:v_elec}.
Since Thomson scattering is independent of photon wavelength, the resulting broad wings appear around all emission lines with similar shapes, as discussed in Section~\ref{sec:thomson}.

The bottom panel of Fig.~\ref{fig:spec_Thomson} shows the dependence on the optical depth $\tau_{\rm Th}$ for a fixed electron temperature $T_{\rm e} = 10^4\, \mathrm{K}$.
In the optically thin case ($\tau_{\rm Th} \leq 1$), where single scattering dominates, the profiles exhibit exponential wings, consistent with the analytical form $\propto \exp(-v / v_0)$ \citep[][]{Sunyaev1980,Lee1999,Laor2006}.
With the single scattering approximation, the analytic solution of the Thomson scattered profile is given by
\begin{equation} \label{eq:Thomson_analytic}
    f(v) = f_0 e^{-v/v_{0}},
\end{equation}
where $v_o \approx 1.1 v_{\rm e, Th}$ \citep{Laor2006}.
This analytic solution (black dashed line) is akin to the Thomson wing at $\tau_{\rm Th} = 0.1$, in the bottom panel of Fig.~\ref{fig:spec_Thomson}.
When $\tau_{\rm Th}$ increases, the profile gets broader as a higher optical depth provides the opportunity for additional scattering.
However, as still single scattering mainly determines the formation of Thomson scattered features, the profile follows the exponential profile well at $\tau_{\rm Th} \leq 1$.
%In the optically thick case, multiple scatterings break exponentail profile d from a purely Gaussian thermal distribution.

In the optically thick case ($\tau_{\rm Th} \gtrsim 2$), multiple scatterings modify the profile shape, deviating from the exponential form as can be seen in the bottom panel of Fig.~\ref{fig:spec_Thomson}. This results in a mismatch between the simulated profiles (solid lines) and the exponential fits (dashed lines). The cumulative effects of repeated velocity shifts and directional variations through multiple scatterings result in a Gaussian-like profile near the line center with a power-law wing \citep{Sunyaev1980,Dessart2009}.

To quantify the broadening, we estimate the full width at half maximum (FWHM) of the Thomson-scattered profiles.
In the top panel Fig.~\ref{fig:Thomson_width}, we show the FWHM as a function of electron temperature. As expected from Eqns.~\eqref{eq:Thomson_analytic} \& \eqref{eq:v_elec}, the FWHM increases as $T_{\rm e}^{1/2}$, independent of $\tau_{\rm Th}$.
In the bottom panel, we show how the spectral broadening depends on the Thomson optical depth. FWHM remains constant for $\tau_{\rm Th} < 1$, where single scattering dominates, but increases approximately linearly with $\tau_{\rm Th}$ in the optically thick regime due to the cumulative effect of multiple scatterings.

In summary, Thomson scattering produces symmetric, broad line wings whose width scales with $T_{\rm e}^{1/2}$ and $\tau_{\rm Th}$.
In the optically thin regime, the line profile exhibits exponential profiles -- a key diagnostic feature that has been identified in broad Balmer emission of some AGNs and LRDs \citep[e.g.,][]{Laor2006, Rusakov2025}. 
However, in the optically thick case ($\tau_{\rm Th} \geq 5$), the Thomson scattered feature becomes non-exponential (as shown in the bottom panel of Fig.~\ref{fig:spec_Thomson}. 
Moreover, if the Thomson-scattering medium is outflowing, bulk motions can cause asymmetry in the broad scattered wings, as demonstrated in Appendix~\ref{sec:thomson_outflow} \citep[see also][]{Auer1972}.
Observationally, the absence of a strong red–blue asymmetry in LRD Balmer wings would constrain $v_{\exp}/v_{\rm th,e}\!\ll\!1$ in the \hii\ scattering layer.
We will further discuss the implications of a Thomson origin for Balmer line broadening in Section~\ref{sec:discussion}.

\section{Discussion}\label{sec:discussion}

\subsection{Observational Signatures of Scattering Mechanisms}

\begin{table*}
    \centering
    \caption{Summary of Scattering Mechanisms in LRDs}
    \label{tab:scattering_summary}
    \begin{tabular}{lcccc}
        \hline
        Process & Medium & Spectral Features & Broadening  \\
        \hline
        Resonance (H$\alpha$, H$\beta$) & H\,\textsc{i} ($n=2$, $2s$) & Double peaks, P-Cygni, variation in the \Ha/\Hb flux ratio & ---  \\
        Raman (UV $\rightarrow$ optical/IR) & H\,\textsc{i} ($n=1$, $1s$) & Broad wings with different widths for each line, up to $\sim10^4$ \kms & $\propto \sqrt{N_{\rm HI}}$ \\
        Thomson & Free electrons & Symmetric broadening with same width for all lines, up to $\sim 10^3 \kms$ & $\propto \sqrt{T_e}$, $\tau_{\rm Th}$ \\
        \hline
    \end{tabular}
    %\par\vspace{0.5em}
    %\raggedright
    %\textit{Note.} $\dagger$: Electron thermal speed $v_{\mathrm{th,e}}$ in Eq~\ref{eq:v_elec}, $\sim 500$ \kms\ at $T_e = 10^4$ K.
\end{table*}

We have investigated three distinct radiative transfer processes that can shape the observed profiles of hydrogen emission lines in LRDs:
\begin{enumerate}
\item Resonance scattering of Balmer lines with hydrogen atoms in the 2s state can produce double-peaked or P-Cygni-like features, influenced by the kinematics of the medium (i.e., \vexp and \sigr in the geometry used here). Furthermore, resonance scattering of \Hb affects the flux ratio of hydrogen lines, as multiple \Hb scatterings convert \Hb photons to Pa$\alpha$ and \Ha via de-excitation to the $n = 3$ state (see Fig.~\ref{fig:Hb_to_Ha} and Section~\ref{sec:monochromatic_balmer}).
\item Raman scattering of UV continuum photons near the Lyman series by ground-state hydrogen atoms can generate broad wings around Balmer lines, with widths up to several tens of thousands of \kms, depending primarily on the HI column density \NHI (with the width scaling as $\NHI^{1/2}$; see Section~\ref{sec:result_Raman} and specifically Fig.~\ref{fig:width_broad_wing}).
\item Thomson scattering of Balmer line photons by free electrons leads to symmetric broadening, with widths proportional to the electron's thermal speed $v_{\rm th,e}$. While mostly this leads to exponential wings ($\propto \exp (-v/v_{\rm e, th})$), in the optically thin case (Thomson optical depth  $\tau_{\mathrm{Th}} < 1$), Thomson-scattered features exhibit potentially observable deviations (cf. Section~\ref{sec:result_Thomson} \& Fig.~\ref{fig:spec_Thomson}).
% , and the width is also proportional to $\tau_{\mathrm{Th}}$.
\end{enumerate}

Fig.~\ref{fig:width_broad_wing} shows a summary of the emergent line width due to Raman and Thompson scattering. Importantly, the broad components of hydrogen emission lines arising from Raman scattering exhibit different widths for each line, while Thomson scattering produces similar broadening across all lines, emitted from the same emission region. 
One should note that if indeed the broad component of LRDs originates from Raman or Thomson scattering, it should be polarized \citep{Lee1998, Lee1999,Kim2007} and will be observable through future instruments and telescopes such as the Habitable Worlds Observatory \citep{Neiner2025}.
These distinctions provide a diagnostic for identifying the dominant scattering mechanism in LRDs. Tables~\ref{tab:scattering_summary} summarize the key properties of each process.

\subsection{Can Resonance Scattering Broaden the Observed \Hb Profile?}\label{sec:discussion_resonance}

\begin{figure}
    \centering
    \includegraphics[width=0.48\textwidth]{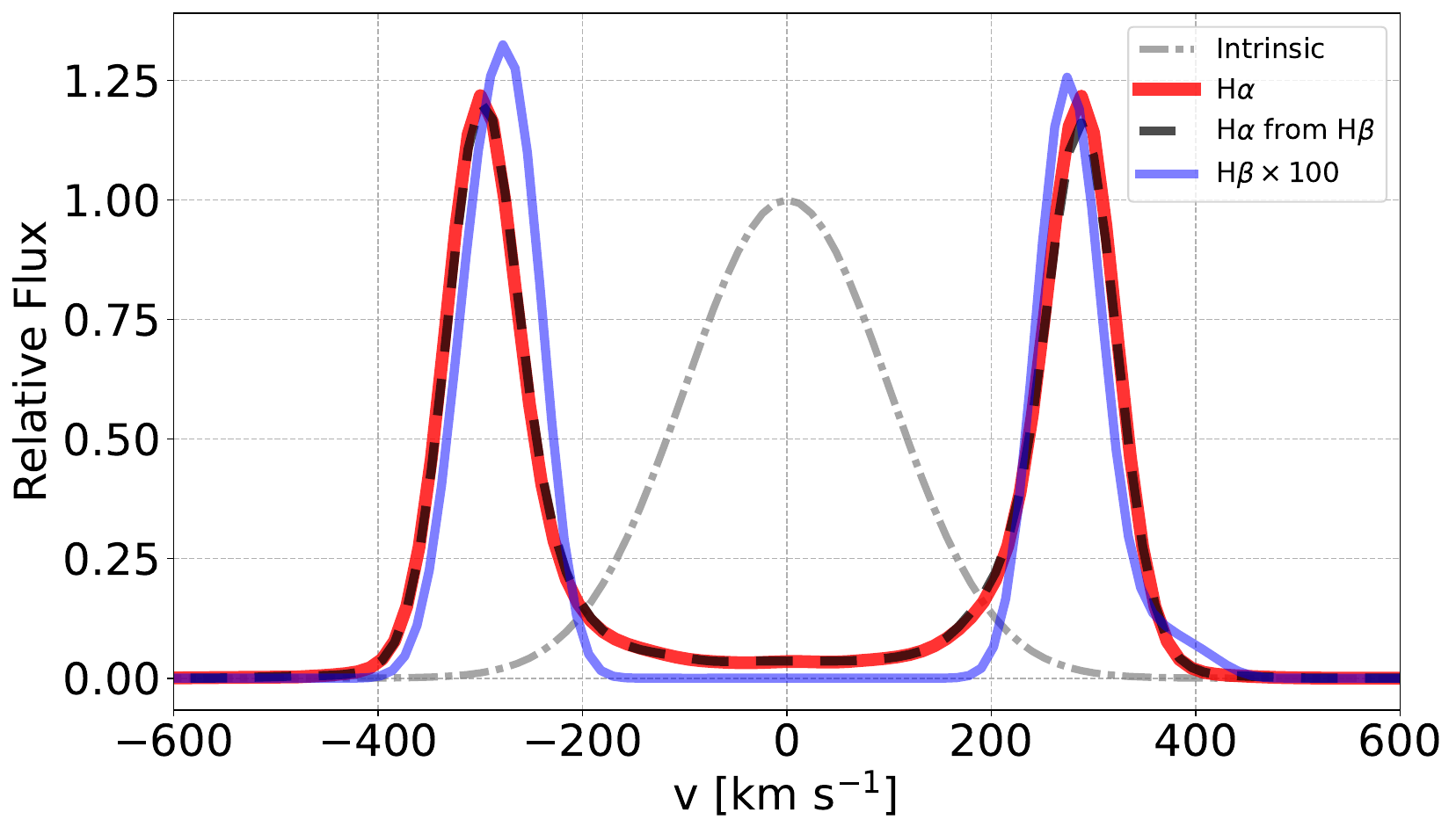}
    \caption{
    \Ha and \Hb spectra at $\NHI =10^{17} \unitNHI$, $\sigran = 100 \kms$, and $\sigsrc = 100 \kms$.
    The intrinsic flux ratio of \Ha and \Hb is fixed at 1 to compare profiles of emergent spectra.
    The solid and dashed red lines represent \Ha spectra from intrinsic \Ha and \Hb, respectively.
    The blue solid line is \Hb spectra multiplied by a factor of 100. 
    At high \NHI and \sigran, \Hb can be broadened by scattering, but its flux is extremely low as most photons convert to \Ha through numerous scatterings.
    }
    \label{fig:Ha_Hb_static}
\end{figure}

Resonance scattering leads to both frequency and spatial diffusion of photons. For example, resonantly scattered \lya emission typically exhibits a double-peaked profile in static media \citep{Neufeld1990}. The absorption features seen in the Balmer lines of LRDs indicate an enhanced $n = 2$ population \citep{Juodzbalis2024a, Juodzbalis2025, Inayoshi2025, Matthee2024,Naidu2025}, suggesting that the gas in these systems is optically thick to Balmer transitions, thus enabling resonance scattering.
\cite{Naidu2025} proposed that the double-peaked \Hb profile in LRDs originates from resonance scattering, analogous to \lya. However, the presence of multiple branching channels at the $n = 4$ level makes \Hb resonance scattering fundamentally different from that of \lya. This section explores whether resonance scattering of \Hb in the static medium can realistically account for the observed \Hb features.

For \Ha, the emergent profile in a static medium shows a characteristic double peak due to resonance scattering (see Fig.~\ref{fig:spec_Ha}), like \lya. However, as discussed in Section~\ref{sec:monochromatic_balmer}, in the optically thick regime ($\tau_{\rm H\beta} \gtrsim 10$), most \Hb photons undergo conversion to P$\alpha$ and \Ha via transitions to the $n = 3$ state. As a result, the emergent \Hb spectrum in these conditions primarily reflects photons that escape directly, with an absorption-like dip at the line center (see the top panel of Fig.~\ref{fig:spec_Hb_Ha_Paa}). Most scattered \Hb photons near the line center thus escape the medium as Pa$\alpha$ or \Ha photons.

To test whether \Hb resonance scattering can be a dominant broadening mechanism, we perform radiative transfer simulations in a static medium ($\vexp = 0 \kms$) with a high column density of $\NHItwo = 10^{17} \unitNHI$ (i.e., $\tau_{\rm 0, H\beta} > 10^3$) and random velocity dispersion $\sigr = 100 \kms$. We inject \Ha and \Hb photons from a central source, each with intrinsic width $\sigsrc = 100 \kms$ and equal flux, ensuring all photons undergo multiple scatterings.
Fig.~\ref{fig:Ha_Hb_static} shows the emergent \Ha spectrum from intrinsic \Ha (solid red line) and from intrinsic \Hb emission (dashed black line). The two spectra are nearly identical, as most \Hb photons are rapidly converted to \Ha in the inner scattering region and then escape via subsequent scatterings.

Although the vast majority of \Hb photons are converted to Pa$\alpha$ and \Ha ($\approx 99\, \%$ in this case) , a small fraction emerges as \Hb. The emergent \Hb spectrum (blue line, scaled by a factor of 100 for visibility) in Fig.~\ref{fig:Ha_Hb_static}) exhibits a clear double-peaked profile. Notably, the peak separation in \Hb is similar to that in \Ha, since $\tau_{\rm 0, H\alpha} \approx \tau_{\rm 0, H\beta}$.
In summary, \Hb spectra can indeed develop a double-peaked shape via resonance scattering. However, due to the efficient conversion to Pa$\alpha$, the emergent \Hb flux is several orders of magnitude weaker than \Ha. Since the flux ratio of \Ha/\Hb in LRDs is not exceeding a hundred, resonance scattering of \Hb is unlikely to be the dominant broadening mechanism\footnote{Note that if the \hi gas is optically thick not only to Balmer lines but also to the Paschen series -- due to a large $n = 3$ population -- then P$\alpha$ photons may also undergo scattering. In this case, the emergent \Hb flux can increase, and P$\alpha$ shows a double-peaked profile.
In gas that is optically thick to both Lyman and Balmer series, the scattering behaviors of Ly$\beta$ and \Ha resemble those of \Hb and P$\alpha$ in gas that is optically thick to Balmer and Paschen series, respectively. \cite{Chang2018a} explored \Ha and Ly$\beta$ emission via resonance scattering and found both can exhibit double-peaked profiles.
}.

\subsection{Detection of Raman Broad Wing}\label{sec:discussion_raman}
Raman-scattered wings around \Ha and \Hb have been observed in a variety of astrophysical environments, such as symbiotic stars \citep{Nussbaumer1989,Chang2018b, Lee2018}, planetary nebulae \citep{Lee2000, Lee2006, Miranda2022}, and \hii regions \citep{Dopita2016, Henney2020}. Theoretical studies have also predicted Raman broad wings in AGNs \citep{Chang2015, Kokubo2024}.
The strength of the Raman scattering wings depends on the input UV radiation. If all incident UV photons are fully Raman scattered, the resulting scattered features have strengths of approximately $\sim 10^{-3}$ relative to the input UV flux (see Fig.~\ref{fig:spec_Raman}) due to less energetic scattered photons (Eq.~\eqref{eq:raman_energy}) and Raman broadening (Eq.~\eqref{eq:Raman_broad}). Therefore, environments where Raman scattering is observed typically require both strong UV radiation and an optically thick \hi medium.

LRDs can provide such conditions, given their strong Balmer breaks and V-shaped SEDs \citep{Kocevski23,Matthee2024,Greene24,Taylor2025,Akins2025}. However, the broad wings around \Ha and \Hb observed in LRDs exhibit similar widths over 1000 \kms \citep{DEugenio2025a,DEugenio2025b,Juodzbalis2025, Brazzini2025}, which contrasts with the basic Raman scattering expectation (see Fig.~\ref{fig:width_broad_wing}) which predicts, e.g., a FWHM ratio of $\sim 3$ between \Ha and \Hb.

Thus, it seems that Raman scattering is unlikely to be important for LRDs. However, it is important to note that Raman scattering to the $n = 2$ state, specifically the transition $1s\to np\to 2s$, leaves an electron in the metastable 2s state and can enhance the $n = 2$ population, increasing the optical depth for resonance scattering of Balmer lines. 
Therefore, although it likely does not shape the emergent spectrum, Raman scattering of continuum photons can still affect the radiative transfer dynamics in this manner.
\\

In addition, as shown in Appendix~\ref{sec:Raman_emission}, Raman scattering of emission lines from the Lyman series can, in fact, produce broad wings of comparable width for both \Ha and \Hb. Assuming the same width of Ly$\beta$ and Ly$\gamma$ emissions, the widths of Raman scattered \Ha and \Hb are $\approx 6.4$ and $5.0$ (cf. Eq.~\eqref{eq:Raman_broad} and Section~\ref{sec:raman}) times broader than those of the Lyman series, resulting in the width ratio of \Ha and \Hb $\approx 1.28$. This is because the Raman-scattered profiles reflect the velocity distribution of the source emission rather than that of the Raman cross-section.
This implies that for observed widths of $\sim 1000\kms$ for LRDs in \Ha, the width of Ly$\beta$ needs to be $\sim 150\kms$, which is well within the observed narrow component of LRDs. 

In this case, to form the broad feature with 10\% of \Ha (\Hb) flux by Raman scattering, the flux ratio of Ly$\beta$/\Ha $\approx 0.64$ (Ly$\gamma$/\Hb $\approx 0.5$) is required, assuming Ly$\beta$ (Ly$\gamma$) photons are fully Raman scattered due to high \NHI. Although a higher Lyman series emission, such as Ly$\beta$ and Ly$\gamma$, is not allowed under case B recombination, the flux ratios of Ly$\beta$/\Ha and Ly$\gamma$/\Hb are $\approx 5$ and 2 under case A recombination \citep{Storey1995}, respectively, which are high enough to cause significant Raman scattered features around \Ha and \Hb. 

Consequently, without strong emission in the higher Lyman series, Raman scattering cannot be the dominant process responsible for the broad wings around Balmer lines in LRDs. However, if Ly$\beta$ and Ly$\gamma$ emission lines are present, Raman scattering could still contribute, producing a characteristic FWHM ratio of \Ha to \Hb of $\gtrsim 1.28$ (cf. Fig.~\ref{fig:Raman_width_Gau} and Appendix~\ref{sec:Raman_emission}). Future higher signal-to-noise observations of larger LRD samples may be able to detect such subtle line width differences.
Additionally, while higher Lyman-series photons are converted to Balmer lines due to Raman scattering, in practice a fraction of Ly$\beta$ and Ly$\gamma$ photons may escape and become observable. Currently, such higher Lyman lines have not been detected in LRDs, but future rest-UV spectroscopy (e.g., \citealp{Tripodi2025}) may uncover them.
If such Lyman emission features are observed, comparing their measured spectral widths with those required for Raman broadening could serve as a test of this scenario. As discussed above, Raman scattering can explain \Ha wings of $\sim1000$~\kms only if the corresponding Ly$\beta$ line has a FWHM of $\sim150$~\kms. Therefore, detecting a higher Ly$\beta$/\Ha width ratio would argue against Raman scattering from higher Lyman series emission as the primary origin of the observed broad wings.

\begin{figure*}
    \centering
    \includegraphics[width=0.98\textwidth]{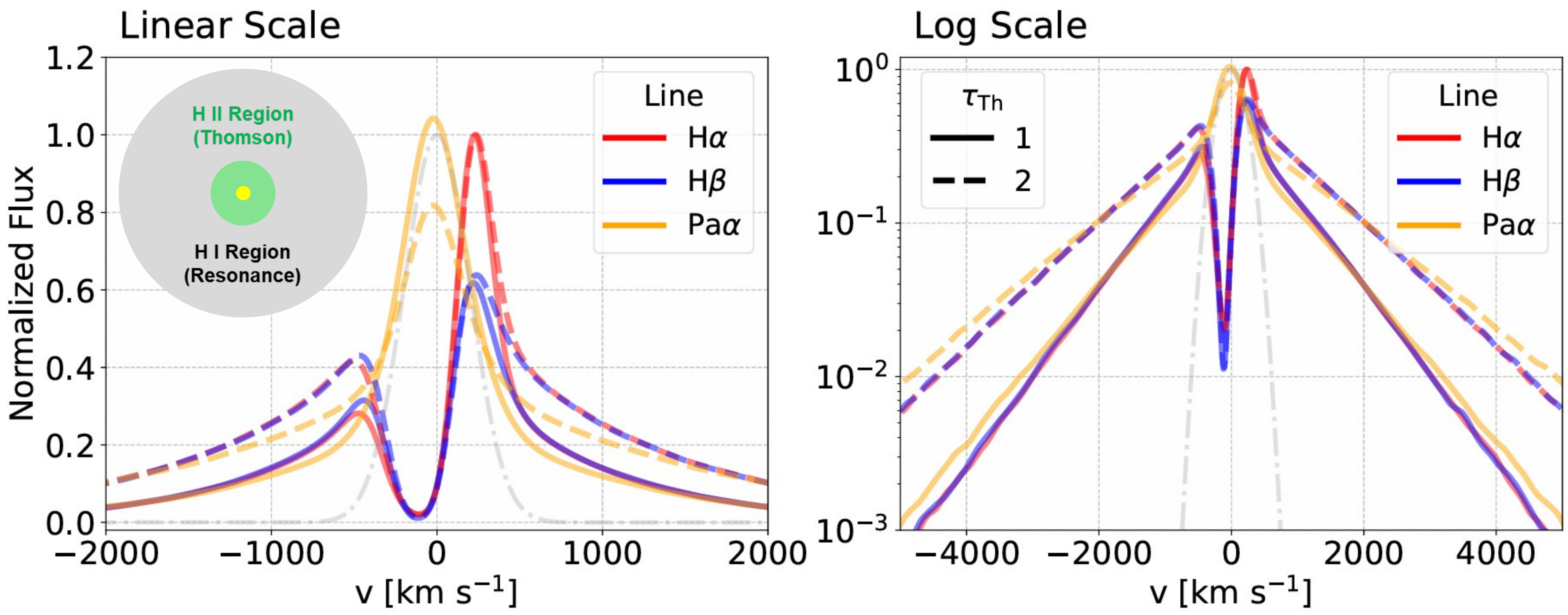}
    \caption{
    Spectra of H$\alpha$ (red), H$\beta$ (blue), and Pa$\alpha$ (orange) for a model including both Thomson and resonance scattering, shown for two Thomson optical depths: $\tau_{\rm Th} = 1$ (solid) and 2 (dashed). Fluxes are normalized at $v = 2000$~km\,s$^{-1}$, assuming a spectral resolution of $R = 3000$ ($\sim$100~km\,s$^{-1}$) and an intrinsic Gaussian ($\sigma_{\rm src} = 200$~km\,s$^{-1}$; shown in gray). The scattering geometry (the sketch in the left panel) consists of a central source surrounded by a spherical region of radius $R$, with an inner ionized (\hii) region ($r < 0.01R$) producing Thomson scattering, and an outer neutral (\hi, with with significant $n = 2$ population; $0.01R < r < R$) producing resonance scattering. The \hii region has a temperature $T_e = 10^4$~K, while the \hi region is characterized by $N_{\rm HI,\,n=2} = 10^{16}$~cm$^{-2}$, an outflow velocity of 100~km\,s$^{-1}$, and a random velocity dispersion of 100~km\,s$^{-1}$. Left and right panels show linear and logarithmic flux, respectively.
    }
    \label{fig:Ha_Hb_Thomson}
\end{figure*}

\subsection{Combining Thomson and Resonance Scattering}\label{sec:Thomson_Resonance}

The spectra of hydrogen emission lines in LRDs exhibit broad components of similar width in both Balmer and Paschen lines (ranging from 1000 to 2000 \kms; \citealp{Juodzbalis2025,Brazzini2025}), along with absorption-like features near the line center, but only for the Balmer lines. As shown in Section~\ref{sec:result_resonance}, resonance scattering of \Ha and \Hb produces P-Cygni-like or symmetric double-peaked profiles in Balmer lines, while Thomson scattering contributes similarly broad wings to both Balmer and Paschen lines (Section~\ref{sec:result_Thomson}). These effects are consistent with several spectral features observed in LRDs.
Therefore, in this section, we explore how the combination of these two scattering mechanisms shapes the emergent hydrogen line profiles.

We adopt a scattering geometry with a central emission source and a spherical scattering medium of radius $R$. The central source emits hydrogen lines -- \Ha, \Hb, and Pa$\alpha$ with intrinsic widths of $\sigsrc = 200 \kms$ and flux ratios under Case B recombination. We assume that the intrinsic width is less than the electron thermal motion, $v_{\rm e,th}$ ($\approx 500 \kms$ at $10^4\, \rm K$), as the line broadening due to Thomson scattering becomes negligible when the width exceeds $v_{\rm e,th}$; we will discuss the dependence on the width of the Thomson broadening in Section~\ref{sec:black_hole_mass}.

The scattering medium is composed of two layers: an inner ionized region responsible for Thomson scattering and an outer neutral region causing resonance scattering. The inner region (within $0.01R$) is fully ionized and characterized by electron temperature $T_e$ and Thomson optical depth $\tau_{\rm Th}$. The outer region is a neutral \hi gas with the 2s population, characterized by column density \NHItwo, outflow velocity \vexp, and turbulent velocity \sigr. Here, we assume a dense and compact \hii region, as a \hii region would be highly ionized to be Thomson thick. Thus, Thomson scattered photons in this compact region act like the broad incident radiation from the central source. Note that the formation of \Ha and \Hb line profiles depends on the radius of the \hii region relevant for Thomson scattering (see Appendix~\ref{sec:Thomson_Resonance_mode}).
In addition, if an outflow velocity of \hii region is comparable to or exceeding the electron thermal speed, the Thomson-scattered wings become asymmetric (Appendix~\ref{sec:thomson_outflow}, see also \citealp{Auer1972,Huang2018}).

Intrinsic photons emitted from the central source first undergo Thomson scattering in the inner \hii region. Then, \Ha and \Hb photons experience additional resonance scattering in the outer \hi layer. In contrast, Pa$\alpha$ photons do not experience resonance scattering, as the $n = 3$ level has a much shorter lifetime than the $n = 2$ state (see Fig.~\ref{fig:energy_level}) and hence is unlikely to be overpopulated.

Fig.~\ref{fig:Ha_Hb_Thomson} shows the emergent spectra of \Ha, \Hb, and Pa$\alpha$ for $\tau_{\rm Th} = 1$ and 2, which correspond to electron column density $\approx 1.5 \times 10^{24}$ and $3.0 \times 10^{24} \unitNHI$, respectively. Both \Ha and \Hb exhibit P-Cygni-like profiles caused by resonance scattering and develop broad wings due to Thomson scattering. Pa$\alpha$ shows a single-peaked profile centered at zero velocity, consistent with no resonance scattering.
In the left panel (linear scale), the core profiles of \Ha and \Hb ($|v| < 500 \kms$) do not depend on $\tau_{\rm Th}$. In the right panel (log scale), the line wings ($|v| > 500 \kms$) are significantly enhanced with increasing $\tau_{\rm Th}$, demonstrating the dominant role of Thomson scattering in producing broad wings in this case.

Interestingly, the red peak of the emergent \Ha is more prominent than that of \Hb (cf. left panel of Fig.~\ref{fig:Ha_Hb_Thomson}). This is because multiple resonance scatterings convert \Hb photons into Pa$\alpha$ and \Ha (as discussed Section~\ref{sec:result_resonance_Hb}). 
This enhanced red peak of H$\alpha$ with respect to H$\beta$ is also seen in observations \citep{Juodzbalis2024a,Brazzini2025}, indicating the importance of Balmer resonance scattering.

Another noteworthy spectral feature is the similarity of the absorption features in the blue parts of the spectra (cf. Fig.~\ref{fig:Ha_Hb_Thomson}). 
This similarity is not seen in observations, where typically the \Ha absorption-like feature is weaker than that of \Hb.
This discrepancy likely results from the fact that \Ha emission may originate not only from regions embedded in optically thick gas but also from additional regions less affected by strong scattering. The absence of a broad component of [O III] forbidden line \citep{Juodzbalis2024a,Juodzbalis2025,Ji2025,Ji2025b,Maiolino2025}, suggests that partial hydrogen emission originates from regions not enshrouded by optically thick gas\footnote{Note that the broad wing feature and P-Cygni-like profile of He I line at 10830 \AA\, observed in LRDs \cite{Juodzbalis2024a,Lin2025} can also be due to scattering processes.}. If these outer regions produce hydrogen lines via collisional excitation and have a high \Ha/\Hb ratio, the resulting \Ha profile will show weaker absorption.
This additional emission can `fill' the line center, thus producing dissimilar spectral features. \\
 
 Quantifying the (dis)similarity in the observations, and the comparison to detailed multi-frequency radiative transfer models will help to characterize the scattering medium of LRDs and help to disentangle the potentially multiple emission sources.
 In future work, we will expand this study using a larger simulation library to fit observed LRD spectra, considering different scattering geometries and additional emission sources. Furthermore, the high-resolution spectroscopy of Balmer lines in LRDs is required to understand their formation.

\subsection{Formation of Hydrogen Balmer Lines in Optically Thick, Dense Gas}\label{sec:balmer_line_formation}

Although our simulations only consider resonance scattering of two Balmer lines, combining Thomson and resonance scattering already produces a wide diversity of line profiles. Under the Case B recombination assumption (on-the-spot approximation), higher Lyman series lines (Ly$\beta$, Ly$\gamma$, etc.) are assumed to be optically thick and to convert into lower-series lines via interaction with neutral hydrogen.
If the \hi gas has a high $n = 2$ population and is optically thick to all Balmer transitions (i.e., on-the-spot approximation for Balmer lines), only \Ha photons can escape, due to the metastable nature of the 2s state. This modifies the Balmer line ratios. Therefore, various \Ha/\Hb ratios observed in LRDs \citep{Taylor2025,Brooks2025} may be due to radiative transfer effects, rather than dust extinction.

The collisional process in a dense gas, such as the broad line region in AGN and stellar atmosphere, affects the formation of Balmer lines \citep{Drake1980,Collin_Souffrin1982,Korista2004,Ferland2017}.
At a high electron number density ($n_{\rm e} > 10^{8}\, \rm cm^{-3}$), the collisional de-excitation is not negligible, thus electrons in the $n = 4$ states are easier de-excited than in the $n = 3$ state. 
It causes a difference between intrinsic \Ha and \Hb profiles, and their emission region is not identical.
Thus, if the Thomson scattering originates from a dense gas (e.g, \citealp{Begelman2025}), \Ha and \Hb emission experience Thomson scattering with different optical depth, resulting in a difference in strength of the broad component around the hydrogen emission lines \citep{Brazzini2025}.

This extreme gas condition may be relevant not only for LRDs and AGN, but also for other objects where \Ha shows strong emission while higher Balmer lines show an absorption-dominated feature. In future studies, we will investigate the formation of Balmer lines under these conditions, including the effects of scattering on higher-order Balmer lines such as H$\gamma$ and H$\delta$.

\subsection{Overestimating $M_{\rm BH}$ due to Thomson scattering}\label{sec:black_hole_mass}

\begin{figure*}
    \centering
    \includegraphics[width=0.98\textwidth]{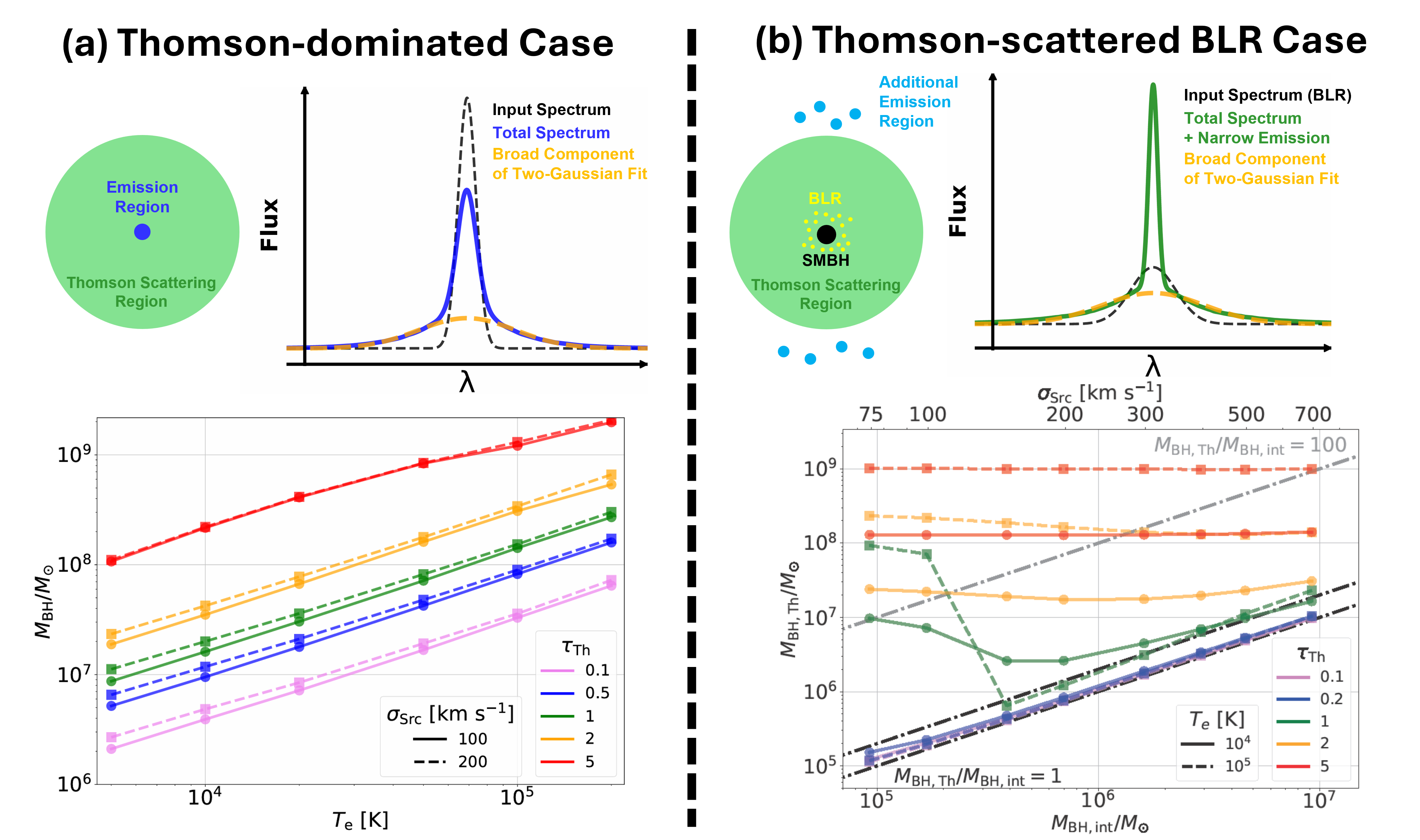}
    \caption{
Schematic illustrations (top) and black hole mass estimates (bottom) for two scenarios involving Thomson scattering. \textit{Left: Thomson-dominated case} — A central source embedded in an ionized, optically thick medium produces electron-scattered profiles with exponential wings. The resulting spectra are fit with two Gaussians to estimate $M_{\rm BH}$ as a function of electron temperature and optical depth. \textit{Right: Scattered-BLR case} — A virialized BLR emits through a Thomson-scattering in \hii region, with an added narrow line component. The bias in $M_{\rm BH}$ due to scattering is shown as a function of intrinsic mass. See text for model details.
    }
    \label{fig:Thomson_fitting}
\end{figure*}

Emission line profiles in AGN spectra are essential for estimating the supermassive black hole mass \massbh with the naive expectation being $\massbh\propto {\rm FWHM}^2$.
 Using samples of (low redshift) black holes where the mass is precisely determined through reverberation mapping, as well as the Balmer line widths are measured, one can calibrate the \massbh-FWHM relation empirically. 
This has been done in numerous studies \citep{Greene2005, Reines2015,Woo2015}. 
Specifically, \citet{Reines2015} found
\begin{equation}
    \log M_{\rm BH} = 6.60 
    + 0.47 \log \left( \frac{L_{\rm H\alpha}}{10^{42} \, \rm erg\, s^{-1}} \right)
    + 2.06 \log \left( \frac{{\rm FWHM}_{\rm H\alpha}}{10^3\, \rm km\, s^{-1}} \right)
\end{equation}
\label{eq:bh_mass}
where $L_{\rm H\alpha}$ and ${\rm FWHM}_{\rm H\alpha}$ are the luminosity and FWHM of the \Ha broad component, respectively.

However, as mentioned above, this relation is calibrated using a local (mostly $z<0.055$) AGN sample and its applicability to LRDs -- presumably AGN at high redshift -- is still uncertain. Moreover, Balmer absorption features commonly seen in LRDs are rare in local AGNs \citep[][]{Hall2007,Lin2025}, suggesting denser or more complex gas environments.
Additionally, the exponential shape of the broad wings observed in LRDs (which is typically absent in local AGNs) suggests Thomson scattering may contribute. Therefore, it is important to evaluate how Thomson scattering could bias \massbh measurements in such systems. 

In this section, we assess how Thomson scattering affects black hole mass estimates under two scenarios: (1) a Thomson-dominated case, and (2) a Thomson-thick broad line region (BLR) case. These two setups are illustrated in Fig.~\ref{fig:Thomson_fitting}.
Note that a similar analysis was recently carried out by \citet{Rusakov2025} who assumed that a narrow line emission region is embedded within Thomson-thick gas. Here, we take a more agnostic approach to the intrinsic emission mechanism of the Balmer lines.

\subsubsection{Thomson-dominated case}

In the first scenario, the Thomson-dominated case, we assume that the broad component of \Ha originates from only Thomson scattering. The \Ha emission region is surrounded by free electrons (i.e., an \hii region). The intrinsic emission undergoes Thomson scattering and develops broad wings, which are similar to the spherical geometry explored in Section~\ref{sec:result_Thomson}.
We consider two intrinsic line widths, \sigsrc = 100 and 200 \kms, and run simulations in a range of electron temperatures $T_e = 5\times 10^3$–$2\times 10^5\, \rm K$ and Thomson optical depths $\tau_{\rm Th} = 0.1-5$ (corresponding to electron densities of $n_{\rm e} \approx 5 \times 10^4  - 2.5 \times 10^6 \, (R / 10\,\mathrm{pc})^{-1}$)\footnote{As LRDs exhibit a blackbody temperature $\approx 5000-6000\, \rm K$, and the dense gas emitting blackbody radiation is proposed as a Thomson scattering medium \citep{Begelman2025}, we consider a lower electron temperature ($<10^4 \, \rm K$).}.

Although the system contains no actual black hole as the broad component originates from Thomson scattering, not AGN broad line region, we fit the emergent line profiles using two Gaussian components to estimate an artificial black hole mass \massbh.
The top left panel of Fig.~\ref{fig:Thomson_fitting} shows an example of this fitting process. The input spectrum (black dashed) is a narrow Gaussian line with width \sigsrc, while the total emergent spectrum (blue solid) includes a broad wing from Thomson scattering. We fit the total spectrum using a two-Gaussian model
\begin{eqnarray}\label{eq:two_gaussian}
    F_{\rm GG}(v) &=& F_{\rm G1} + F_{\rm G2} \\ \nonumber
    &=& \frac{A_1}{\sqrt{2 \pi \sigma_1^2}} \exp\left(-\frac{v^2}{2\sigma_1^2} \right)
    + \frac{A_2}{\sqrt{2 \pi \sigma_2^2}} \exp\left(-\frac{v^2}{2\sigma_2^2} \right),
\end{eqnarray}
where $F_{\rm G1}$ and $F_{\rm G2}$ represent the narrow and broad components, respectively, and we impose $\sigma_2 > \sigma_1$ in the fitting process. An example fit for the broad component $F_{\rm G2}$ is shown as the orange dashed line in the top part of Fig.~\ref{fig:Thomson_fitting}.

To estimate the black hole mass \massbh from the broad spectral component, we fix the total line luminosity $L_{\rm H\alpha} = 10^{42}\, \rm erg\, s^{-1}$ consistent with typical LRDs \citep[$L_{\rm H\alpha} \sim 10^{42-43} \, \rm erg\, s^{-1}$,][]{Matthee2024,Juodzbalis2025}. The luminosity of the broad component is then $A_2 / (A_1 + A_2) \times L_{\rm H\alpha}$. Assuming a Gaussian, we calculate the FWHM of the broad component as $2.355\, \sigma_2$.
Combining this with Eq.~\eqref{eq:bh_mass}, the black hole mass estimate becomes:
\begin{equation} \label{eq:bhmass_fit}
    \log M_{\rm BH} \approx 6.60 
    + 0.47 \log \left( \frac{A_2}{A_1 + A_2} \right)
    + 2.06 \log \left( \frac{2.355\, \sigma_2}{10^3\, \rm km\, s^{-1}} \right).
\end{equation}
We apply this equation across simulations with various $T_e$ and $\tau_{\rm Th}$, for both \sigsrc = 100 and 200 \kms.

The bottom left panel of Fig.~\ref{fig:Thomson_fitting} shows the resulting inferred \massbh values. Note that since there is no intrinsic broad component in this model, there is no black hole and thus no correct black hole mass. The inferred \massbh increases with both $T_e$ and $\tau_{\rm Th}$, while dependence on \sigsrc is negligible. When $T_e > 10^5\, \rm K$ or $\tau_{\rm Th} > 1$, \massbh exceeds $10^7\, M_{\odot}$, despite no intrinsic virial motion.
This result demonstrates that Thomson scattering alone can mimic a broad component, leading to significant overestimation of \massbh through Gaussian fitting. For reference, if the narrow component represents emission from a genuine BLR, an intrinsic velocity of \sigsrc = 100 \kms would correspond to $M_{\rm BH,int} \approx 10^5\, M_{\odot}$. Therefore, if the broad Balmer components of LRDs are shaped by Thomson scattering in ionized gas ($T_e > 10^4\, \rm K$; \citealp{Rusakov2025}), the inferred \massbh may be overestimated by a factor of $> 10$.

\subsubsection{Thomson-scattered broad line region case}

In the second scenario -- the Thomson-scattered BLR case -- we consider a geometry similar to the first, with a central emission region surrounded by an \hii medium. However, the emission region is treated as the broad line region (BLR) of an AGN, where the gas is virialized by a central supermassive black hole.

The top-right panel of Fig.~\ref{fig:Thomson_fitting} shows an example spectrum for this setup. To cover a wide range of intrinsic \massbh, we adopt $\sigsrc = 75$–$700 \kms$ (black dashed line), corresponding to $M_{\rm BH,int} = 10^{4-7} M_{\odot}$ at fixed $L_{\rm H\alpha} = 10^{42} \, \rm erg\, s^{-1}$, based on the \massbh scaling relation. We set $T_e = 10^4$ and $10^5 \, {\rm K}$ and explore $\tau_{\rm Th} = 0.1–5$.
We also include an additional emission region located outside the Thomson scattering medium, representing a narrow-line region, star-forming region, or outflow. This component has a fixed width of $50 \kms$ and contributes half the flux of the intrinsic BLR emission. The total emergent spectrum (green solid line in the top right panel of Fig.~\ref{fig:Thomson_fitting}) thus consists of Thomson-scattered BLR emission and a narrow Gaussian component.
As previously, we fit the total spectrum with two-Gaussian profiles in Eq.~\eqref{eq:bh_mass} and estimate \massbh using Eq.~\ref{eq:bhmass_fit}, assuming the total \Ha luminosity fixed at $10^{42} \, \rm erg\, s^{-1}$.

The bottom right panel of Fig.~\ref{fig:Thomson_fitting} shows the resulting \massbh estimates $M_{\rm BH,Th}$ as a function of the intrinsic black hole mass $M_{\rm BH,int}$. The intrinsic masses are obtained by fitting spectra with no scattering (i.e., $\tau_{\rm Th} = 0$).
The Thomson scattering fraction is given by $1 - \exp(-\tau_{\rm Th})$. In the optically thin regime ($\tau_{\rm Th} < 1$), the scattering fraction is small, and $M_{\rm BH,Th} \approx M_{\rm BH,int}$; most scattered photons contribute only weakly to the observed profile. However, even in this case, an exponential wing extending beyond $1000 \kms$ is present due to scattering.

In the optically thick regime ($\tau_{\rm Th} \gtrsim 1$), the scattering fraction becomes significant ($\approx 0.86$ at $\tau_{\rm Th} = 2$), and Thomson scattering dominates the broad component formation. This leads to a nearly flat $M_{\rm BH,Th}$ curve, with mass overestimation by more than an order of magnitude.
At intermediate optical depth ($\tau_{\rm Th} \approx 1$), the overestimation depends on the intrinsic line width. For narrow intrinsic width ($\sigsrc < 200 \kms$), the scattered photons are interpreted as the primary broad component. For a broader width ($\sigsrc > 200 \kms$), the scattering causes additional broadening, leading to \massbh overestimates by factors of $\sim 2$.
Consequently, if Thomson scattering occurs in the gas surrounding the BLR, the observed line width can be significantly broadened by thermal electron motions, resulting in overestimated black hole masses.

\subsubsection{Implications for Line Profiles and \massbh Estimates}

In summary, this section investigates the effect of Thomson scattering on black hole mass measurements.
We explored two scenarios: (1) a Thomson-dominated wing and (2) a Thomson-scattered BLR. In both cases, we estimated \massbh using the standard scaling relation calibrated from reverberation mapping of local AGN.
In the first case, we showed that pure Thomson scattering can generate broad wings around Balmer lines, resulting in fictitious black hole masses exceeding $10^6 M_{\odot}$ when $T_e \ge 10^4 \, \rm K$ and $\tau_{\rm Th} \ge 0.1$.
In the second case, we demonstrated that scattering of BLR emission broadens the observed profile and leads to overestimation of \massbh, particularly when $\tau_{\rm Th} \ge 1$.
Although Thomson scattering has a weaker impact on \massbh estimates in the optically thin regime ($\tau_{\rm Th} < 1$), it still produces exponential wings in the far line wings, as seen in local AGNs \citep[e.g.,][]{Laor2006}.

The role of Thomson scattering in LRDs remains debated. \citet{Rusakov2025} proposed that the broad \Ha features in LRDs are dominated by Thomson scattering, suggesting that intrinsically narrow emission lines with widths $\sim100-200 \kms$ are broadened through scattering in dense, ionized gas, which is similar to our first scenario. In contrast, \citet{Brazzini2025} argued that the broad hydrogen line features in LRDs are inconsistent with this Thomson-dominated scenario. They suggest that if scattering dominates, the \Ha scattering fraction should be higher than that of \Hb due to dust extinction. However, as discussed in Section~\ref{sec:balmer_line_formation}, the Thomson optical depth of \Hb can be higher than that of \Ha in dense environments, complicating this interpretation.
These contrasting views highlight the need for detailed radiative transfer modeling to understand the formation of broad emission features in LRDs.

Our scenarios have important implications for LRDs, where Balmer absorption features suggest dense gas environments. Thomson scattering can naturally produce exponential wings and apparent line broadening, potentially biasing \massbh estimates. In addition, if the bulk velocity of the scattering medium is faster than the electron thermal speed, Thomson scattering can provide asymmetric wings on emission lines, depending on the velocity structure of the gas.  However, line profile analysis alone is insufficient to distinguish between kinematic and scattering origins of broad wings. Since Thomson scattering induces strong polarization \citep{Lee1999,Kim2007}, future spectropolarimetric observations of LRDs are a potential avenue to test this hypothesis.

\section{Conclusion}
In this study, we explored the physical origin of the hydrogen line profiles observed in Little Red Dots, which are compact, high-redshift sources identified through JWST spectroscopy. These systems exhibit broad wings (FWHM $\gtrsim 1000$~\kms) in Balmer and helium emission lines, along with Balmer absorption features and a strong Balmer break, indicating a dense, neutral hydrogen medium in the $n = 2$ state.
Under such conditions, radiative transfer effects may affect the formation of hydrogen lines.

We performed 3D Monte Carlo radiative transfer simulations to examine three key scattering mechanisms that shape hydrogen line profiles: (1) resonance scattering by hydrogen in the $n = 2$ state, (2) Raman scattering of UV photons by ground-state hydrogen ($n = 1$), and (3) Thomson scattering by free electrons. For \Hb, we explicitly include multi-branching transitions in the resonance scattering process, accounting for both $n = 4 \rightarrow 2$ and $n = 4 \rightarrow 3$ decay channels. These pathways allow \Hb photons to convert into Pa$\alpha$ and \Ha, significantly suppressing the emergent \Hb flux. For Raman scattering, we adopt wavelength-dependent cross sections and branching ratios near the Lyman series. For Thomson scattering, we explore a range of electron temperatures ($T_e$) and optical depths ($\tau_{\rm Th}$).

Our results from each scattering process show that:
\begin{itemize}

\item Resonance scattering explains the distinct profile differences between \Ha, \Hb, and Pa$\alpha$ and leads to photon conversion from \Hb to Pa$\alpha$ and \Ha, altering flux ratios (i.e., leading to large deviations from the value under the case B recombination, see Fig.~\ref{fig:ratio}) and absorption depths. We have demonstrated how these effects imprint the gas geometry and kinematics on the emergent line shapes (Section~\ref{sec:result_resonance}), and will explore their diagnostic power systematically in future work.

\item While resonant scattering can significantly alter (and for very large optical depths) widen the emergent \Ha spectrum significantly, this is not the case for \Hb due to the efficient conversion to Pa$\alpha$ (Section~\ref{sec:discussion_resonance}). 

\item Raman scattering can produce extremely broad wings (up to $\sim 10^4$~\kms), with widths scaling approximately as ${\rm FWHM} \propto \sqrt{N_{\rm HI}}$ and different degrees of broadening for each Balmer line (Fig.~\ref{fig:width_broad_wing} and Section~\ref{sec:result_Raman}).
We show that the observed Balmer line widths are not consistent with these continuum Raman scattered features. In addition,  the UV continuum in LRDs is likely too faint, effectively ruling out Raman scattered continuum photons to be the origin of the broad Balmer lines observed (Section~\ref{sec:discussion_raman}).

\item
However, we also show that if higher Lyman series with the width $\sim 150 \kms$ are fully Raman scattered, the width of the Raman wing around \Ha and \Hb is over 1000 \kms, with a FWHM ratio of \Ha and \Hb is $\approx 1.28$ (Fig.~\ref{fig:Raman_width_Gau} and Appendix~\ref{sec:Raman_emission}).

\item Thomson scattering produces symmetric broad wings with similar widths across all hydrogen lines. In the optically thin regime, the wings exhibit exponential profiles, with widths scaling as ${\rm FWHM} \propto \sqrt{T_e}$ (Figs.~\ref{fig:width_broad_wing} and \ref{fig:spec_Thomson}). At higher optical depths ($\tau_{\rm Th} \gtrsim 2$), the wings deviate from the exponential shape due to multiple scatterings, and their width further increases with $\tau_{\rm Th}$ (Figs.~\ref{fig:spec_Thomson} and \ref{fig:Thomson_width}).
\end{itemize}

We show that a combination of Thomson and resonant scattering can explain most spectral features observed in LRDs, specifically the broad wings and the central asymmetric double peak (Section~\ref{sec:Thomson_Resonance}).
However, to understand different absorption features and broad wings of Balmer lines, the additional radiative transfer effects, such as on-the-spot approximation of Balmer lines, collisional excitation, and collisional de-excitation, are required (Section~\ref {sec:balmer_line_formation}).

Importantly, we demonstrate that if the broad component of Balmer lines is shaped by Thomson scattering rather than intrinsic broad emission, then black hole masses estimated using methods for local AGNs can be significantly overestimated (Section~\ref{sec:black_hole_mass}). For optically thick cases ($\tau_{\rm Th} \ge 2$), the overestimation can exceed a factor of 10 (Fig.~\ref{fig:Thomson_fitting}).

Our results show the importance of detailed radiative transfer modeling when interpreting hydrogen line profiles, not only for LRDs and AGN, but also for other astrophysical systems exhibiting broad hydrogen wings and absorption features. In future work, we will apply our simulation library to analyze observed spectra of LRDs and expand our models to include more complex environments, such as those combining resonance and Thomson scattering in multi-phase media.

\begin{comment}

\section{Notes}
missing:
\begin{itemize}
    \item general RT effects of Ha,Hbeta line: kinematics, column density variation (-->like fig 2 now but also v=0)
    \item widening of Ha line --> maybve later?
    \begin{itemize}
        \item proof of concept with fc~fcrit, large sigmacl; can widen line purely RT
        \item clumpy medium with fc>>1 (~fccrit) with large motion but no LOS motion --> narrow absorption but widening
    \end{itemize}
    \item Hbeta %(small point of paper, 1 fig example)
        \item fix fig 1
        \item predictions for Pa$\alpha$
        \item ratio Ha/Hbeta changed?
        \item RT -> Hakinematics, interesting
    \item proper level 2 population --> not taken into account here, we do calculations with $N_{2s}$ only. here: need to discuss, future work. 
\end{itemize}
\end{comment}

\section*{Acknowledgements}
The authors thank the anonymous referee for constructive comments, which improved the clarity of this paper.
SJC acknowledges support from the ERC synergy grant 101166930 - RECAP. 
MG thanks the Max Planck Society for support through the Max Planck Research Group, and the European Union for support through ERC-2024-STG 101165038 (ReMMU). JM acknowledges funding by the European Union (ERC, AGENTS,  101076224).
CAM acknowledges support by the European Union ERC grant RISES (101163035), Carlsberg Foundation (CF22-1322), and VILLUM FONDEN (37459).
Computations were performed on HPC systems Freya and Orion at the Max Planck Computing and Data Facility.

%%%%%%%%%%%%%%%%%%%%%%%%%%%%%%%%%%%%%%%%%%%%%%%%%%
\section*{Data Availability}
The data of this study is available upon reasonable request to the first author.

%%%%%%%%%%%%%%%%%%%% REFERENCES %%%%%%%%%%%%%%%%%%

% The best way to enter references is to use BibTeX:

\bibliographystyle{mnras}
\bibliography{example} % if your bibtex file is called example.bib

% Alternatively you could enter them by hand, like this:
% This method is tedious and prone to error if you have lots of references
%\begin{thebibliography}{99}
%\bibitem[\protect\citeauthoryear{Author}{2012}]{Author2012}
%Author A.~N., 2013, Journal of Improbable Astronomy, 1, 1
%\bibitem[\protect\citeauthoryear{Others}{2013}]{Others2013}
%Others S., 2012, Journal of Interesting Stuff, 17, 198
%\end{thebibliography}

%%%%%%%%%%%%%%%%%%%%%%%%%%%%%%%%%%%%%%%%%%%%%%%%%%

%%%%%%%%%%%%%%%%% APPENDICES %%%%%%%%%%%%%%%%%%%%%

\appendix

\section{Raman scattering cross section}\label{sec:Raman_scattering_cross_section}

\begin{figure*}
    \centering
    \includegraphics[width=0.98\textwidth]{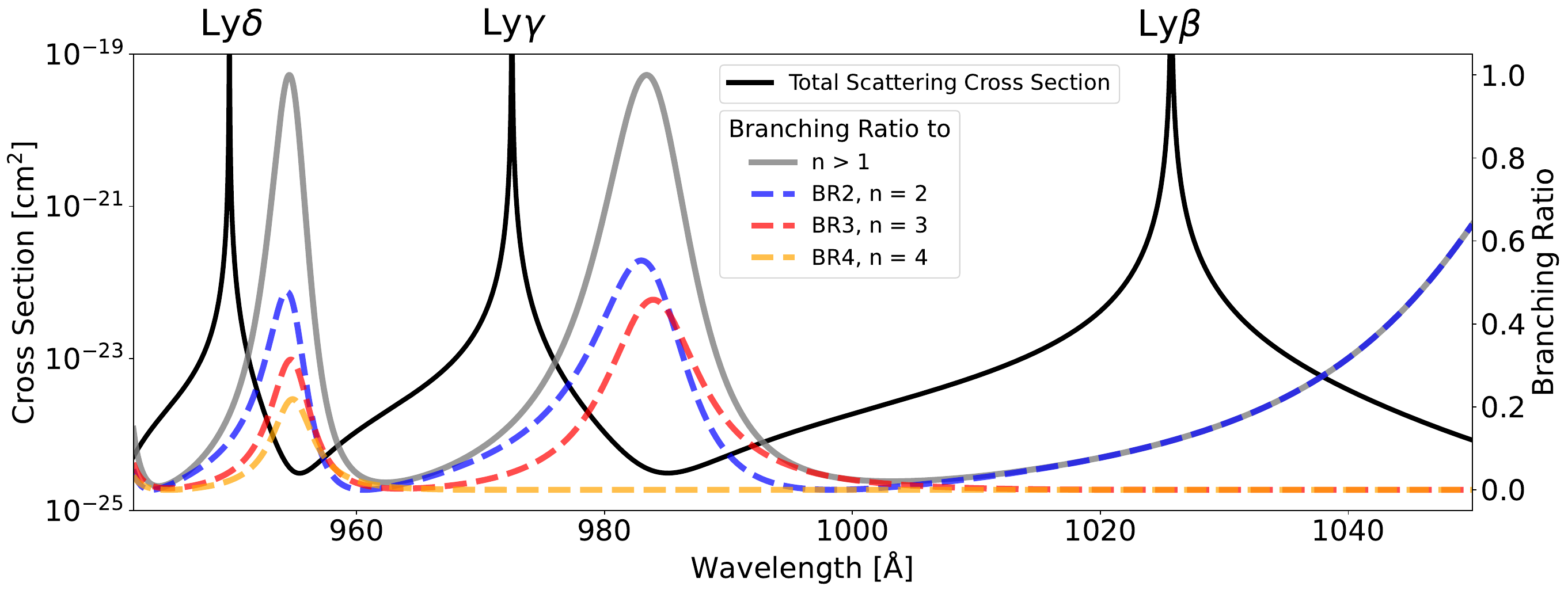}
    \caption{Scattering cross section and branching ratios of UV radiation near Ly$\beta$, Ly$\gamma$, and Ly$\delta$ as functions of wavelength. The black solid line represents the total scattering cross section computed using the Kramers--Heisenberg formula. The colored dashed lines show the branching ratios for Raman scattering to $n=2$ (blue), $n=3$ (red), and $n=4$ (orange) states.}
    \label{fig:Raman_cross_section}
\end{figure*}

Fig.~\ref{fig:Raman_cross_section} shows the total scattering cross section for both Rayleigh and Raman scatterings near Ly$\beta$, Ly$\gamma$, and Ly$\delta$, as well as the branching ratios for Raman scattering to $n=2$, $3$, and $4$ states, computed using the Kramers--Heisenberg formula \citep{Lee2013,Chang2015}.

\section{Raman Broad Features by Lyman Series Emission}\label{sec:Raman_emission}

\begin{figure}
    \centering
    \includegraphics[width=0.48\textwidth]{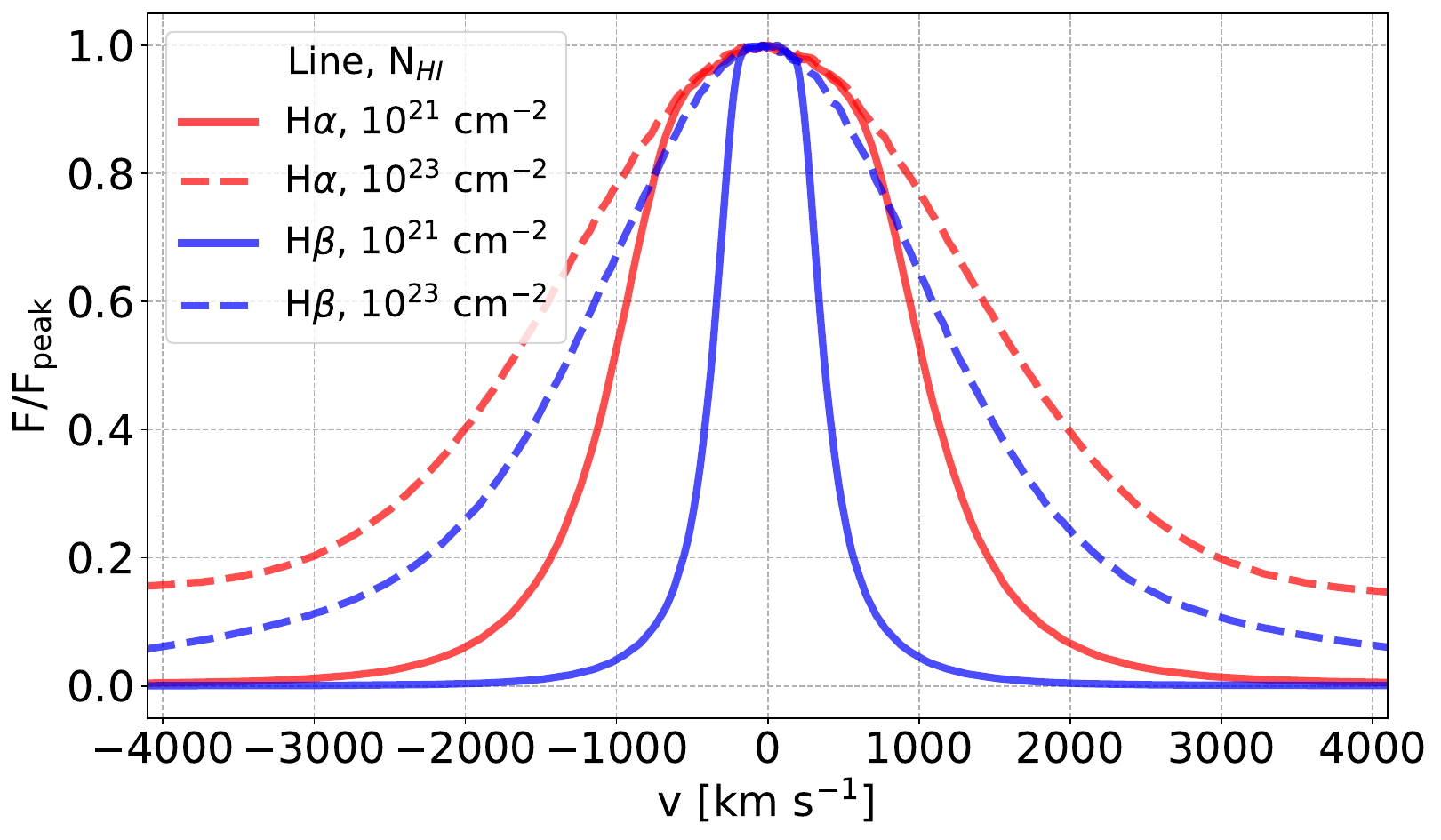}
    \caption{
    Raman-scattered features around \Ha (red) and \Hb (blue) produced by Ly$\beta$ and Ly$\gamma$ emission lines. The input Lyman-series lines are assumed to have intrinsic widths of $200 \kms$ and an equivalent width of 10~\AA. Solid and dashed lines correspond to low ($10^{21} \unitNHI$) and high ($10^{23} \unitNHI$) column densities, respectively.
    }
    \label{fig:Raman_wing_Gau}
\end{figure}

\begin{figure}
    \centering
    \includegraphics[width=0.48\textwidth]{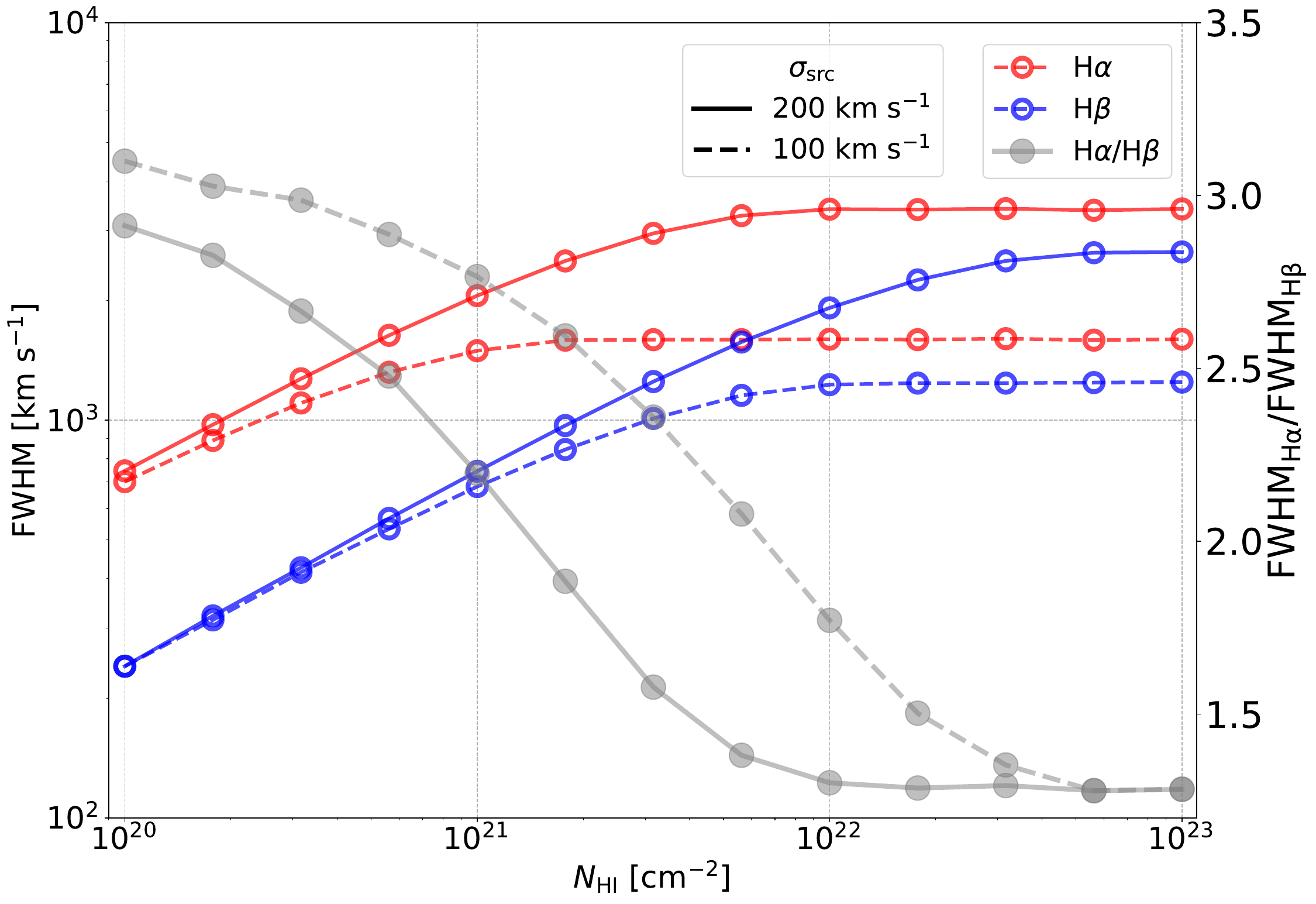}
    \caption{
    FWHM of Raman features around \Ha and \Hb, and their ratio, as a function of \NHI. At high column density ($\sim 10^{23} \unitNHI$), where Lyman-series photons are fully converted into Raman features, the \Ha/\Hb width ratio approaches the theoretical value of $\sim 1.28$ (from Eq.~\ref{eq:Raman_broad}). At low \NHI ($\sim 10^{20} \unitNHI$), the ratio is $\sim 3$, reflecting the wavelength dependence of the Raman scattering cross section, as seen in the flat continuum case (Fig.~\ref{fig:width_broad_wing}).
    }
    \label{fig:Raman_width_Gau}
\end{figure}

As discussed in Section~\ref{sec:result_Raman} (cf. Fig.~\ref{fig:width_broad_wing}), Raman-scattered \Ha wings are typically $\sim 3$ times broader than those of \Hb when the input radiation is a flat UV continuum. This difference arises from the wavelength-dependent scattering cross sections near Ly$\beta$ and Ly$\gamma$. However, when the input radiation contains strong Ly$\beta$ and Ly$\gamma$ emission lines, the width ratio between Raman \Ha and \Hb features can be significantly lower. In particular, if both lines are fully converted into Raman-scattered photons at high \NHI, the resulting widths are determined by the intrinsic line profile and the Raman wavelength shift, as given by Eq.~\eqref{eq:Raman_broad}. In this case, the width ratio between \Ha and \Hb converges to $\sim 1.28$.

Fig.~\ref{fig:Raman_wing_Gau} shows Raman-scattered \Ha and \Hb from Lyman series emissions for two representative column densities: $10^{21}$ and $10^{23}\unitNHI$. We assume a Gaussian input profile for both Ly$\beta$ and Ly$\gamma$ with \sigsrc = 200~\kms. At low \NHI, the \Ha Raman wing is $\sim 2-3$ times broader than that of \Hb due to cross-section differences, resembling the flat continuum case shown in Fig.~\ref{fig:spec_Raman}. At high \NHI, the \Ha and \Hb profiles become comparable in width, as nearly all input photons are scattered.

To quantify the trend of Raman \Ha and \Hb widths, Fig.~\ref{fig:Raman_width_Gau} shows the full width at half maximum (FWHM) of the Raman \Ha and \Hb features, and their ratio, as a function of \NHI, for two input widths: \sigsrc = 100 and 200~\kms. At low \NHI ($< 10^{21} \unitNHI$), the scattering cross section primarily controls the profile width and favors broader \Ha wings. As \NHI increases, the conversion probability becomes saturated, and the intrinsic width \sigsrc begins to dominate the emergent profile. At high \NHI ($> 10^{22} \unitNHI$), the \Ha/\Hb FWHM ratio converges to the theoretical value of $\sim 1.28$, determined by the Raman wavelength shifts.

When Lyman-series emission lines of equal intrinsic width are fully Raman scattered, the width of the resulting Raman features is set by the broadening factor in Eq.~\eqref{eq:Raman_broad}.
As noted in Section~\ref{sec:Raman_scattering_mechanism}, the additional broadening in velocity space is
\begin{equation}
 \gamma = \frac{\Delta v_{\rm sc}}{\Delta v_{\rm inc}} = \frac{\lambda_{\rm sc}}{\lambda_{\rm inc}},
\end{equation}
where $\lambda_{\rm sc}$ and $\lambda_{\rm inc}$ are the wavelengths of the scattered and incident photons, respectively.
Table~\ref{tab:Raman_broadening} lists $\gamma$ for each transition, along with the H~I column density $N_{\rm HI, 100}$ at which an incident Lyman line of 100~\kms\ width is fully Raman scattered.
For column densities below $N_{\rm HI, 100}$, the FWHM ratio between Raman \Ha and \Hb reverts to the UV continuum case ($\approx 3$ as shown in Fig.~\ref{fig:width_broad_wing}).
Notably, since both Raman \Hb and Pa$\alpha$ originate from an inelastic scattering of Ly$\gamma$, their FWHM ratio is fixed at $\approx 3.86$, which is the ratio of their respective broadening factors.

In summary, Raman scattering of Ly$\beta$ and Ly$\gamma$ emission lines leads to \Ha and \Hb broad wings with similar widths when the \hi column density is high, but different widths of Balmer and Paschen lines. This effect can mimic the symmetric wing profiles produced by Thomson scattering and must be considered when interpreting the origin of broad components in Balmer lines.

\begin{table}
    \centering
    \caption{
Raman broadening factor, $\gamma = \lambda_{\rm sc} / \lambda_{\rm inc}$, for various transitions.
The first and third columns list the incident (Lyman-series) and corresponding Raman-scattered photons, respectively.
The second column gives the H~I column density $N_{\rm HI, 100}$ required for an incident Lyman line with a width of 100~\kms\ to be fully Raman scattered.
The fourth column shows the velocity broadening factor $\gamma$, which relates the FWHM of the scattered and incident profiles.}
    \label{tab:Raman_broadening}
    \begin{tabular}{llcc}
        \hline
        Input Photon & $N_{\rm HI, 100}^*$ & Scattered Photon & $\gamma$ \\
        \hline
        \multirow{2}{*}{Ly$\beta$} & \multirow{2}{*}{$\approx 2 \times 10^{21} \unitNHI$} & \multirow{2}{*}{\Ha} & \multirow{2}{*}{6.4} \\
        & & & \\
        \hline
        \multirow{2}{*}{Ly$\gamma$}
        & \multirow{2}{*}{$\approx 10^{22} \unitNHI$}
         & \Hb & 5.0 \\
        & & Pa$\alpha$ & 19.3 \\
        \hline
        \multirow{3}{*}{Ly$\delta$}
        & \multirow{3}{*}{$\approx 5\times 10^{22} \unitNHI$}
        & H$\gamma$ & 4.6 \\
        & & Pa$\beta$ & 13.5 \\
        & & Br$\alpha$ & 42.7 \\
        \hline
    \end{tabular}
     \\
    \footnotesize $*$: \NHI for fully Raman scattered case.
\end{table}

\section{Thomson scattering in outflowing gas}
\label{sec:thomson_outflow}

\begin{figure}
    \centering
    \includegraphics[width=0.47\textwidth]{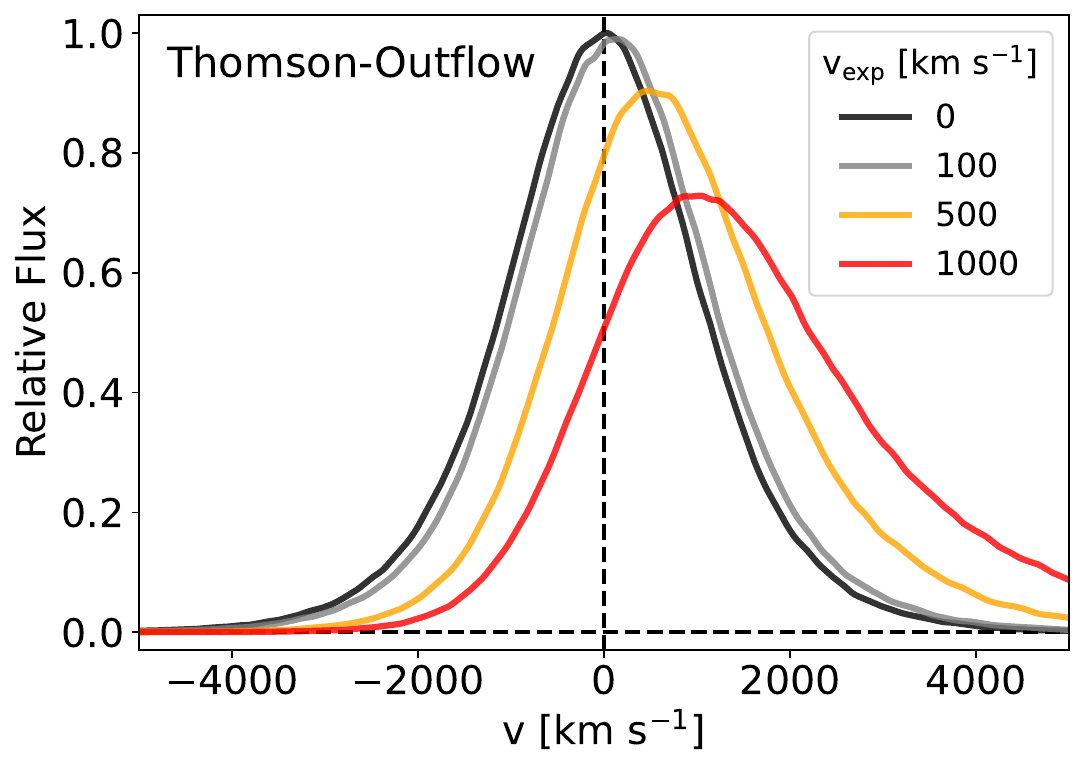}
    \caption{
    Spectra composed of Thomson-scattered photons for various outflow velocities $\vexp$, normalized by the peak flux of the static case ($\vexp = 0~\kms$).
    The line colors correspond to $\vexp = 0$ (black), 100 (grey), 500 (orange), and $1000~\kms$ (red).
    As $\vexp$ increases, the spectra become asymmetric, with enhanced red wings.
    }
    \label{fig:thomson_outflow}
\end{figure}

As discussed in Section~\ref{sec:result_Thomson}, Thomson scattering in a static medium produces symmetric broad wings around emission lines, with a width proportional to $\sqrt{T_{\rm e}}$.
However, when the scattering medium exhibits a bulk outflow comparable to or exceeding the electron thermal velocity, $v_{\rm th,e} \approx 548~\kms$ at $T_{\rm e}=10^4~{\rm K}$, Doppler shifts by the bulk motion can break the symmetry of the Thomson-scattered wings \citep[e.g.][]{Auer1972,Huang2018}. 
For this reason, we consider an outflow in the spherical \hii geometry for Thomson scattering and assume a constant radial outflow velocity, which is similar to the outflow velocity in the geometry for resonance scattering in Section~\ref{sec:resonance}.

Fig.~\ref{fig:thomson_outflow} shows Thomson-scattered spectra for various outflow velocities, $\vexp = 0$--$1000~\kms$, at an electron temperature $T_{\rm e}=10^4~{\rm K}$ and Thomson optical depth $\tauth = 1$.
At $\vexp \lesssim 100~\kms$, the wings remain nearly symmetric since the bulk velocity is much smaller than $v_{\rm th,e}$.
At higher velocities ($\vexp \gtrsim 500~\kms$), the emergent spectrum develops a pronounced red asymmetry, reflecting the Doppler shift of photons scattered by outflowing electrons.
This effect becomes significant as $\vexp / v_{\rm th,e}$ approaches unity or higher, and may serve as a diagnostic of fast, outflowing \hii regions in systems where the Balmer wings show measurable red–blue asymmetry.
Conversely, when the bulk velocity of the ionized gas is much smaller than $v_{\rm th,e}$, the scattered wings remain nearly symmetric, as shown in the spectroscopic study of P~Cygni \citep[e.g.][]{Bernat1978}.

\section{Dependence on radii of \hii and \hi regions, combining Thomson and Resonance Scattering }\label{sec:Thomson_Resonance_mode}

\begin{figure*}
    \centering
    \includegraphics[width=0.95\textwidth]{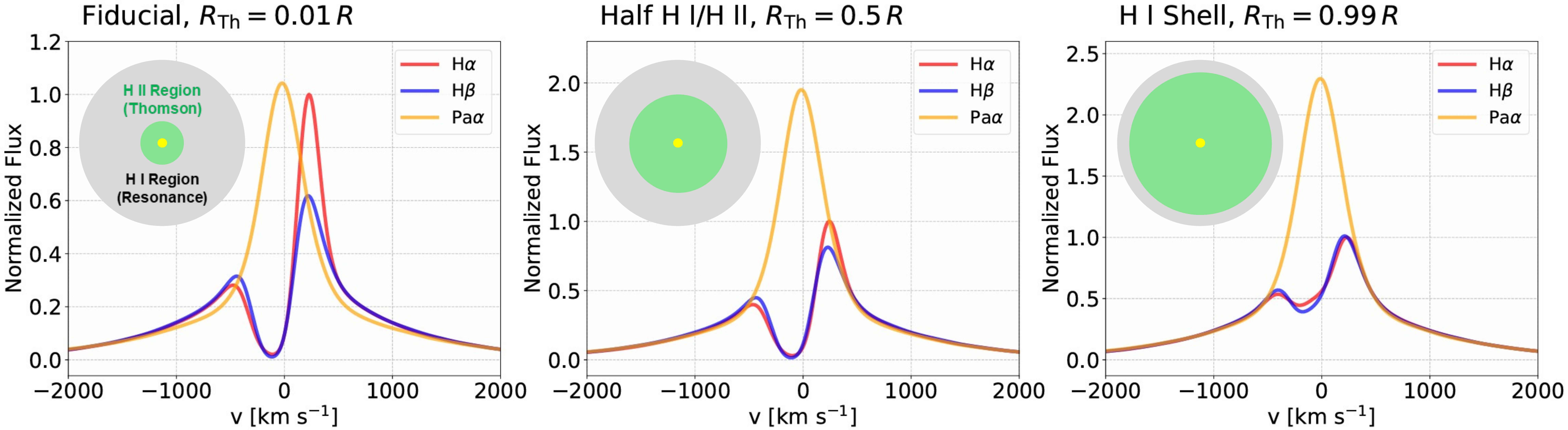}
    \caption{
    Emergent \Ha (red), \Hb (blue), and Pa$\alpha$ (orange) line profiles for different sizes of the Thomson scattering region.
    All models assume a spherical geometry with an inner ionized (\hii) region for Thomson scattering and an outer neutral (\hi) region with $n = 2$ hydrogen enabling resonance scattering.
    The flux normalization and other simulation parameters are identical to those used in Fig.~\ref{fig:Ha_Hb_Thomson}; all spectra are normalized at $v = 2000$~\kms.
    From left to right, the Thomson region extends to $R_{\rm Th} = 0.01R$ (fiducial, matching the $\tau_{\rm Th} = 1$ model in Fig.~\ref{fig:Ha_Hb_Thomson}), $0.5R$, and $0.99R$.
    As the Thomson-scattering region becomes larger, the broad wings become more prominent, and the differences between the \Ha and \Hb profiles driven by resonance scattering are reduced.
    }
    \label{fig:Ha_Hb_Thomson_model}
\end{figure*}

In this section, we examine how the size of the inner ionized region for Thomson scattering affects emergent line profiles when both Thomson and resonance scattering are active, as discussed in Section~\ref{sec:Thomson_Resonance}. Fig.~\ref{fig:Ha_Hb_Thomson_model} shows the resulting \Ha, \Hb, and Pa$\alpha$ spectra for three different Thomson region radii: $R_{\rm Th} = 0.01R$, $0.5R$, and $0.99R$, all at a fixed Thomson optical depth $\tau_{\rm Th} = 1$. The outer neutral region, enabling resonance scattering, extends from $R_{\rm Th}$ to $R$ and contains hydrogen in the $n = 2$ state. Other parameters are fixed: outflow velocity \vexp = 100~\kms, turbulent velocity \sigr = 100~\kms, intrinsic width \sigsrc = 200~\kms, and \hi column density in the 2s state \NHItwo = $10^{16} \unitNHI$, which are identical to those of spectra in Fig.~\ref{fig:Ha_Hb_Thomson}.

In Fig.~\ref{fig:Ha_Hb_Thomson_model}, the profile difference between \Ha and \Hb diminishes as the Thomson region expands. This trend arises from the interaction between geometry and scattering behavior. Since the Thomson optical depth is fixed, the emergent energy distribution from the inner \\hi region remains similar regardless of its size. However, the spatial path taken by photons after emission differs significantly.

In the small Thomson region case (e.g., $R_{\rm Th} = 0.01R$), photons quickly exit the ionized core and enter the neutral \hi shell, where they undergo resonance scattering. Because the \hii region is small, photons that scatter in the \hi shell are unlikely to re-enter the Thomson region; the effects of resonance and Thomson scattering remain spatially separated. 
In contrast, when the ionized region is large (e.g., $R_{\rm Th} = 0.99R$), photons undergoing resonance scattering near the boundary between \hi and \hii regions can easily scatter back into the ionized zone. These photons are then subject to additional Thomson scattering before escaping. This back-and-forth between the two regions effectively reduces the impact of resonance scattering, as many photons escape via Thomson scattering rather than multiple resonance scattering \hi region.

Since the line profile differences between \Ha and \Hb originate primarily from resonance scattering, especially due to branching in \Hb, this effect becomes suppressed as $R_{\rm Th}$ increases. As a result, in the large ionized region case (right panel of Fig.~\ref{fig:Ha_Hb_Thomson_model}), the \Ha and \Hb profiles appear nearly identical.

%%%%%%%%%%%%%%%%%%%%%%%%%%%%%%%%%%%%%%%%%%%%%%%%%%

% Don't change these lines
\bsp	% typesetting comment
\label{lastpage}
\end{document}